\documentclass[twocolumn]{aastex62}

\usepackage{hyperref}
\received{February 26 2021}
\revised{May 27, 2021}
\accepted{June 17 2021}
\submitjournal{ApJ}

\shorttitle{Neutral Outflow at z=2.09}
\shortauthors{Butler et al.}
\watermark{ }

\def\kms{\ifmmode \mbox{\rm km\ s}^{-1}
	\else {\rm km\ s}$^{-1}$\ignorespaces         
	\fi}
\def\mum{\ifmmode \mu\mbox{\rm m}  
	\else $\mu${\rm m}         
	\fi}
\def\Htwo{\ifmmode \mbox{\rm H}_2   
	\else {\rm H}$_2$         
	\fi}
\def\OHp{\ifmmode \mbox{\rm OH}^+   
	\else {\rm OH$^+$}         
	\fi}
\def\CHp{\ifmmode \mbox{\rm CH}^+   
	\else {\rm CH}$^+$         
	\fi}
\def\Hp{\ifmmode \mbox{\rm H}^+   
	\else {\rm H}$^+$         
	\fi}
\def\Hthreep{\ifmmode \mbox{\rm H}^+_3   
	\else {\rm H}$^+_3$         
	\fi}
\def\mHI{\relax                                      
	\ifmmode {\rm m_{\mbox{\scriptsize\rm H\,\sc I}}} 
	\else $m_{\mbox{\scriptsize\rm H\,\sc I}}$
	\fi}
\DeclareRobustCommand{\HI}{%
	\mbox{H\check@mathfonts\fontsize\sf@size\z@\selectfont I}%
}
\begin{document}

\title{Resolved Neutral Outflow from a Lensed Dusty Star-Forming Galaxy at z=2.09}

\correspondingauthor{Kirsty May Butler}
\email{kirstymaybutler@gmail.com}

\author[0000-0001-7387-0558]{Kirsty M. Butler}
\affil{Leiden Observatory, Leiden University, PO Box 9513, 2300 RA Leiden,
	the Netherlands}

\author{Paul P. van der Werf}
\affiliation{Leiden Observatory, Leiden University, PO Box 9513, 2300 RA Leiden,
	the Netherlands}

\author{Matus Rybak}
\affiliation{Leiden Observatory, Leiden University, PO Box 9513, 2300 RA Leiden,
	the Netherlands}
\affiliation{THz Sensing Group, Faculty of Electrical Engineering, Mathematics and Computer Science, TU Delft, the Netherlands}

\author{Tiago Costa}
\affiliation{Max-Planck-Institut f\"{u}r Astrophysik, Karl-Schwarzschild-Stra{\ss}e 1, D-85748 Garching b. M\"{u}nchen, Germany}

\author{Pierre Cox}
\affiliation{Sorbonne Universit{\'e}, UPMC Universit{\'e} Paris 6 \& CNRS, UMR 7095, 
Institut d'Astrophysique de Paris, 98b boulevard Arago, 75014 Paris, France}

\author{Axel Wei{\ss}}
\affiliation{Max-Planck-Institut f\"{u}r Radioastronomie, Auf dem H\"{u}gel 69 D-53121 Bonn, Germany}

\author{Micha\l{} J. Micha\l{}owski}
\affiliation{Astronomical Observatory Institute, Faculty of Physics, Adam Mickiewicz University, ul. S\l{}oneczna 36, 60-286, Pozna\'n, Poland}

\author{Dominik A. Riechers}
\affiliation{Department of Astronomy, Cornell University, Space Sciences Building, Ithaca, NY 14853, USA}

\author{Dimitra Rigopoulou}
\affiliation{Astrophysics, Department of Physics, University of Oxford, Keble Road, Oxford OX1 3RH, UK}

\author{Lucia Marchetti}
\affiliation{Department of Astronomy, University of Cape Town, Private Bag X3, 7701 Rondebosch, Cape Town, South Africa}
\affiliation{INAF - Institute for Radio Astronomy, Via Gobetti 101, 40129, Bologna, Italy}

\author{Stephen Eales}
\affiliation{School of Physics and Astronomy, Cardiff University, The Parade, Cardiff CF24 3AA, UK}

\author{Ivan Valtchanov}
\affiliation{Telespazio UK for ESA, European Space Astronomy Centre, Operations Department, E-28691 Villanueva de la Ca\~nada, Spain}

\begin{abstract}
		We report the detection of a massive neutral gas outflow in the z=2.09 gravitationally lensed Dusty Star-Forming Galaxy HATLASJ085358.9+015537 (G09v1.40), seen in absorption with the $\OHp\!(1_1-1_0)$ transition using spatially resolved ($0.5^{\prime\prime}\times0.4^{\prime\prime}$) Atacama Large Millimeter/submillimeter Array (ALMA) observations. The blueshifted \OHp line is observed simultaneously with the CO(9-8) emission line and underlying dust continuum. These data are complemented by high angular resolution ($0.17^{\prime\prime}\times0.13^{\prime\prime}$) ALMA observations of $\CHp\!$(1-0) and underlying dust continuum, and Keck 2.2 $\mu$m imaging tracing the stellar emission. The neutral outflow, dust, dense molecular gas and stars all show spatial offsets from each other. The total atomic gas mass of the observed outflow is $6.7\times10^9 {\rm M_{\odot}}$, $>25\%$ as massive as the gas mass of the galaxy. We find that a conical outflow geometry best describes the \OHp kinematics and morphology and derive deprojected outflow properties as functions of possible inclination ($0.38^\circ-64^\circ$). The neutral gas mass outflow rate is between $83-25400\ {\rm M_{\odot}\ yr}^{-1}$, exceeding the star formation rate ($788\pm300\ {\rm M_{\odot}\ yr^{-1}}$) if the inclination is $>3.6^\circ$ (mass-loading factor = 0.3-4.7). Kinetic energy and momentum fluxes span $4.4-290\times10^{9}\ {\rm L_\odot}$ and $0.1-3.7\times10^{37}$ dyne, respectively (energy-loading factor = 0.013-16), indicating that the feedback mechanisms required to drive the outflow depend on the inclination assumed. We derive a gas depletion time between 29 and 1 Myr, but find that the neutral outflow is likely to remain bound to the galaxy, unless the inclination is small, and may be re-accreted if additional feedback processes do not occur.  
	
\end{abstract}



\section{Introduction} \label{sec:intro}
	
	The formation and evolution of galaxies are intrinsically linked to the cosmic web. Dark matter halos accrete gas from the intergalactic medium (IGM) which, by the dissipation of energy, cools and condenses to form a central galaxy \citep{Rees1977,White1978}. Gas within galaxies may then collapse to form stars or accrete onto supermassive blackholes, injecting energy back into the interstellar medium (ISM) via stellar winds, radiation pressure, supernovae (SNe) explosions, or through strong feedback associated with an active galactic nucleus (AGN), respectively. Mild forms of these feedback processes heat and disturb the surrounding ISM, prolonging its collapse into new stars whilst in their extremes, eject gas from the galaxy. Ejected gas is either recycled through the circumgalactic medium (CGM) where it can be re-accreted back onto the galaxy at a later time or be lost to the IGM, removing the fuel for star formation (SF) altogether \citep{Nelson2019,Mitchell2020,Bregman1980,Bregman2013,Fluetsch2019,Spilker2020b}.
	
	This model of self-regulated galaxy growth became evident in early cosmological simulation work, which failed to reproduce disky galaxy morphologies without invoking sufficiently strong supernova feedback processes capable of removing low angular momentum material from the centers of galaxies (later coined the angular momentum catastrophe; \citealt{Governato2010}). Today's state-of-the-art theoretical models (e.g., EAGLE, \citealt{Schaye2015}; Illustris-TNG \citealt{Nelson2019,Pillepich2019}; L-GALAXIES \citealt{Henriques2020}) similarly rely on a series of feedback processes that tap into the energy released by stars and active galactic nuclei (AGN) in order to regulate stellar mass growth (the overcooling problem, e.g., \citealt{Somerville1999,Cole2000,Benson2003,Keres2009,Bower2012}), reproduce metallicity gradients, the galaxy mass function and to pollute the CGM/IGM with metals \citep{Veilleux2005}.
	
	In fact, the identification of metals in the low-density CGM/IGM via absorption in QSO sight lines (\citealt{Meyer1987,Simcoe2004}) provided the first observational evidence that some fraction of the enriched matter within galaxies must be ejected. Optical and X-ray observations led to the first direct evidence of outflowing material from galaxies in the form of ionised gas (\citealt{Heckman1990,Strickland2004}) which has since been complemented by observations of the molecular and atomic phases probed by IR/submm wavelengths (e.g., Walter et al. 2002; Sturm et al. 2011; Bolatto et al. 2013), corresponding to an enormous range in temperatures ($10-10^8$K, \citealt{Veilleux2005}) and densities ($\sim10-10^5 {\rm cm^{-3}}$, \citealt{Shopbell1998,Aalto2015}). 
	
	The bright emission and absorption lines associated with the ionised gas phase have led the majority of galaxy outflow observations to focus on this component and have successfully shown that galaxy outflows are ubiquitous in the local universe from dwarf galaxies to luminous infrared galaxies (LIRGs, ${\rm L_{IR}}$ $> 10^{11} L_\odot$). Over this range, outflow velocities have been found to correlate with star formation rate (SFR), stellar mass (${\rm M}_*$), and SFR surface density (e.g., \citealt{Lehnert1996,Rupke2002,Martin2005,Westmoquette2012,Rubin2014,Heckman2016,Chisholm2016}), suggesting a close connection between galaxy outflows and the ongoing evolution of their host galaxies. The hot phase, however, only dominates the thermal and kinetic energy of the outflow, whilst the cooler, denser molecular and neutral phases are believed to dominate the mass and momentum budget (\citealt{Walter2002,Rupke2005,Feruglio2010,Alatalo2011,Rupke2013,Rupke2017,Fluetsch2020,Herrera-Camus2020}). 
	
	The first detections of molecular outflows from local Ultraluminous infrared galaxies (ULIRGs, ${\rm L_{IR}}$ $> 10^{12} L_\odot$) were achieved almost simultaneously using very deep ground-based CO(1-0) spectra (\citealt{Feruglio2010}), and spectra of OH lines obtained with the Herschel satellite (\citealt{Fischer2010}) of the ULIRG/AGN Mrk 231. Now, with the addition of new facilities such as the Atacama Large Millimeter Array (ALMA) and the Northern Extended Millimeter Array (NOEMA), outflows have been detected in a large number of local LIRGs and ULIRGs, using both CO and OH rotational lines \citep{Cicone2014,Spoon2013,Sturm2011,Veilleux2013}. In all cases, mass outflow rates of the order or even significantly larger than the star formation rate in the galaxy were derived, suggesting that galactic winds regulate star formation in these systems.
	
	At redshifts $z=1-3$ where the cosmic star formation and black hole accretion peak \citep{Madau2014}, outflows are expected to be ubiquitous. However, observing molecular gas outflows using CO observations in high-$z$ galaxies is extremely challenging. In local (U)LIRGs, the CO emission from the outflowing gas typically represents only a few percent of the total CO emission of the galaxy, and requires high S/N observations at high spatial resolution, in order to observationally separate the outflowing gas from the bulk CO emission \citep{Cicone2014,GarciaBurillo2014,GarciaBurillo2015,PereiraSantaella2018,PereiraSantaella2020}. Such observations at high-$z$ have provided mostly tentative results and only in galaxies hosting an AGN (see, e.g., \citealt{Weiss2012,Feruglio2017,Carniani2017,Vayner2017,Fan2018,Brusa2018,Herrera-Camus2019}), and is generally beyond present observational capabilities for other classes of galaxies, even with ALMA\null. 
	
	High-velocity wings in [CII] 158 $\mu$m spectra have provided strong evidence of outflowing gas in one main-sequence star-forming galaxy at $z\sim5$ \citep{Herrera-Camus2021} and in a handful of particularly extreme high redshift QSO spectra \citep{Maiolino2012,Cicone2015}. In the stacked [CII] spectra of somewhat less extreme systems, the evidence of high-velocity wings range from suggestive \citep{Gallerani2018} to undetected \citep{Decarli2018} in QSOs, and strong \citep{Ginolfi2020} to undetected (even considering only galaxies with known molecular outflows, \citealt{Spilker2020}), in star-forming galaxies. It is further uncertain to what extent [CII] traces ionised, neutral and molecular gas in these outflows and at $z\sim2$ the high-frequency observations needed to observe this fine structure atomic line further complicate its use. 
	
	High-excitation water transitions present a promising probe of the dense warm molecular outflowing component but have so far only been observed in one starburst galaxy at z=5.656 \citep{Jones2019}. Alternatively, one can utilize the blue-shifted absorption features of outflowing gas situated between the observer and the host galaxy. The OH $119$\mum doublet absorption line provides promising strength based on low (e.g., \citealt{Sturm2011,Spoon2013,Veilleux2013,Stone2016,GonzalezAlfonso2017,Calderon2016}), and high redshift investigations \citep{Zhang2018} but is only observable with ground-based facilities at redshifts z>4 (e.g., \citealt{Spilker2018,Spilker2020,Spilker2020b}). 
	
	\textit{Herschel} SPIRE spectra of \OHp in local (U)LIRGs also reveal blue-shifted absorption lines or (in a minority of cases) even complete P-Cygni profiles (\citealt{vdWerf2010,Rangwala2011,GonzalezAlfonso2018}). \OHp spectral lines lie at much lower frequencies than the OH $119$\mum line, allowing us to probe them with ALMA Bands 3-7 at ${\rm z>1.75}$ or NOEMA Band 3 at ${\rm z>2.74}$, and at even lower z with higher-frequency weather-sensitive bands. Additionally, the $\OHp\!(1_1-1_0)$ line at 1033.1 GHz lies closely to the CO(9-8) line at 1036.9 GHz, which traces warm dense gas in the host galaxy disk. Both $\OHp$ and CO(9-8) can be observed with a single ALMA tuning, providing simultaneous velocity measurements of the host galaxy disk (in CO(9-8)) and any outflowing gas (through $\OHp$ absorption, if blue-shifted). Early observations of \OHp at high redshift have detected the line in absorption towards the massive starburst galaxy HFLS3 at z=6.34 \citep{Riechers2013}. More recent observations of \OHp and ${\rm H_2O^+}$ in two $z\sim2.3$ lensed SMGs SMM J2135-0102 and SDP 17b have been used to constrain the cosmic ray ionisation rate within these galaxies \citep{Indriolo2018}, finding rates much lower than predicted for the star-forming regions in these galaxies. The consequence of this finding being that \OHp likely traces the diffuse, turbulent and predominantly neutral gas halos, also seen in \CHp \citep{Falgarone2017}, surrounding high z galaxies.

	Recent observations of \OHp have further demonstrated the importance of this molecular ion to trace fuelling and feedback in high-z galaxies with the detection of inflowing gas via the three ground-state transitions of \OHp (together with $\CHp\!$) in the star-forming galaxy HerBS-89a at z-2.95 \citep{Berta2021} and a powerful outflow traced by a P-Cygni profile in the hyper-luminous ($>10^{13}\ {\rm L_{\odot}}$) starbursting merger ADFS-23 at z=5.655 \citep{Riechers2021}.  As discussed in \citep{Berta2021} and shown by the results of \cite{Riechers2021b} who observed a sample of 18 starburst galaxies at z=2-6, most \OHp measurements to date have revealed cases of outflow with only a few examples displaying clear detections of infall activity. In all 18 galaxies studied by \cite{Riechers2021b}, \OHp is detected in either absorption (14), emission (10) or both (8).

	 Even in the brightest high redshift sources, however, detecting and spatially resolving \OHp outflows remains observationally expensive (on the order of days) and thus limits analysis to unresolved studies. Fortunately, observing time can be significantly reduced with the aid of strong gravitational lensing, which magnifies the light emitted by the background source, increasing its on sky size whilst maintaining surface brightness and thereby improving the source plane resolution and total observed flux.
	
	This paper presents spatially resolved ALMA band 6 observations of the z=2.0924 gravitationally lensed Dusty Star-Forming Galaxy (DSFG; a galaxy selected at infrared or submillimeter wavelengths, see \citealt{Casey2014}) HATLASJ085358.9+015537 (hereafter G09v1.40), revealing a large scale neutral outflow traced by \OHp (and \CHp) in addition to the warm dense gas component and dust in the host galaxy traced by CO(9-8) and underlying dust continuum emission, respectively. Identified in the Herschel Astrophysical Terahertz Large Area Survey, H-ATLAS \citep{Negrello2010,Negrello2017}, G09v1.40 and its lens have previously been modelled in the submm \citep{Bussmann2013,Enia2018} and near-infrared (NIR) \citep{Calanog2014}, providing excellent input parameters for gravitational modelling with our high-resolution ALMA observations. Additional studies of G09v1.40 include: accurate redshift determination and CO SLED modelling using multiple CO transitions \citep{Yang2017}, analysis of the turbulent halo of diffuse gas surrounding the galaxy seen in \CHp absorption \citep{Falgarone2017}, and rest-frame optical spectral energy distribution modelling including \textit{Spitzer}/IRAC imaging at 3.6 and ${\rm4.5\mu\ m}$ \cite{Ma2015}. 
	
	Throughout our work we assume a flat $\Lambda$CDM cosmology with $\Omega_{\rm m}= 0.307$ and ${\rm H_0 = 67.7\ km\ s^{-1}\ Mpc^{-1}}$ \citep{Planck2016}. At the redshift of G09v1.40, z=2.0924, one arcsecond corresponds to 8.53 kpc.
	
\section{Observations and data reduction} \label{sec:Obs}
		\begin{figure*}
			\begin{center}
				\includegraphics[width=\textwidth]{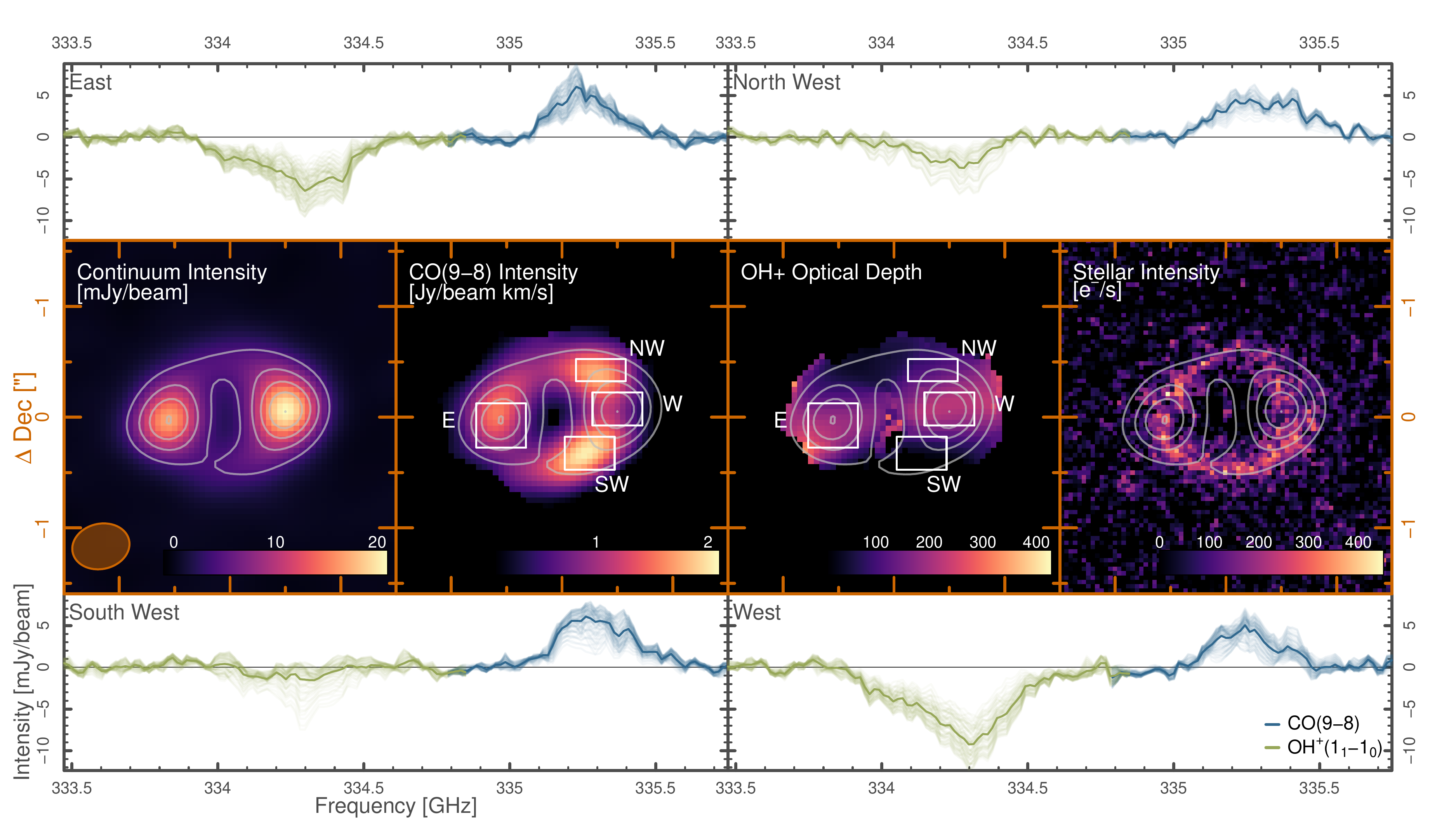}
				\caption{\textbf{Middle Row - Far Left Panel:} rest-frame  1034 GHz dust continuum emission observed with ALMA. The peak of the dust emission is lensed into a prominent double image configuration with a brighter western image, accompanied by a complete Einstein ring. The synthesized beam ($0.52^{\prime\prime}\times0.41^{\prime\prime}$) is shown by the faded orange ellipse in the bottom left. The grey contours indicate continuum levels of 20\%, 40\%, 60\%, 80\% and 100\% that of the peak and are overlaid in the following three panels. \textbf{Middle Row - Center Left Panel:} Intensity map of the rest-frame  1036.9 GHz CO(9-8) emission line observed with ALMA, produced via a single Gaussian fitting procedure. The CO(9-8) emission reveals itself in 3 distinctly different peaks compared to the continuum but again accompanied by a complete Einstein ring. \textbf{Middle Row - Center Right Panel:} Integrated optical depth map of the blue-shifted component of the $\OHp\!(1_1-1_0)$ absorption line observed with ALMA, produced via a double Gaussian fitting procedure. The peak \OHp optical depth is approximately cospatial with the peaks in the continuum emission albeit with an elongated morphology stretching from East to West. \textbf{Middle Row - Far Right Panel:} Keck Ks ($\lambda=2.2\ \mu m$) stellar intensity map with the lens subtracted \citep{Calanog2014}. As in the continuum and CO(9-8) emission, a distinct Einstein ring is observed, but with gaps at the positions of peak FIR continuum emission. \textbf{Outer 4 Panels:} Each of the four outer panels correspond to a region with a matching label (E: East, NW: North West, SW: South West, W: West) indicated by the solid white boxes in the central two panels and are placed over peaks in the continuum and CO(9-8) emission. The outer panels display continuum subtracted CO(9-8) and \OHp spectra of the individual spaxels (pixel area = $0.05^{\prime\prime}\times0.05^{\prime\prime}$) of each region in transparent blue and green, respectively. The average spectra of each region are shown by the solid line. The CO(9-8) and \OHp spectra do not display significant variation in spectral shape or central velocity between the four regions. }
				\label{fig:MapStack}
			\end{center}
		\end{figure*}		
	
	\subsection{ALMA Band 6 Observations and Reduction}\label{sec:ObsALMA}
		In this paper, we present ALMA Band 6 data of the DSFG G09v1.40. The ALMA Band 6 observations of G09v1.40 were taken in ALMA Cycle 3 as a part of project 2015.1.01042.S (PI: P. van der Werf). The observations were taken on 2016 April 22 using 36 antennas of the 12-meter array with baseline lengths spanning 15-462 m. The uv-plane coverage provides sensitivity down to spatial scales of 3.5 kpc (and lower in the source plane) at z=2.0924. The average precipitable water vapor level was 1.00 mm and average system temperature of 168.8 K. G09v1.40 was observed for a duration of 21.60 min, with an additional 35.82 min allocated to phase calibration (J0909+0121), atmosphere and water vapour radiometry calibration (J0854+2006, J0909+0121 and J085358.9+015537), bandpass, flux and pointing calibration (J0854+2006).
		
		The Band 6 receivers were tuned to observe the $\OHp\!(1_1-1_0)$ and CO(9-8) lines simultaneously in two slightly overlapping spectral windows to ensure continuous coverage of both lines without a decrease in sensitivity through the intermediate frequencies. This provided a 3.24 GHz bandwidth equating to $\sim2890$ \kms. One more 2 GHz wide spectral window was placed to detect rest-frame 1034 GHz continuum at high sensitivity. All spectral windows were configured with a channel resolution of 15.625 MHz. 
		
		The data were reduced with the ALMA Cycle 3 pipeline using Common Astronomy Software Applications ({\tt CASA}: \citealt{McMullin2007}) version 4.53. We use Briggs weighting with a robust parameter 0, resulting in a beam with dimensions of $0.52''\times0.41''$ and position angle $-78.2^\circ$. This choice of weighting provides the optimal combination between sidelobe suppression and surface brightness sensitivity. The rest-frame 1034 GHz continuum map was created using 112 channels resulting in an ${\rm RMS=0.12}$ mJy beam$^{-1}$. The \OHp($1_0-1_1$) and CO(9-8) lines were separated into two data cubes with velocity resolutions of 14.5 \kms\ and ${\rm RMS=0.56,0.58}$ mJy beam$^{-1}$, respectively.
		
	\subsection{Ancillary NIR data}\label{sec:ObsKeck}
		We include ancillary NIR imaging of G09v1.40, captured by the Keck II Near-Infrared Camera 2 (NIRC2) using the ${\rm K}$ ($\lambda = 2.2\mu m$) filter with laser guide star adaptive optics. The observation and reduction of this data are presented in detail by \cite{Calanog2014}.
		
	\subsection{Ancillary \CHp data}\label{sec:ObsCHp}
		To supplement our analysis of G09v1.40 we incorporate two ancillary $\CHp\!$(1-0) and rest-frame 836 GHz continuum data sets observed with ALMA during Cycles 2 and 4. The Cycle 2 observations (ALMA project 2013.1.00164.S, P.I. E. Falgarone) have been discussed and analysed by \cite{Falgarone2017} with focus on the \CHp line. In this study, we simply utilize the pipeline product available from the ALMA archive, reduced in CASA version 4.2.2 and imaged with a Briggs weighting of 0.5, providing a beam size of $0.59''\times0.46''$. 
		
		The Cycle 4 data was observed as part of the 2016.1.00282.S ALMA program (P.I. E. Falgerone), aiming to measure the \CHp(1-0) and dust continuum emission at a higher spatial resolution. We reduce the data using CASA version 4.7.0-1 with a Briggs weighting of 0, resulting in a beam size of $0.17''\times0.13''$. The \CHp(1-0) line is detected in both absorption and emission, and the detected continuum emission provides the highest spatial resolution data of the dust profile in our analysis.
		
		We refer to these data sets as the 'low' and 'high' spatial resolution $\CHp\!$(1-0) line and underlying rest-frame 836 GHz continuum data, respectively, throughout the paper.

\section{Results} \label{sec:Results}
	\begin{figure}
		\begin{center}
			\begin{tabular}{c}
				\includegraphics[width=\linewidth]{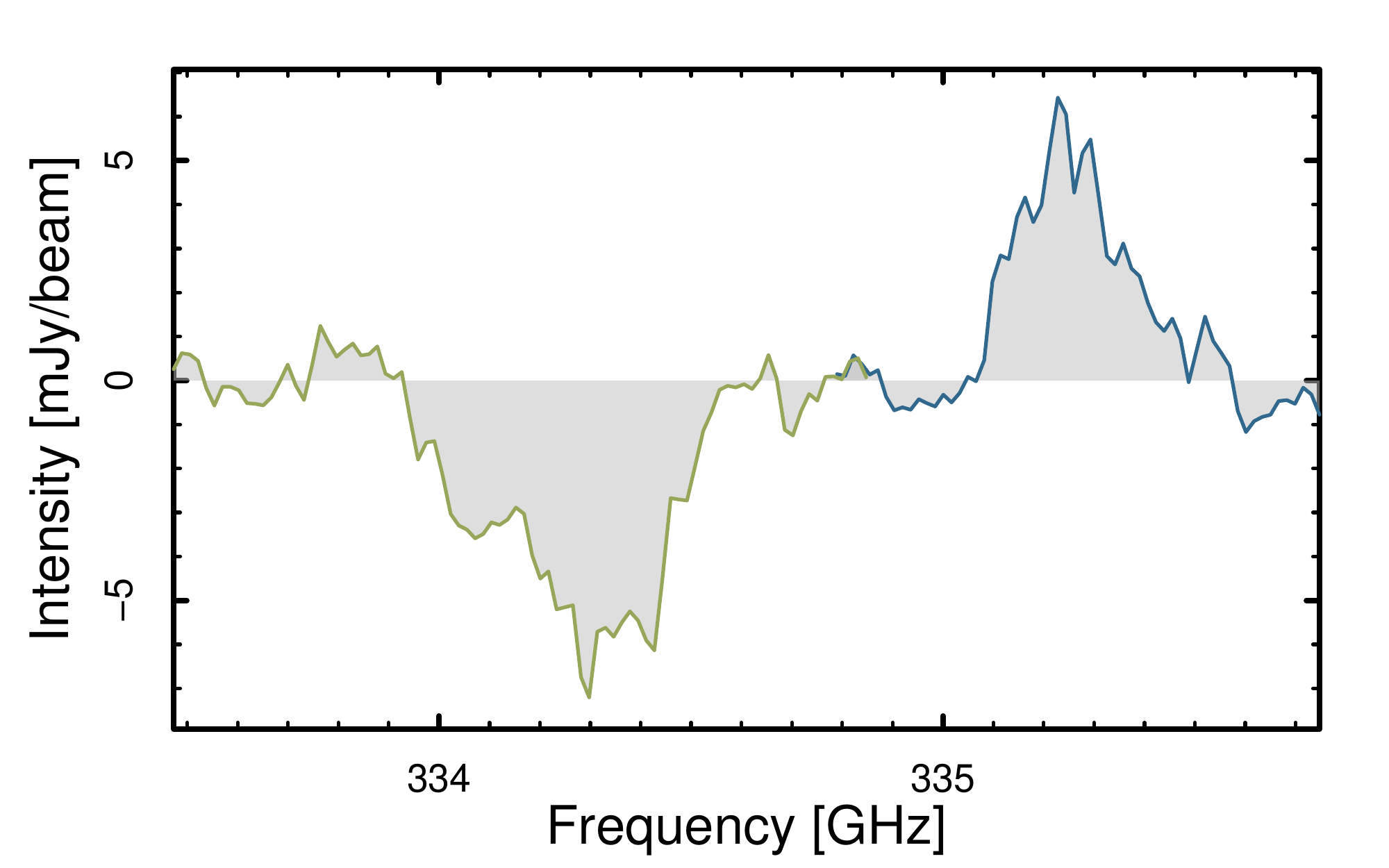}\\
				\\[-25pt]
				\includegraphics[width=\linewidth]{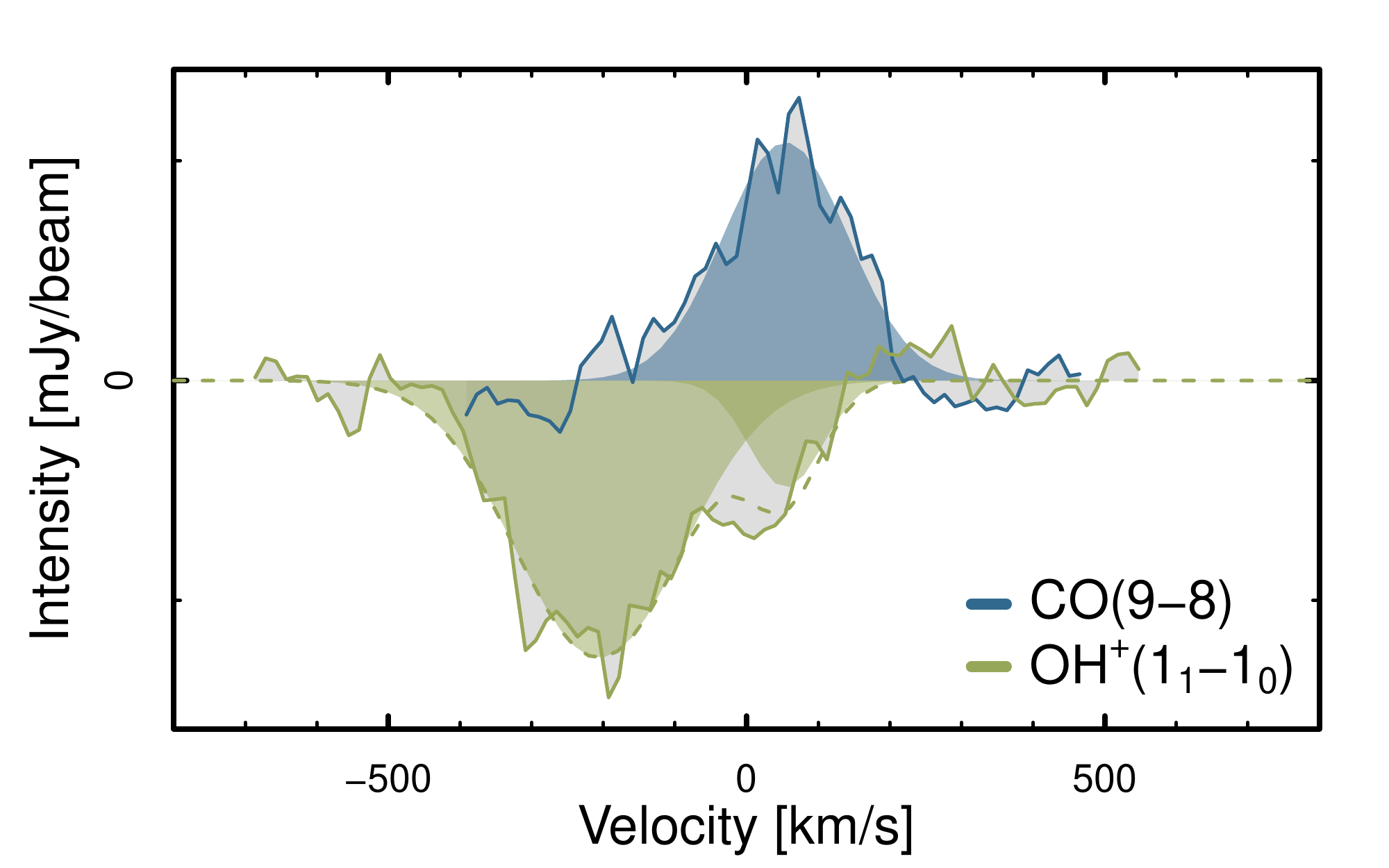}
			\end{tabular}
			\caption{Continuum subtracted $\OHp\!(1_1-1_0)$ absorption (green) and CO(9-8) emission (blue) spectra in a single ALMA spaxel (pixel area = $0.05^{\prime\prime}\times0.05^{\prime\prime}$) observed simultaneously in the DSFG G09v1.40 \textbf{Top:} Spectra plotted as a function of observed frequency. \textbf{Bottom:} Spectra plotted as a function of velocity with respect to the galaxy's systemic velocity determined from the z=2.0924 redshift measurement by \cite{Yang2017}. We fit the \OHp($1_0-1_1$) absorption and CO(9-8) emission lines with double and single Gaussian functions, respectively; the shaded regions of the same colour indicate the individual Gaussian components to each spectral line, whilst the dashed green line presents the full \OHp fit. The main \OHp component is blue-shifted $\sim200\ \kms$ with respect to the bulk molecular gas as traced by the CO(9-8) emission, revealing a neutral outflow at these velocities. Note also the larger width of the \OHp line.}
			\label{fig:CO98OHvelfit}
		\end{center}
	\end{figure}		

	The ALMA Band 6 imaging successfully resolves and detects the 1034 GHz dust continuum, CO(9-8) line emission and the $\OHp\!(1_1-1_0)$ line seen in absorption. The dust continuum reveals a complete Einstein ring and two bright images, one to the East and a brighter one to the West (Fig. \ref{fig:MapStack}). This configuration is indicative of a single extended source with a bright central dust region that is lensed by a single,  almost perfectly aligned foreground galaxy. The brighter Western image indicates that the source sits just to the West of the foreground lens.
	
	The CO(9-8) similarly displays a complete Einstein ring, however in contrast to the continuum, the CO(9-8) emission reveals three distinct peaks. The South-Western, and brightest peak is separated into two emission peaks by the deconvolved model provided by the cleaning procedure (see Fig. \ref{fig:BeamSmearingCO98} in Appendix \ref{sec:beam} for more discussion). The spectra observed across all peaks (outer panels of Fig. \ref{fig:MapStack}) do not display significant variation in shape or central velocity, suggesting that we are seeing the same CO(9-8) component lensed into four images, of which two are blended together in the South-West. This requires the peak of the CO(9-8) component to be lying directly over a section of the inner caustic, where magnification of the source goes to infinity (diamond in bottom two rows of Fig. \ref{fig:GridMaps}).
	
	The two peaks in \OHp optical depth (and hence column density) are approximately co-spatial with the two continuum images, albeit with elongated morphologies stretching from East to West and dropping off in intensity to the North and South. As for CO(9-8), the \OHp spectra do not display significant variation in spectral shape or central velocity across its images. Unlike both the dust continuum and CO(9-8), there is no discernible Einstein ring in the \OHp optical depth, indicating already in the image plane that the alignment of the \OHp component in the source plane does not lie, even partially, over the inner caustic and must therefore be located fully to the West of it.
	
	For comparison, we also include the lens subtracted Keck near-IR stellar intensity map from \cite{Calanog2014}. There are no distinct peaks but a clear Einstein ring indicates that the stellar component is lying directly over the inner caustic.
	
	The close spectral proximity of the $\OHp\!(1_1-1_0)$ and CO(9-8) lines allows us to capture both transitions simultaneously, with a single ALMA tuning. This is highlighted in Fig. \ref{fig:CO98OHvelfit}, displaying a single, typical spaxel in our data set. Remarkably, at every position, the \OHp absorption peak is blue-shifted by ${\rm \sim200\ \kms}$, with respect to the central velocity of the CO(9-8) emission line. Since the redshift determined from the CO(9-8) emission is consistent with the multi-line redshift, ${\rm z=2.0924\pm0.0001}$, precisely determined by \citealt{Yang2017}), and since we detect the \OHp line in absorption and are therefore tracing gas located in front of the dust continuum, the blueshift of the \OHp absorption indicates that it is tracing gas outflowing from the host galaxy towards us. 
		
	To create the CO(9-8) emission and $\OHp\!(1_1-1_0)$ optical depth maps presented in Fig. \ref{fig:MapStack}, we feed the CO(9-8) and \OHp data cubes through a robust Gaussian spectral line fitting algorithm whereby each spaxel is individually fitted. We find, by examining the residuals produced by this fitting process, that a Gaussian, or combination of Gaussian profiles provide a good fit to the observed spectra.  The CO(9-8) and $\OHp\!(1_1-1_0)$ spectral lines do not overlap in frequency at any source location and so this process is performed separately for each of the spectral line data cubes. The fitted intensity (S), velocity (V) and velocity dispersion ($\sigma_v$) of each spaxel then form a 2D data map in the RA-Dec. plane. This technique is favoured over the use of moment maps as spectral fitting is capable of cleanly disentangling separate overlapping velocity components, as discussed below.
	
	We attempted to fit one, two and three Gaussian functions to the CO(9-8) spectra in order to fit the more complex spectral shape, as seen in Fig. \ref{fig:MapStack}. The multi-Gaussian fits, however, did not produce smooth velocity or intensity fields, indicating that this method was not capable of extracting separate kinematic CO components. We therefore apply single Gaussian fits over the full source, finding this to best trace the bulk gas component. The intensity, velocity and velocity dispersion of the CO(9-8) were left as free parameters in this process.
	
	The \OHp spectrum is relatively consistent across the entire source with absorption typically peaking at velocities $\sim200-300\ \kms$ blue-shifted with respect to the source's systemic velocity (Fig. \ref{fig:CO98OHvelfit}). Additional absorption at systemic velocities is responsible for the skewed spectra at all source locations (Fig. \ref{fig:MapStack}) and is cleanly disentangled from the outflowing component by our spectral fitting described below.
	
	We first fit a double Gaussian function to the \OHp absorption spectra where one Gaussian is fixed at the systemic velocity and the velocity of the blueshifted component is left free (intensity and velocity dispersion are also left free for both lines). This provides the central velocity and velocity dispersions of the \OHp in each spaxel. We then convert the \OHp absorption data cube into an optical depth cube, $\tau$, via
	
	\begin{equation}
		\tau_{i,j}(\nu) = -\ln\Big(\frac{S_{abs,i,j}(\nu)}{S_{cont,i,j}}\Big)
	\end{equation}
	
	where $S_{abs,i,j}(\nu)$ is the absolute value of the \OHp flux density in pixel [i,j] at frequency $\nu$ and $S_{cont,i,j}$ is the continuum flux in the same pixel. A double Gaussian function is then fitted to the \OHp optical depth profiles where again, one Gaussian is fixed at the systemic velocity and the other is fixed at the central velocity found in the previous fit to the absorption spectra. This second fitting step provides us with integrated optical depths for each spaxel.  
	
\begin{figure}
		\begin{center}
			\includegraphics[width=\linewidth]{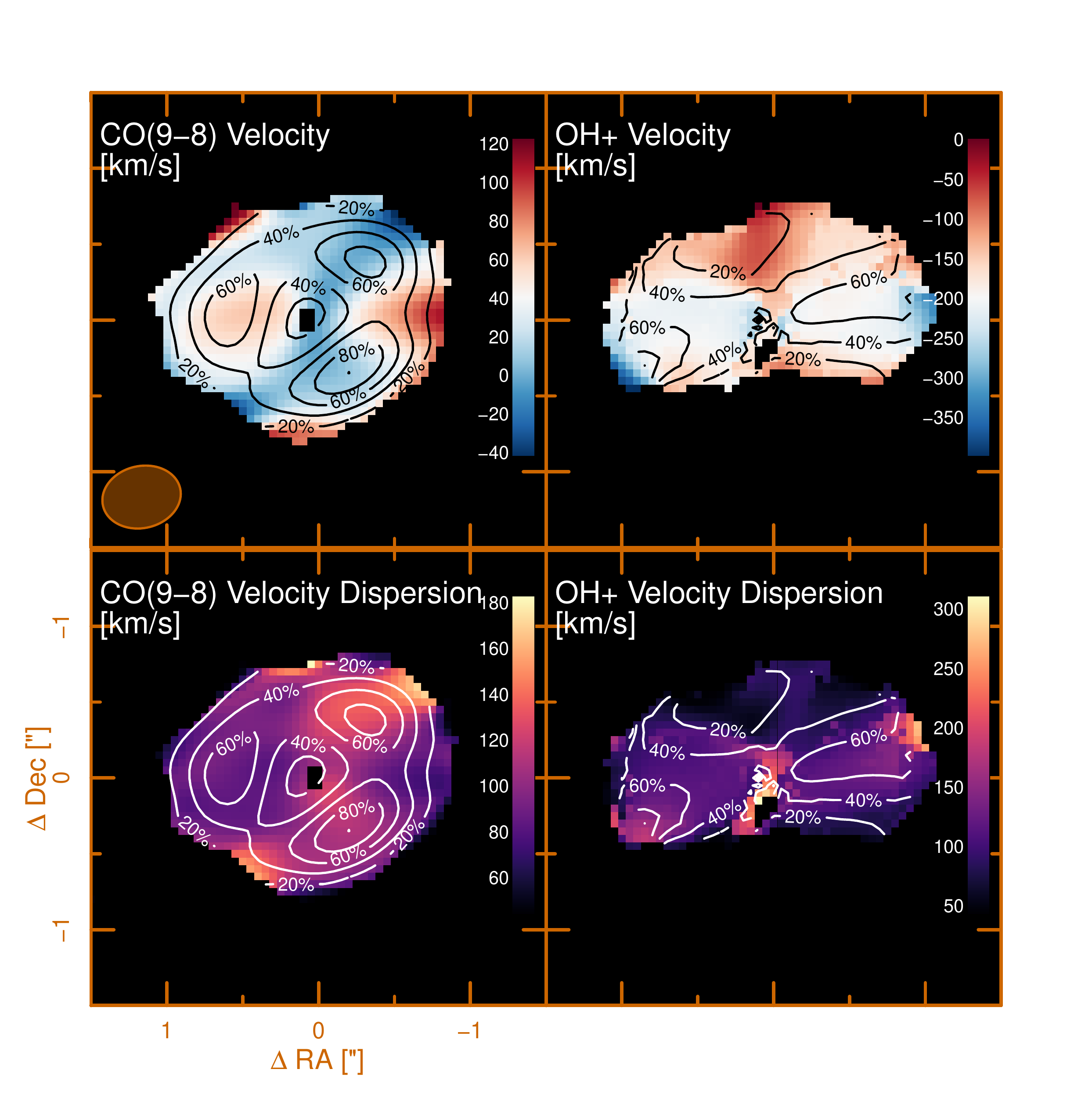}
			\caption{ \textbf{Top Left:} Image plane velocity map of the CO(9-8) emission produced via a single Gaussian fitting procedure described in the text. \textbf{Top Right:} Image plane velocity map of the $\OHp\!(1_1-1_0)$ blue-shifted absorption component, produced via a double, Gaussian fitting procedure described in the text. \textbf{Bottom Left:}  Image plane velocity dispersion map of the CO(9-8) emission produced via a single Gaussian fitting procedure described in the text. \textbf{Bottom Right:}  Image plane velocity dispersion map of the  $\OHp\!(1_1-1_0)$ blue-shifted absorption component produced via a double Gaussian fitting procedure described in the text. The beam ($0.52^{\prime\prime}\times0.41^{\prime\prime}$) is shown by the faded orange ellipse in the bottom left corner of the upper left panel. Contours indicate flux levels at 20\%, 40\%, 60\% and 80\% of the peak flux and optical depth of the CO(9-8) and \OHp spectral lines, respectively.}
			\label{fig:VelVeldisp}
		\end{center}
	\end{figure}

	We discard CO(9-8) and \OHp spaxel fits that return flux amplitudes below the noise level of each cube (with integrated fluxes reaching S/N much higher than 1), velocities outside the observed bandwidth, velocity dispersion narrower than two channels or wider than the full velocity bandwidth. \OHp spaxels are further rejected if the continuum flux is below $5\sigma$. The remaining CO(9-8) and \OHp spectral fits (e.g., shaded Gaussians in the bottom panel of Fig. \ref{fig:CO98OHvelfit}) are then used to make clean CO(9-8) intensity and blueshifted \OHp optical depth maps as presented in Fig. \ref{fig:MapStack} along with their respective velocity and velocity dispersion maps, shown in Fig. \ref{fig:VelVeldisp}. Note that we only present and analyse the blueshifted component of the \OHp since we aim to study the outflow.
	
	The maximum blueshift in the \OHp line is found along the peak in \OHp optical depth (indicated by the contours in Fig. \ref{fig:VelVeldisp}). The velocity dispersion similarly peaks along the main outflow component, with both the projected outflow velocity and velocity dispersion increasing mildly towards the far Western and Eastern edges, as seen in the optical depth. These regions are likely where the outflow is at its widest, however it must be noted that the background continuum in these extreme regions is faint and care should be taken when interpreting the \OHp absorption.
	
\section{Gravitational Lens modelling and Source Plane Reconstruction} \label{sec:Grav}
	\subsection{Lens modelling: VISILENS}\label{sec:visilens}
		\begin{deluxetable}{clc}[]
			\tablecaption{\texttt{visilens} parameters of the best-fit SIE Lens and Sersic source models, using the underlying dust continuum of the high spatial resolution $\CHp\!$(1-0) observations.} \label{tab:LensPars}
			\tablecolumns{3}
			\tablewidth{0pt}
			\startdata
			\\
			\multicolumn{3}{c}{\bf Lens (SIE)}\\
			\hline
			{\boldmath$x_L$}&$[\,^{\prime\prime}\,]$&$2.824\pm0.677$\\
			{\boldmath$y_L$}&$[\,^{\prime\prime}\,]$&$2.194\pm0.042$\\
			{\boldmath$M_L$}&$[10^{11}\rm{M_\odot}]$&$0.459\pm0.119$\\
			{\boldmath$e_L$}&&$0.037\pm0.042$\\
			{\boldmath$\theta_L$}&$[\rm{deg}]$CCW from E&$162.933\pm23.426$\\
			\hline\hline
			\multicolumn{3}{c}{\bf Source (Sersic)}\\
			\hline
			{\boldmath$\Delta x_S$}&$[\,^{\prime\prime}\,]$&$0.087\pm0.029$\\
			{\boldmath$\Delta y_S$}&$[\,^{\prime\prime}\,]$&$0.001\pm0.001$\\
			{\boldmath$F_S$}&$[\rm{mJy}]$&$2.971\pm1.097$\\
			{\boldmath$a_S$}&$[\,^{\prime\prime}\,]$&$0.058\pm0.009$\\
			{\boldmath$b_S/a_S$}&$[\,^{\prime\prime}\,]$&$0.679\pm0.120$\\
			{\boldmath$n_S$}&&$4.805\pm1.638$\\
			{\boldmath$\phi_S$}&[deg] CCW from E&$164.264\pm52.123 $\\
			\hline
			\enddata
			\tablecomments{Lens positions are given with respect to the ALMA phase center: (J2000) 08:53:58.68 +01:55:35.45 and source positions with respect to the lens position. Parameter descriptions are as follows: {\boldmath$x_L$}, lens position in right ascension, {\boldmath$y_L$}, lens position in declination, {\boldmath$M_L$}, lens mass inside the Einstein radius, {\boldmath$e_L$}, lens elipticity, {\boldmath$\theta_L$}, lens position angle, {\boldmath$\Delta x_S$}, source position in right ascension, {\boldmath$\Delta y_S$}, source position in declination, {\boldmath$F_S$}, source flux, {\boldmath$a_S$}, source major axis, {\boldmath$b_S/a_S$}, source axis ratio, {\boldmath$n_S$}, source sersic index, {\boldmath$\phi_S$}, source position angle.}
		\end{deluxetable}

		Since we are interested in the intrinsic properties of G09v1.40 and its outflow, we must first model and remove the effects of gravitational lensing. ALMA, as an interferometer, observes the Fourier transform of the sky intensity distribution over a range of two-dimensional spatial frequencies in the uv-plane. Noise properties and resolution effects are much better understood in the uv-plane than in the inverted images where uncertainties become correlated and may bias further measurements. To avoid such biases affecting our lens modelling, we have chosen to employ the parametric reconstruction code \texttt{visilens} \citep{Spilker2016,Hezaveh2013}, which fits a lens model directly to visibility measurements. We invoke the \texttt{modelcal} option available in \texttt{visilens} which corrects for calibration errors caused by, e.g., uncertain antenna positions and atmospheric conditions, allowing for multiplicative amplitude re-scaling and astrometric drift. If present and not corrected for, these calibration errors can result in a shift of the model parameters away from their intrinsic values. See \cite{Hezaveh2013} and \cite{Spilker2016} for more in-depth discussions of the code.
		
\begin{figure}
			\centering
			\includegraphics[width=\linewidth]{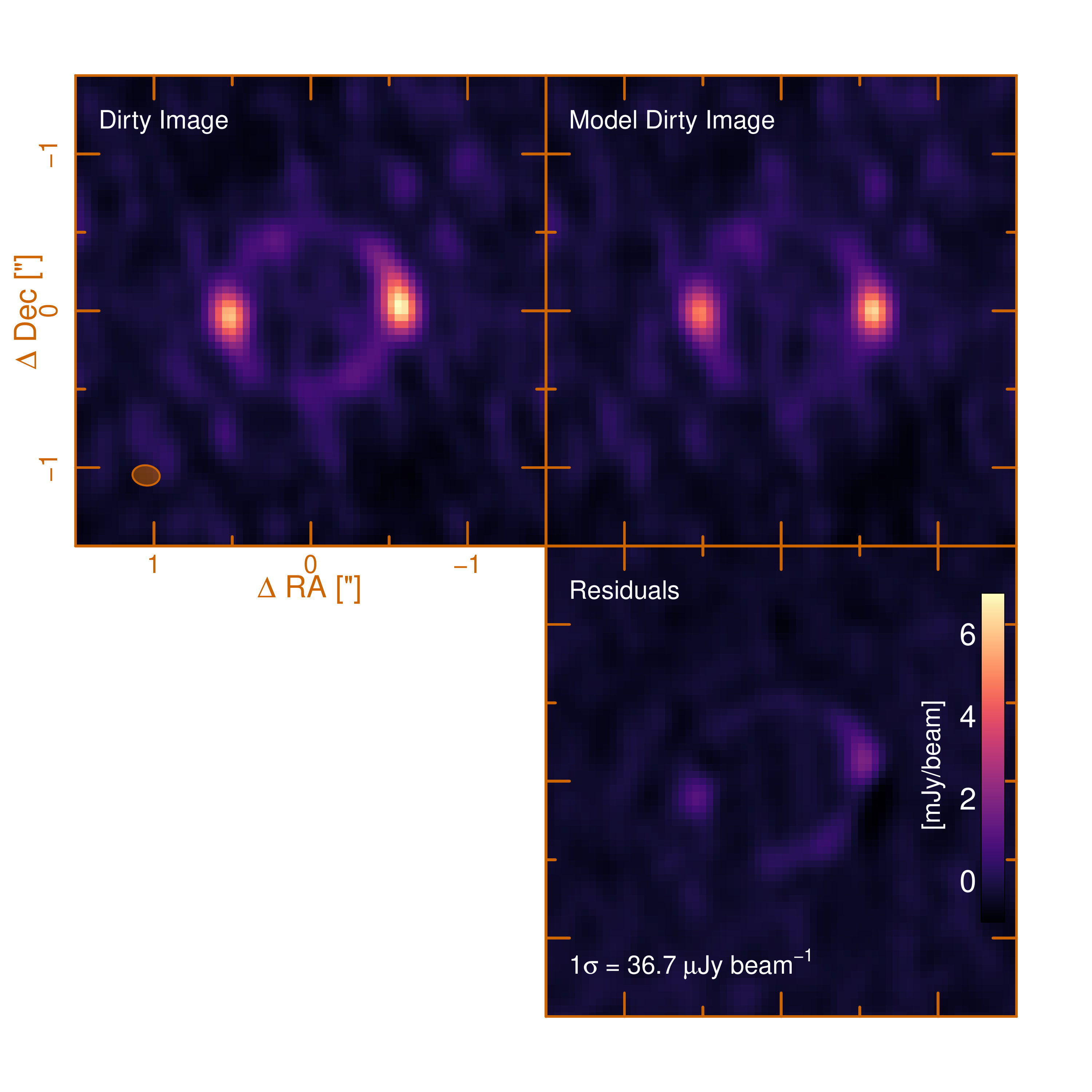}
			\caption{\texttt{visilens} modelling of the high spatial resolution $\CHp\!$(1-0) continuum emission in G09v1.40. \textbf{Top Left:} Dirty image of the data with beam (0.17"x0.13") shown as the orange ellipse in the bottom left corner. \textbf{Top Right:} Dirty image of the \texttt{visilens} model, recovering 85\% of the flux in the dirty image. \textbf{Bottom:} Residuals, with the rms beam$^{-1}$ shown in the bottom left corner.}
			\label{fig:visilenspanels}
		\end{figure}

		To improve the accuracy of our lens model further we take advantage of the underlying 836 GHz dust continuum emission in the `high' spatial resolution $\CHp\!$(1-0) data. The higher spatial resolution of this data provides \verb|visilens| more information on the dust distribution in G09v1.40 from which a more accurate model can be derived (Fig. \ref{fig:visilenspanels}). 
		
		We use a single S\'ersic profile to represent the continuum emission of G09v1.40, characterised by a S\'ersic index $n_S$, half-light radii $a_S$ with axis ratio $b_S/a_S$, position angle East of North $\phi_S$, flux density $F_S$ and position $\Delta x_S$, $\Delta y_S$ with respect to the lens. We model the mass profile of the lens with a Singular Isothermal Ellipsoid (SIE), fitting for the lens position $x_L$, $y_L$ with respect to the ALMA phase center, mass $M_L$ (and corresponding Einstein radius, $\theta_{E,L}$, within which the mass is parameterised), ellipticity $\epsilon_L$, and position angle, East of North $\phi_L$. We do not invoke any external shear in our model. 
		
		\texttt{visilens} begins by creating a 2D source plane parametric model of the source dust emission and lenses this model into the image plane for a given lens model. The two-dimensional lensed emission is then Fourier transformed into the uv-plane where it is directly compared to our observed interferometric data. We initiate this procedure with values taken from \cite{Bussmann2013}, and a Markov Chain Monte Carlo (MCMC) sampling algorithm explores the model parameter space of both the source emission and lens mass profiles, using the \verb|emcee| code \citep{Foreman-Mackey2013}. For each point in the parameter space a source plane, image plane and uv-plane visibilities are generated and checked for quality of fit to the data using a $\chi^2$ metric, where the best-fit parameters minimise the $\chi^2$ value. Note that the \verb|emcee| routine is known to underestimate uncertainties in some circumstances and may be the cause of the extremely small uncertainty provided for our source position.
		
		From the high-resolution data, we extract 280 channels of continuum over the three available spectral windows to model, for which \verb|visilens| finds the best-fit lens and source parameters presented in Tab. \ref{tab:LensPars}. The model recovers the bulk (85\%) of the dirty image flux, as shown in Fig. \ref{fig:visilenspanels}. The residuals exhibit structure that are not consistent with the rms, suggesting that the dust continuum of the source contains more complex structure than can be captured by a single S\'ersic profile. We attempted to model the source with 2 S\'ersic profiles but did not find significant changes to the lens model. Since we are primarily interested in the lens model and do not use the parametric models of the source in our following analysis, we opt to use the single-source model.
		
	\subsection{Source Reconstruction: LENSTOOL}\label{sec:lenstool}
		\begin{figure}
			\centering
			\includegraphics[width=\linewidth]{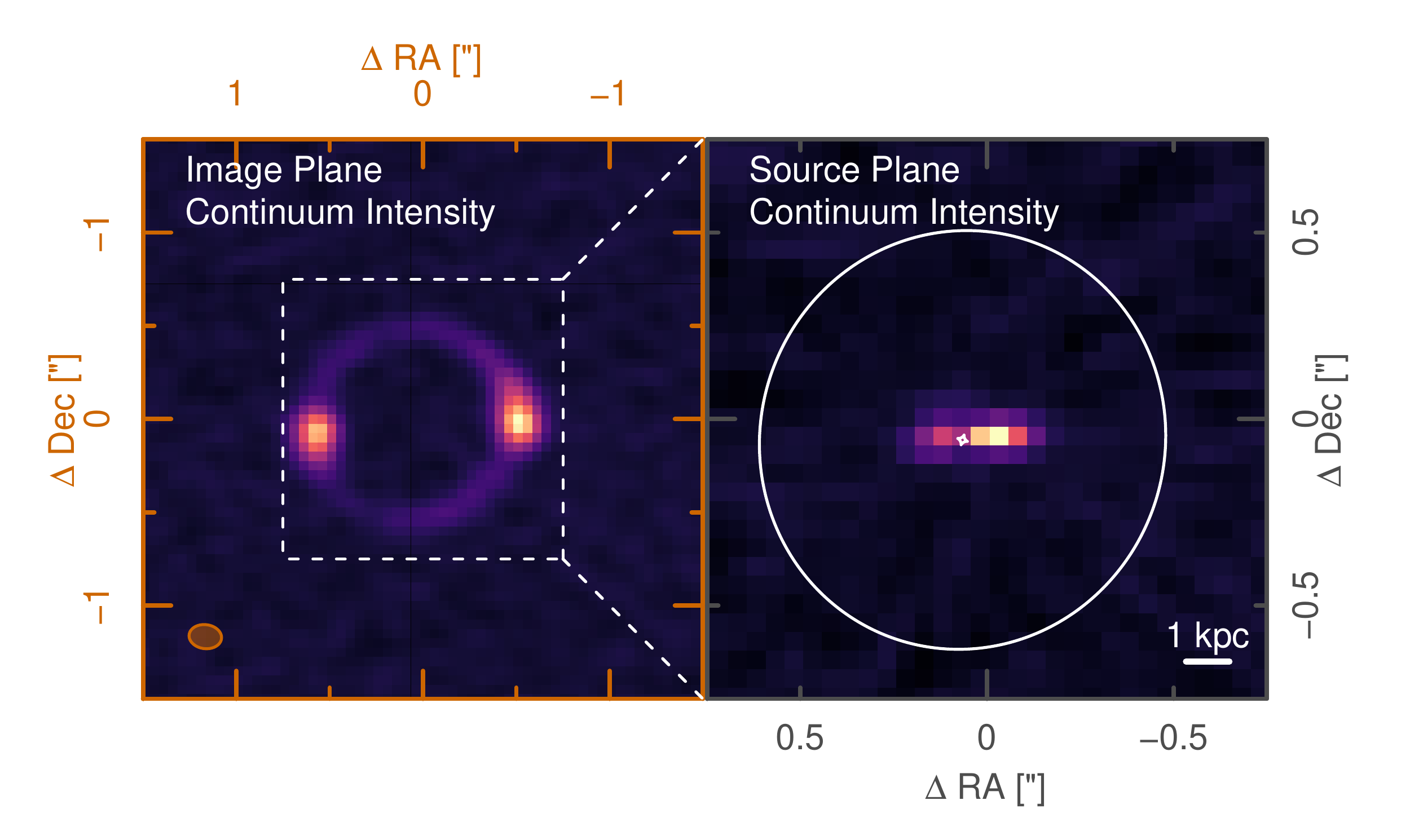}
			\caption{Source plane reconstruction of dust continuum emission from the 'high' spatial resolution $\CHp\!$(1-0) data set using \texttt{LENSTOOL}. The \textbf{left panel} displays the observed image plane dust continuum intensity with the beam (0.17"x0.13") shown in the lower left by the shaded orange ellipse. The \textbf{right panel} is a zoom in of the white dashed region in the left panel, displaying the source plane reconstruction of the dust continuum. The lens caustics are indicated by the solid white lines and a physical scale of 1kpc is given in the lower right corner. A single compact and elongated source is revealed directly to the West of the inner caustic. The weak eastern feature is an artifact of beam smearing over the outer caustic in the image plane.}
			\label{fig:highrescontsing}
		\end{figure}

		\begin{figure*}
			\begin{center}
				\includegraphics[width=\textwidth]{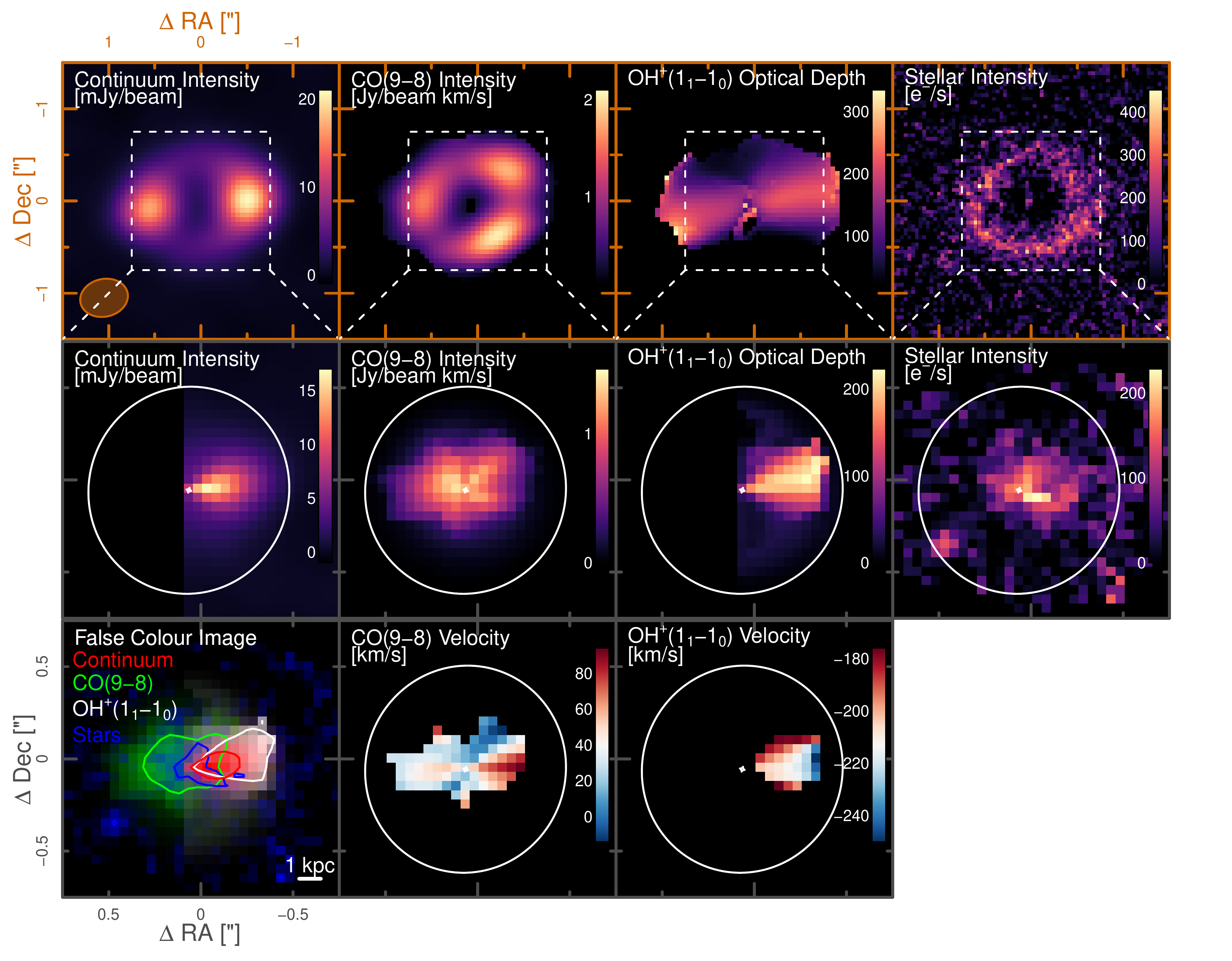}
				\caption{ALMA Cycle 3 observations of the gravitationally lensed galaxy H-ATLAS J085358.9+015537 at redshift z = 2.09. 
					\textbf{Row 1:} Image plane intensity maps of the dust continuum, CO(9-8) emission tracing the warm dense gas in the disk, $\OHp\!(1_1-1_0)$ optical depth tracing neutral outflowing gas and Keck Ks band stellar emission. The beam ($0.52^{\prime\prime}\times0.41^{\prime\prime}$) is shown in the bottom left corner of the left-hand panel. 
					\textbf{Row 2:} Reconstructed source plane intensity maps of the components shown in row 1. Lensing caustics are shown in white. 
					\textbf{Row 3 Left panel:} False colour image of all the source plane intensity maps of the dust (red), CO(9-8) (green), \OHp (white) and stars (blue). A single contour of each distribution is drawn at 60\% of their respective peak values.
					\textbf{Row 3:} Reconstructed source plane velocity maps. The second panel shows the velocity map of the CO(9-8) emission and the third panel shows the \OHp velocity field. The first panel is the \OHp velocity map minus the CO(9-8) velocity map which, if the outflow was a flat sheet of gas lifting off the galaxy disk, should be uniform. This is clearly not the case and this scenario is therefore ruled out (See Sec. \ref{Sheet}). Pixel area ($0.05^{\prime\prime}\times0.05^{\prime\prime}$) is the same in both the image and source planes and a source plane physical scale of 1kpc is given in the lower right corner of the last panel.}
				\label{fig:GridMaps}
			\end{center}
		\end{figure*}		

		From the best-fit lens model obtained in Sec. \ref{sec:visilens} we can derive a lens operator that maps our data from the image plane into the source plane. For this task we employ the pixelated reconstruction code \verb|LENSTOOL| \citep{Kneib1996,Jullo2007,Jullo2009}. A pixelated reconstruction has the advantage of capturing details that can not be easily modelled by a parametric model. This is of particular interest in the case of the outflowing component which need not follow a Sersic profile, and which is implied by the unusual \OHp optical depth morphology.
		
		The reconstructed dust continuum emission, from the `high' spatial resolution $\CHp\!$(1-0) data set used to model the gravitational lens, reveals a single elongated and compact dust continuum profile (Fig. \ref{fig:highrescontsing}). The weak feature directly to the East of the inner caustic is a result of beam smearing across the outer caustic in the image plane. When reconstructed, the flux that has been smeared to the opposite side of the outer caustic, is traced to the wrong side of the inner caustic in the source plane. This effect becomes more severe with larger beam sizes (Fig. \ref{fig:BeamSmearingCont}) and is more difficult to interpret when the galaxy component is lying directly over the caustic line (Fig. \ref{fig:BeamSmearingCO98}; see Appendix \ref{sec:beam} for a detailed discussion).  
				
		We use Lenstool to reconstruct the ${\rm OH^+}$, CO(9-8), underlying 1034 GHz dust continuum and ancillary Keck NIR data using the same lens model (Fig. \ref{fig:GridMaps}). As in the high-resolution continuum data, the rest-frame 1034 GHz dust continuum emission reveals a single compact and elongated source. The eastern artifact seen in the high-resolution continuum reconstruction is also present here but at higher significance (39\% of the source plane flux as opposed to 9\% for the high-resolution data). The more prominent eastern artifact in the 1034 GHz dust continuum reconstruction is due to the larger beam size and thus enhanced beam smearing (see Appendix \ref{sec:beam}). We examine the model intensity maps of the 1034 GHz dust continuum produced during cleaning, which when reconstructed reveal only a single western source (no eastern artifact). This is consistent with our initial interpretation of the image plane dust continuum emission as a single, double-imaged extended source. The brighter Eastern image indicates that the peak in the dust continuum lies outside and to the West of the inner caustic, with the eastern edge of the extended emission crossing the caustic and forming the faint Einstein ring. Since the continuum peaks outside of the inner caustic, the effects of beam smearing on the source plane reconstruction appear to be straight forward and can be rectified by simply masking the eastern artifact (maps shown in Fig. \ref{fig:GridMaps} display maps with the eastern artifact removed). We do this for all the dust continuum maps (including those from other data sets) in the rest of our analysis. 
		
		In agreement with the distinct CO(9-8) image plane morphology, a distinct source plane morphology is found. The CO(9-8) emission, forms a 'three-pronged' morphology connecting over the inner caustic. The image plane quadruply imaged CO(9-8) emission indicated that the CO(9-8) component crosses the inner caustic, which is indeed seen in the reconstruction. The striking source plane morphology however is not already evident in the image plane and is likely a manifestation of beam smearing of the image plane. We investigate the effects of beam smearing on the source plane reconstruction by taking the model produced by the cleaning procedure and convolving it with different beam sizes (see Appendix \ref{sec:beam}, Fig. \ref{fig:BeamSmearingCO98}). In particular, the North Western 'prong' of the source plane morphology is deemed to be an artifact of beam smearing, but overall the effects of the beam on the small scale source plane structure is complex. For this reason, we do not attempt any source plane masking of the reconstructed CO(9-8) emission. The more extended distribution and spatial offset of the CO(9-8) emission with respect to the dust continuum is already evident in the image plane, and can be interpreted as intrinsic differences between the CO(9-8) and dust components. A more detailed analysis of the source structure would require higher spatial resolution observations.
		
		By construction the \OHp absorption is only observed where there is background continuum, however the reconstructed \OHp optical depth (and thus column density) exhibits a distinctively different source plane morphology than that of all the other components. We find an elongated triangular \OHp distribution, with its central axis running from the east and flaring out towards the West. Unlike the CO(9-8) emission, the elongated morphology of the \OHp optical depth is already evident in the image plane distribution. In particular, the steep fall off in optical depth to the North and South, compared to the background dust continuum, is indicative of a sharp physical transition and is interpreted as a true feature. Despite the background continuum displaying a full Einstein ring, this feature is absent in the \OHp optical depth, indicating that, unlike the dust, CO(9-8) and stellar components, the \OHp distribution does not cross the inner caustic. This is further evidence that the Eastern artifact seen in the continuum reconstruction, and which is also found in the \OHp reconstructions, is an artifact caused by beam smearing in the image plane. We mask this feature in the source plane \OHp maps in the same fashion as for the continuum. Lastly, the \OHp optical depth appears to increase towards and peak in the West. Higher sensitivity and spatial resolution observations are required to show if this trend is indeed true, particularly in the extreme Western edge where the background continuum, and therefore \OHp spectral S/N, decreases significantly. It is likely however that the \OHp distribution extends past the background continuum in this direction, causing a sharp cut-off in the \OHp distribution visible to us via absorption. 
		
		The Keck NIR emission, which shows a similar image plane morphology to the CO(9-8) emission but with a more prominent Einstein ring, displays a similarly extended source plane component, partially crossing the inner caustic. The peaks in stellar and CO(9-8) emission are not cospatial, with the stellar peak lying to the South West of the inner caustic, below that of the peak in dust continuum and \OHp optical depth. We do not apply any source plane mask to the NIR emission as beam smearing is not an issue with this data set. 
		
		In the lower left panel of Fig. \ref{fig:GridMaps} we display a false colour image of all the reconstructed intensity maps. We highlight the bulk offsets of each distribution, in particular, the dip in stellar light at the position of peak dust continuum. Spatial offsets between the gas, dust and stellar components of galaxies at high redshift have previously been observed in many cases (e.g., \citealt{Riechers2010,Chen2017,Rybak2015,Hodge2015,Fujimoto2017,Simpson2017,CalistroRivera2018,Cochrane2021}). The most likely explanation for the offset between stellar and dust components is that the optical component does indeed extend into the dusty regions but experiences high extinction. This scenario is consistent with the dip in stellar intensity observed in G09v1.40 (evident already in the image plane distributions). Additional observations would be needed, however, to concretely justify this claim and we can not reject the possible scenario of true physical misalignments between gas, dust and stars or the possible scenario of two merging systems, one with extreme dust extinction and another optically bright galaxy.
	
		For each component we measure the image plane and source plane luminosities, providing a value for the magnification specific to each component (Tab. \ref{tab:Sprops}). Source plane luminosities are measured with the eastern artifacts removed for the continuum and \OHp maps. We additionally provide the SFR and SFR surface density derived from the $L_{IR}$ measured by \cite{Bussmann2013} and corrected using our new dust continuum magnification factor and source size. Since the source size is the same for both our model and that of \cite{Bussmann2013}, this results in a change of 6\% for both values.

		\begin{deluxetable}{clc}
			\tablecaption{Intrinsic Source Properties.}\label{tab:Sprops}
			\tablecolumns{3}
			\tablewidth{0pt}
			\startdata
			\\
			{\boldmath${\rm L_{FIR}}$}&$[10^{11}\rm{\ L_\odot}]$&$45.5\pm25$\\
			{\boldmath${\rm L'_{CO(9-8)}}$}&$[10^{9}\ {\rm K\ km\ s^{-1}\ pc^2}]$&$3.1\pm1.2$\\
			{\boldmath${\rm L_{NIR}}$}&[$\mu$Jy]&$1.7\pm0.7$\\
			{\boldmath${\rm\mu_{\rm cont.\ model}}$}&&$11.6\pm4.5$\\
			{\boldmath${\rm\mu_{\rm high\ res\ cont}}$}&&$14.5\pm5.6$\\			
			{\boldmath${\rm\mu_{\rm836\ GHz}}$}&&$11.1\pm4.3$\\		
			{\boldmath${\rm\mu_{CO(9-8)}}$}&&$7.7\pm3.0$\\ 			
			{\boldmath${\rm\mu_{OH^+}}$}&&$9.3\pm3.6$\\			
			{\boldmath${\rm\mu_{NIR}}$}&&$11.4\pm4.4$\\	
			{\boldmath${\rm\mu_{CH^+}}$}&&$10.0\pm3.9$\\
			{\boldmath${\rm r_{eff,\ cont.\ model}}$}&[pc]&408$\pm$7.0\\
			{\boldmath${\rm r_{eff,\ NIR}}$}&[pc]&1100$\pm2.0$\\
			\textbf{SFR}&$[\rm M_\odot\ {\rm yr^{-1}}]$&788$\pm300$\\
			{\boldmath$\Sigma_{\rm SFR}$}&$[\rm{M_\odot\ yr^{-1}\ kpc^{-2}}]$&753$\pm290$\\
			\enddata
			\tablecomments{Parameter descriptions are as follows: {\boldmath${\rm L_{FIR}}$}, lens corrected total infrared luminosity (${\rm8-1000\ \mu m}$) using the ${\rm\mu L_{FIR}}$ from \cite{Bussmann2013} and the magnification factor of the high spatial resolution $\CHp\!$(1-0) dust continuum.  {\boldmath${\rm L_{CO(9-8)}}$}, de-lensed CO(9-8) line luminosity. {\boldmath${\rm M_{NIR}}$}, de-lensed NIR magnitude using the apparent NIR luminosity from \cite{Calanog2014} and NIR magnification factor derived in this work. {\boldmath$\mu_{\rm cont,\ model}$}, magnification factor provided by the \texttt{visilens} model of the \OHp underlying continuum, {\boldmath${\rm \mu_{high\ res\ cont}}$}, magnification factor of the spatial resolution $\CHp\!$(1-0) dust continuum. {\boldmath${\rm \mu_{\rm836\ GHz}}$}, magnification factor of the \OHp underlying dust continuum. {\boldmath${\rm \mu_{CO(9-8)}}$}, magnification factor of the CO(9-8) component. {\boldmath${\rm \mu_{OH^+}}$}, magnification factor of the masked, outflowing \OHp component. {\boldmath${\rm \mu_{NIR}}$}, magnification factor of the NIR stellar component. {\boldmath${\rm \mu_{CH^+}}$}, magnification factor of the low spatial resolution $\CHp\!$(1-0) component. {\boldmath${\rm r_{eff,cont,model}}$}, effective radius of the dust continuum derived from the \texttt{visilens} model parameters, where ${\rm r_{eff}=a_S\sqrt{b_S/a_S}}$. {\boldmath${\rm r_{eff,cont,model}}$}, the NIR effective radius from \cite{Calanog2014}. \textbf{SFR}, lens corrected SFR derived from the ${\rm L_{IR}}$ using the \cite{Kennicutt1998} calibration, SFR=$1.73\times10^{-10}\ {\rm L_{IR}\ M_\odot\ yr^{-1}}$ and assuming a Salpeter IMF,  {\boldmath$\Sigma_{\rm SFR}$ lens corrected SFR surface density.}}
		\end{deluxetable}

		We reconstruct the CO(9-8) and blueshifted \OHp velocity and velocity dispersion maps. The \OHp line is significantly blue-shifted with respect to the CO(9-8) at all locations across the source and displays an opposite velocity gradient across the North-East to South-West axis as expected from the image plane velocity maps. Further insight into the intrinsic velocity structure of the CO(9-8) emission should not be read into from the reconstructed CO(9-8) velocity map due to beam smearing, blending components of the CO(9-8) emission on opposite sides of the inner caustic together. Indeed the source plane CO(9-8) velocity field appears comparatively more chaotic than what would be expected from the smooth image plane velocity field (Fig. \ref{fig:VelVeldisp}). This may in turn disguise signatures of rotation in the host galaxy, if present. Fortuitously, this is not an issue in the \OHp maps since the entire \OHp component lies to the West of the inner caustic, and the negative gradient to the West and positive gradients to the North and South are interpreted as real kinematic features.
		
	\subsection{Comparison with Previous Lens Models}
		\begin{table*}
			\centering
			\caption{Best fit parameters from previous gravitational lens models by \cite{Bussmann2013} and \cite{Calanog2014}.} \label{tab:OldPars}
			\begin{tabular}{cccccccccc}
				\hline\hline
				{\boldmath${\rm \theta_E}$}&{\boldmath${\rm e_L}$}&{\boldmath${\rm \phi_L}$}&{\boldmath${\rm n_S}$}&{\boldmath${\rm a_S}$}&{\boldmath${\rm e_S}$}&{\boldmath${\rm r_{eff}}$}&{\boldmath${\rm \phi_S}$}&{\boldmath${\rm \mu_{\rm source}}$}\\
				$[\,^{\prime\prime}\,]$&&[deg] E of N&&$[\,^{\prime\prime}\,]$&[deg]&[kpc]&[deg] E of N&\\
				\hline
				\multicolumn{9}{l}{\bf \cite{Bussmann2013} (SMA $880\ {\rm \mu m}$ source \& lens model)}\\
				$0.553\pm0.004$&$0.06\pm0.02$&$70\pm12$&$2\pm0.7$&$0.06\pm0.01$&$0.33\pm0.14$&$0.41\pm0.08$&$83\pm17$&$15.3\pm3.5$\\
				\multicolumn{9}{l}{\bf \cite{Calanog2014} (NIR $2.2\ {\rm \mu m}$ source \& lens model)}\\
				$0.56^{+0.01}_{-0.02}$&$0.0^{+0.1}_{-0.2}$&$-57^{+4}_{-1}$&$0.51^{+0.02}_{-0.04}$&$0.18^{+0.01}_{-0.01}$&$0.49^{+0.02}_{-0.06}$&$1.1\pm0.002$&$87^{+6}_{-4}$&$11.4^{+0.9}_{-1}$\\
				\multicolumn{9}{l}{\bf \cite{Calanog2014} (NIR source model using SMA lens model)}\\
				&&&&$0.18^{+0.01}_{-0.01}$&$0.51^{+0.03}_{-0.1}$&$1.1\pm0.002$&&$10^{+1}_{-1}$\\
				\hline
			\end{tabular}
			\tablecomments{Parameter descriptions are as follows: {\boldmath${\rm \theta_E}$}, Einstein radius of the lens. {\boldmath${\rm e_L}$}, lens elipticity. {\boldmath${\rm \theta_L}$}, lens position angle. {\boldmath${\rm a_S}$}, source major axis. {\boldmath${\rm e_S}$}, source elipticity (where ${\rm e_S=1-b_S/a_S}$ and ${\rm b_S/a_S}$ is the axis ratio). {\boldmath${\rm n_S}$}, source sersic index. {\boldmath${\rm \phi_S}$}, source position angle. {\boldmath${\rm \mu_S}$}, magnification of the source.}
		\end{table*}

		Gravitational lens models of G09v1.40 have been previously derived by \cite{Bussmann2013} and \cite{Calanog2014} using ${\rm880\mu m}$ Submillimeter Array (SMA) and Keck II Near-Infrared Camera 2 (NIRC2) $2.2\mu m$ observations, respectively. Our lens parameters are consistent with those measured by \cite{Bussmann2013} and \cite{Calanog2014}, summarised in Tab. \ref{tab:OldPars}. The contrast in image plane morphology between the NIR and submm SMA imaging (Einstein ring and double image, respectively) was interpreted by \cite{Calanog2014} as a consequence of poor spatial resolution in the submm data compared to the Keck AO. With our high spatial resolution analysis of the dust continuum in this paper, it is now clear that contrasts in image plane morphology are due to intrinsic differences in the source plane morphology of these components. The almost perfect alignment of G09v1.40 with the lensing galaxy means that small variations and offsets of the stellar, dust and gas components in the source plane, produce strikingly different image plane morphologies and magnifications.
		
\section{Outflow Geometry} \label{sec:Geo}
	With the source-plane reconstructed maps in hand, we now investigate possible geometries of the outflowing gas. Given the limitations in spatial resolution of our data, we compare the suitability of three simple outflow geometries: a flat sheet lifting off a star-forming disk, a spherical outflow originating and expanding from a single location in the galaxy, and a conical outflow. 
	\subsection{Sheet}\label{Sheet}
		For a galactic disk with extended star formation, it is easy to imagine a flat sheet of outflowing gas lifting off perpendicularly from the disk. In this scenario the velocity signature of the disk can be imprinted onto that of the outflow. Both velocity maps will therefore exhibit the same velocity gradients, albeit offset in the RA and Dec plane depending on the inclination of the disk and height of the outflow. It is obvious directly from the image plane CO(9-8) and $\OHp\!(1_1-1_0)$ velocity maps (Fig. \ref{fig:VelVeldisp}) that the molecular gas in the host galaxy and outflowing neutral gas display opposite velocity gradients across their 2D projections. The uncertainty in the velocity fields is likely less than the velocity resolution ($<14.5\ \kms$) of the spectra, and much less than the observed velocity gradients, given the high integrated S/N of each spaxel. The observed velocity gradients seen in the CO(9-8) and \OHp are therefore believed to be truly disparate, and we dismiss this outflow geometry. 
		
	\subsection{Spherical}\label{Spherical}
		In this scenario, we explore a spherical shell expanding uniformly from a single location, such as an active galactic nucleus or central star-forming region. To explore this geometry we have created a simple toy model that converts an expanding 3D spherical shell into a 2D projected velocity map. As we are using an absorption line in our analysis and therefore only probe gas in front of the galaxy, we consider only the front-facing hemisphere in our model. The process of creating and comparing the model outflow with the observed \OHp outflow is as follows.
		
\begin{figure}
			\begin{center}
				\includegraphics[width=\linewidth]{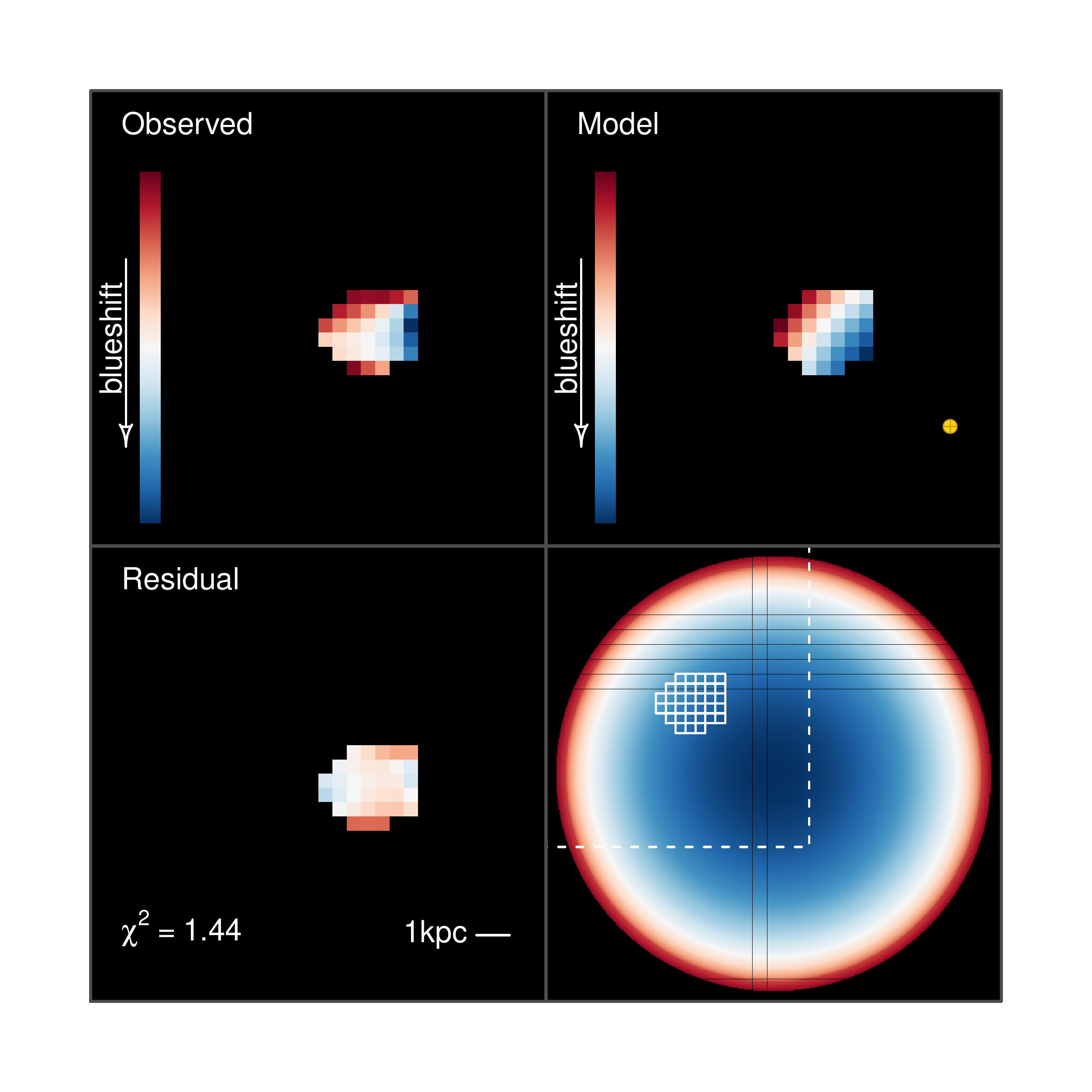}
				\caption{Example spherical outflow model compared to the observed $\OHp\!(1_1-1_0)$ velocity map. The model suggests the outflow is ejected from outside the galaxy and is therefore ruled out. {\bf Top Left:} Reconstructed and normalised velocity map of the observed outflow in $\OHp\!$. {\bf Top Right:} Spatially averaged model velocity map from region in the dashed box in the panel below. The yellow marker indicates the ejection location predicted by this model. {\bf Bottom Left:} Residual velocity map of the normalised model velocity map and normalised \OHp velocity map. {\bf Bottom Right:} Full 2D LOS velocity map of the spherical outflow model with a dashed box outining the edges of the panel above.}
				\label{fig:BestFitModel16}
			\end{center}
		\end{figure}

		To build a hemispherical outflow with radius R and finite thickness dR, we first create a 3D box with dimensions $2(R+1)\times2(R+1)\times(R+1)$ and grid it with 10 times the spatial resolution of our observed source plane data maps (0.05'' or $\sim0.43$ kpc). In this model the third, shorter axis is parallel with the Line Of Sight (LOS) and the center of the outflow is placed at the farthest distance along this axis from the observer and in the center of the other two axes such that the full outflow fits within the box. Grid elements of the box that lie within R and R-dR are then assigned values equal to their LOS positions. The box is then collapsed and averaged along the LOS axis to create a 2D map of the average LOS position for each location across the face of the outflow. Since we assume the outflow is moving radially outwards, this 2D map has an identical gradient to the average deprojected outflow velocity and can thus be used to compared to the 2D deprojected velocity map of the observed outflow (see bottom right panel of Fig. \ref{fig:BestFitModel16}). 
		
		Since in reality, we can only observe the outlflowing gas situated in front of the dust continuum we must then select small regions of the model outflow velocity field to compare with our data. This is done systematically: moving pixel by pixel across the face of the projected model velocity field, we cut regions with matching pixel dimensions of the observed reconstructed \OHp velocity map. This allows us to determine where the observed outflow may lie with respect to the ejection point. Once cropped, the model velocity field is spatially averaged to match the resolution of the observed field and both maps are normalised such that only the gradient of the fields are compared. A residual map and $\chi^2$ value are derived for each region selected from the model velocity map, allowing us to find the most probable position of our observed outflow with respect to the ejection location (see Fig. \ref{fig:BestFitModel16} for an example). 
		
		We find that increasing the radius of the toy model monotonically reduces the $\chi^2$ value of the best-fit velocity map. This in turn predicts a larger and larger distance of the outflow ejection point with respect to the observed outflow, placing it far outside the host galaxy (Fig. \ref{fig:BestFitModel16}). Adopting a spherically symmetric model would therefore imply accepting the unphysical situation that the origin of the outflow lies far outside the host galaxy and the model is therefore rejected. We further note that a spherical outflow with an ejection point directly at the peak of the dust continuum would produce a deprojected velocity field with the peak in projected outflow velocity at the same position. This is not the case for the observed \OHp velocity field (see Fig. \ref{fig:GridMaps}) and is obvious already in the image plane velocity fields (Fig. \ref{fig:VelVeldisp}). 
		
	\subsection{Conical}\label{sec:Cone}
		Following the failure of the sheet and spherical geometries in characterising the observed outflow morphology in G09v1.40, we consider a conical outflow geometry. In this scenario, gas is ejected from a localised region in the host galaxy and expands as it flows radially away from the galaxy. We therefore expect to observe the vertex of the conical outflow co-spatial with signatures of the ejection mechanism, i.e., the peak in the dust continuum where star formation is assumed to be at a maximum. 
		
		As discussed in Sec. \ref{sec:lenstool}, the absence of an Einstein ring in the image plane \OHp optical depth map indicates that the outflow does not extend over the inner caustic towards the East. Similarly, the \OHp optical depth decreases towards the North and South (evident in both the image and source planes) away from the peak of the dust continuum intensity, indicating that we are observing the true edge of the \OHp distribution to the East, North and South. This can not be said for the \OHp distribution in the Western direction where the \OHp component likely extends farther out than the dust component but becomes invisible to us without the background continuum to absorb. 
		
		With these points in mind, the elongated triangular morphology revealed in the source plane \OHp optical depth map can be interpreted as the 2D projection of a 3D conical structure viewed from outside the opening angle (i.e., not observed 'down the barrel'). The vertex of this rough isosceles triangle sits co-spatial with the peak of the dust continuum (Fig. \ref{fig:GridMaps}), and flares out and away from the dust continuum peak towards the West. Additionally, there is a mild negative gradient in LOS velocity observed along the East to West axis, and an increase in velocity dispersion. This may be indicative of an outflow driven over an extended period of time, as opposed to a single ejection event. 
		
\begin{figure}
			\begin{center}
				\includegraphics[width=\linewidth]{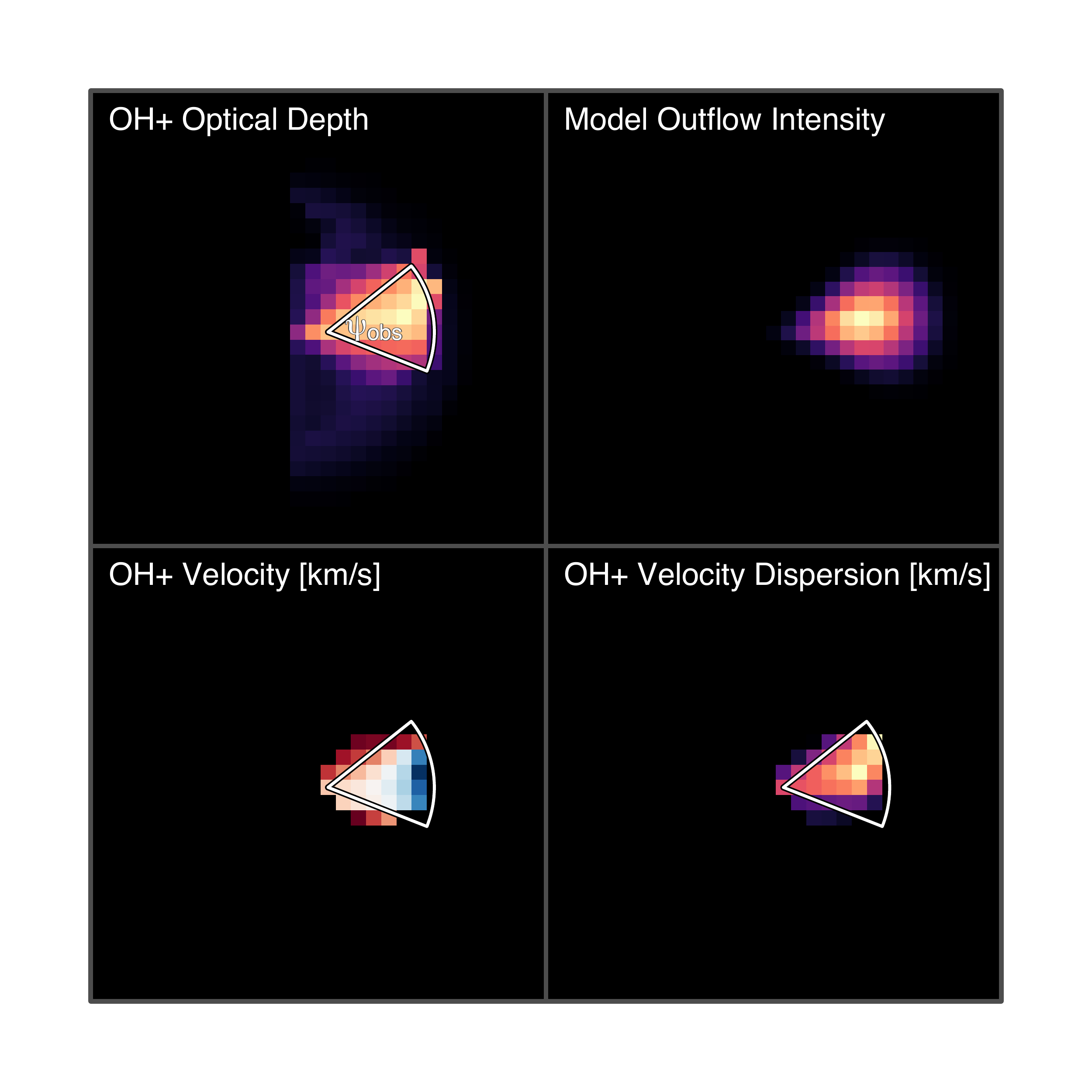}
				\caption{Example Conical outflow model compared to the observed \OHp optical depth map. {\bf Top Left:} Source plane \OHp optical depth map with measured 2D projected cone geometry overlaid in white. The projected radius is denoted by $R_{obs}$ and 2D projected opening angle by $\Delta\phi$. {\bf Top Right:} 2D projection of a model conical outflow with constant radial density. {\bf Bottom Left:} Source plane \OHp line of sight velocity map with measured 2D projected cone geometry overlaid in white. Outflow velocity increases with radius. {\bf Bottom Right:} Source plane \OHp velocity dispersion map with measured 2D projected cone geometry overlaid in white. Outflow velocity dispersion increases with radius. }
				\label{fig:ConeModel}
			\end{center}
		\end{figure}

		To illustrate this model we employ our simple toy model introduced in Sec. \ref{Spherical}, to construct multiple 3D cone models over a range of inclinations, such that they all have the same 2D projected radius, $R_{obs}$, opening angle, $\Delta\phi_{\rm obs}$, and position angle $\alpha_{\rm obs}$ observed in the source plane \OHp optical depth map (Fig. \ref{fig:ConeModel}). We note that the radius of the outflow measured directly from the source plane \OHp optical depth map is larger than its true value due to the effects of beam smearing and is used in this toy model for illustrative purposes only.
		
		We find that all conical models produce similar projected morphologies (Fig. \ref{fig:ConeModel}), and reproduce the general characteristics observed in the \OHp optical depth map. The conical outflow geometry is therefore the most suitable geometry to describe the observed \OHp outflow in G09v1.40 and is therefore adopted throughout the rest of our analysis, and expanded upon in Sec. \ref{sec:MOFR}. 
		
		An obvious difference between the data and model projected outflow morphologies is the offset in peak optical depth, with the optical depth in the data peaking towards the `edge' of the cone. If real, this can be explained by an outflow extending past the background continuum, causing the fainter far edge of the cone to become invisible to the observer and thus bringing the observed edge closer to the peak in observed optical depth. We stress however that such detailed interpretation of the source plane structure should be taken with extreme care, given the effects of the beam. Higher resolution observations are needed to investigate the structure within the outflow.
		
		Whilst we do not attempt to model the velocity structure of the conical outflow we note that there is a slight negative radial gradient in LOS velocity, indicating an increase in outflow speed at larger radii, which is coupled with an increase in velocity dispersion. Assuming the cone geometry is the correct choice for this outflow, the positive trend of outflow velocity and velocity dispersion with radius is indicative of an outflow driven over an extended period of time, as opposed to a single ejection event. This interpretation however should be taken simply as a suggestion given the velocity difference from one end of the outflow to the other is a mere $\sim40\kms$. Additional observations and analysis are required to confirm this scenario.
		
\section{Chemical Properties of the Gas} \label{sec:Chemistry}
	In addition to dynamical and morphological information, molecular spectral lines offer insight into the physical state of their media through an understanding of the chemical and physical requirements to form them. We first address the state of the ISM in the host galaxy as traced by the CO(9-8) emission line observed at systemic velocities in Sec. \ref{sec:CO}. We then discuss the formation processes and necessary environmental conditions required to produce the \OHp($1_0-1_1$)  and $\CHp\!$(1-0) transitions separately (Sec. \ref{sec:OHp} and \ref{sec:CHp}), before commenting on the physical state of the outflowing gas as traced by both of these transitions. 
	
	\subsection{CO(9-8)}\label{sec:CO}
		High redshift DSFGs contain large molecular gas reservoirs \citep{Carilli2013} which fuel their rapid ongoing star formation. CO rotational lines can be used to constrain the kinetic temperature and gas density when observed over multiple-J lines. Non-local thermodynamic equilibrium (non-LTE) models of the CO spectral line energy distributions (SLEDs) in high redshift heavily star-forming galaxies suggest there are likely two (or more) excitation components dominating the CO emission (e.g., \citealt{Ivison2010,Danielson2011,Yang2017}) in these galaxies. The low excitation component corresponds to a cooler extended molecular gas reservoir, dominating the global CO SLED at low J transitions. The high excitation component needed to explain the mid/high J transitions, on the other hand, is produced by a warmer, denser, and more compact molecular gas reservoir believed to closely trace the ongoing star formation.
		
		Fifteen $z\sim2$ SMG CO SLEDs, including that of G09v1.40, were analysed by \cite{Yang2017} using a large velocity gradient (LVG) statistical equilibrium method. Fitting the CO(2-1), (4-3), (6-5) and (7-6) transition lines, G09v1.40 required a two-component model, indicating that the emission in CO(9-8) traces the warm, dense and more compact molecular gas dominating the high excitation component likely associated with ongoing star formation. 
		
		High CO excitation is also observed in galaxies habouring powerful AGNs capable of dominating the IR luminosity. Significant boosting of the high CO($J>10$) transition lines is believed to be caused by AGN heating within the central hundreds of pc around the nucleus \citep{vdWerf2010} and may contribute significantly to the excitation of CO(9-8). From our data we derive an intrinsic CO(9-8) luminosity in G09v1.40 of ${\rm L'_{CO(9-8)}} =3.1\times10^{9}\ {\rm K\ km\ s^{-1}\ pc^2}$, which is consistent with the two-component CO SLED presented by \cite{Yang2017} and does not indicate obvious signs of an AGN boost. Furthermore, in a scenario where the AGN is significantly contributing to the thermal dust continuum and high J CO transition lines, we would expect these two-components to be co-spatial. This is not the case for G09v1.40 where the CO(9-8) emission is both offset and more extended than that of the dust continuum (Fig. \ref{fig:GridMaps}). We therefore maintain our assumption that the CO(9-8) emission observed in G09v1.40, is excited predominantly via mechanisms associated with ongoing star formation.
		
	\subsection{\OHp}\label{sec:OHp}
		The observed \OHp absorption line is clearly blue-shifted with respect to the systemic velocity of G09v1.40 (Fig. \ref{fig:CO98OHvelfit},\ref{fig:VelVeldisp},\ref{fig:GridMaps}) and therefore must trace outflowing gas in this system. It is important however to further constrain what phase of the outflowing gas is traced by \OHp, in order to accurately derive properties and interpret the multiphase outflow as a whole. We do this by addressing the chemistry required to produce \OHp absorption.
		
		In the cool diffuse neutral ISM, neutral-neutral reactions advance slowly, allowing ion-neutral reactions to dominate the chemistry when in the presence of an external ionisation field. In these conditions, chemical species with a first ionisation potential less than that of neutral hydrogen (13.6 eV) will be predominantly ionised by the incident far-ultraviolet (UV) radiation, and species with a first ionisation potential $>13.6$ eV, such as oxygen, O, are shielded by the abundant atomic hydrogen, H. Many of the reaction networks of the latter are therefore kicked off by \Hp and $\Hthreep\!$, predominantly formed via cosmic ray (CR) ionisation.
		
		Indeed, the dominant formation pathway of \OHp in the cool diffuse ISM begins with the ionisation of neutral H by a CR: 
			\begin{eqnarray} \label{CROHp}
				{\rm H}+{\rm CR}\rightarrow&{\rm H}^++e^-+{\rm CR'}
				\label{OH1}
			\end{eqnarray}
		${\rm O}^+$ can then be formed via an endothermic charge transfer between O and \Hp,
			\begin{eqnarray}  	
				{\rm H}^++{\rm O}+\Delta{\rm E}\leftrightarrow &{\rm O}^++{\rm H}
				\label{OH2}
			\end{eqnarray}	
		which proceeds backwards, uninhibited, in an exothermic reaction. The neutralisation of \Hp via $e^-$ capture and charge transfer with Polycyclic Aromatic Hydrocarbons (PAHs) also counteract the production of ${\rm O}^+$. 
		
		The remaining ${\rm O}^+$ can react with H$_2$, 
			\begin{eqnarray}
				{\rm O}^++{\rm H}_2\rightarrow&{\rm OH}^++{\rm H}
				\label{OH6}
			\end{eqnarray}
		to produce \OHp, which can then be destroyed by dissociative recombination, photodissociation or proceed further along the oxygen chemistry network through abstraction reactions with H$_2$,
			\begin{eqnarray}
				{\rm OH}^++{\rm H}_2&\leftrightarrow{\rm H_2O^+}+{\rm H}.
				\label{OH7}
			\end{eqnarray}
		The rapidity by which this abstraction process proceeds creates a sensitive relationship between the \OHp and ${\rm H_2}$ abundances. 
		
		\OHp can alternatively form via,
			\begin{eqnarray}
				{\rm O}+{\rm H}_3^+\rightarrow&{\rm OH}^++{\rm H}_2.
				\label{OH8}
			\end{eqnarray}
		However, this reaction requires a significantly higher molecular hydrogen fraction in combination with low $e^-$ abundances (fuller discussions on oxygen chemistry in the ISM can be found in \citealt{Hollenbach2012,Indriolo2015}).
		
		As the first oxygen bearing ion to form after the ionisation of H, \OHp is a key ingredient in constraining the physical and chemical properties of the ISM, including the cosmic ray density and molecular hydrogen fraction. Analyses of Milky Way sight lines have shown that \OHp predominantly traces the cool diffuse gas in the ISM where hydrogen is primarily neutral \citep{Gerin2016}. It is therefore expected that the bulk of \OHp present in the ISM forms via Eq.\ref{OH6}, with formation via Eq.\ref{OH8} only dominating within the opaque and predominantly molecular interiors of molecular clouds \citep{Hollenbach2012,Indriolo2018}. 
		
		\cite{Bialy2019} further investigated the large scatter in \OHp-to-neutral hydrogen column density ratios, N($\OHp\!$)/N(H), measured in Milky Way sightlines, in the context of a turbulent medium. The abstraction of hydrogen in Eq.\ref{OH7} means \OHp is highly sensitive to the abundance of \Htwo which in turn is sensitive to density fluctuations in the underlying turbulent medium \citep{Bialy2017}. Using magnetohydrodynamic (MHD) simulations they modelled increasingly turbulent density fields and post-processed them with chemical models to obtain probability density functions of the abundances. The model that best reproduced the observations required high levels of turbulence suggesting that turbulence is an important factor in the production of \OHp in the cool diffuse ISM. 
		
		Observational studies analysing \OHp (and ${\rm H_2O^+}$) absorption in the high-redshift galaxies SMM J2135-0102 and SDP 17b (${\rm z}\sim2.3$, \citealt{Indriolo2018}) similarly conclude that the bulk of the \OHp resides in cool diffuse gas, surrounding the galaxies in massive extended haloes. 
		
		We therefore interpret the blueshifted \OHp absorption measured in our observations of G09v1.40 as a tracer of the predominantly atomic gas phase component in the outflow. 
		
	\subsection{\CHp}\label{sec:CHp}
		To further constrain our analysis of the neutral outflowing gas as traced by ${\rm OH^+}$, we include an analysis of the two ancillary observations of the $\CHp\!$(1-0) transition in G09v1.40. Before comparing the two light hydride data sets directly, we first explore the chemistry required to form \CHp and the physical conditions needed to produce its spectral lines. 
		
		\CHp can form via the endothermic reaction between ionised C and molecular hydrogen,
			\begin{eqnarray}
			{\rm C}^++{\rm H_2}+h\nu\rightarrow&{\rm CH}^++{\rm H},
			\label{CHroute}
			\end{eqnarray}
		but requires temperatures $\gtrsim10^3$K.
		
		In the diffuse ISM, such temperatures can be reached locally via the intermittent dissipation of turbulent energy \citep{Godard2009}. Due to the high critical density of the J=1-0 transition, ${\rm n_{crit}\sim 10^7\ cm^{-3}}$, most of the \CHp in this diffuse environment will be in the ground state, causing high J=1-0 line opacities. \CHp absorption in the MW has been observed in several hundreds of sight lines, with abundances of the molecule scaling positively with the turbulent energy transfer rate of the diffuse molecular gas, supporting the scenario where \CHp is formed predominantly via the reaction shown in Eq.\ref{CHroute} \citep{Godard2014}. Local extragalactic observations similarly find \CHp absorption in turbulent environments, such as the medium surrounding the supernova SN 2014J, in M82 \citep{Ritchey2015}, and the ISM of the starburst galaxy Arp 220 \citep{Rangwala2011}. 
		
		In regions of dense gas ($n_H\!>\!10^5{\rm\ cm^{-3}}$), sufficiently illuminated by ultraviolet radiation and undergoing suprathermal heating, \CHp can be observed in emission \citep{Godard2013}. Star-forming regions such as DR21 are well modelled by a C-shock scenario where the approaching magnetic field causes sudden heating of the upstream neutral gas via ion-neutral friction, resulting in very wide emission signatures \citep{Falgarone2010}. \CHp emission observed in the nearby ULIRG Mrk231 \citep{vdWerf2010} is also likely due to the very strong UV field and photon dense regions present in this source.
			
		\begin{figure}
			\begin{center}
				\begin{tabular}{c}
					\includegraphics[width=\linewidth]{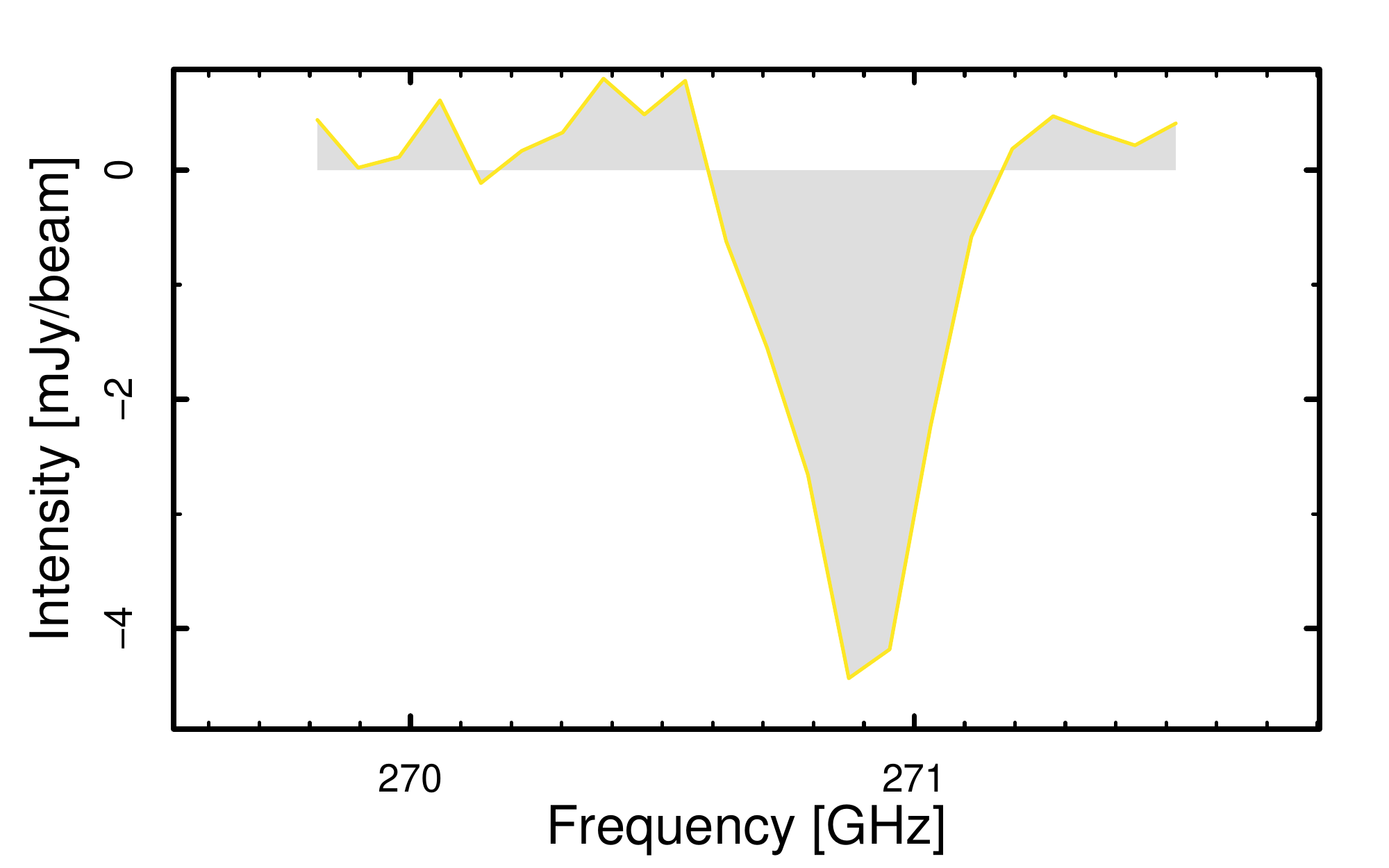}\\
					\\[-25pt]
					\includegraphics[width=\linewidth]{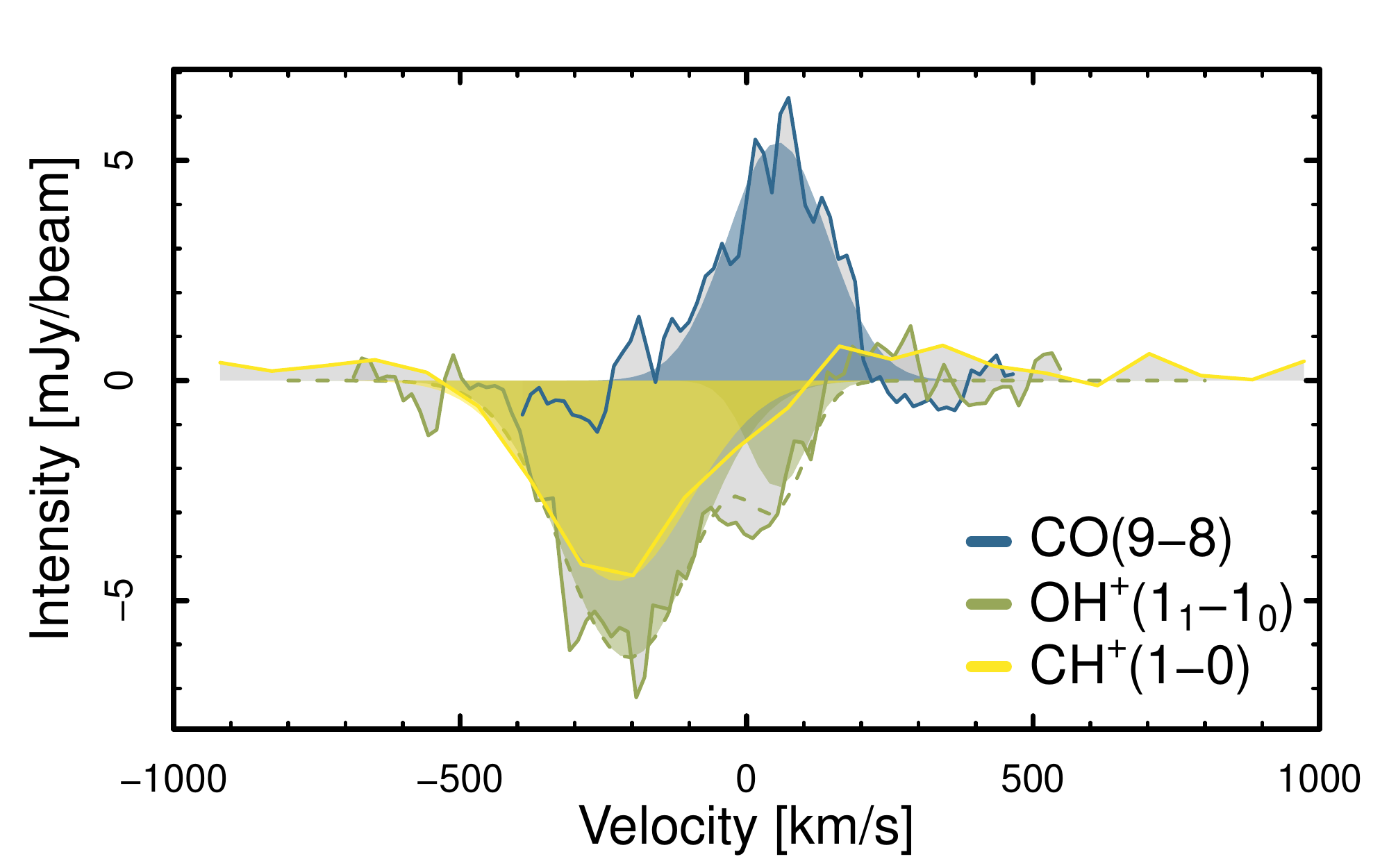}
				\end{tabular}
				\caption{Continuum subtracted $\CHp\!$(1-0) spectra of a single ALMA spaxel (pixel area = $0.05^{\prime\prime}\times0.05^{\prime\prime}$) in the low spatial resolution \CHp data set, compared with the $\OHp\!(1_1-1_0)$ and CO(9-8) spectra previously presented in Fig. \ref{fig:CO98OHvelfit}, at the same physical location in G09v1.40. \textbf{Top:} \CHp spectrum plotted as a function of observed frequency. \textbf{Bottom:} \CHp spectrum (yellow solid line) plotted over the \OHp (green) and CO(9-8) (blue) lines as a function of velocity with respect to the galaxy's systemic velocity \citep{Yang2017}. The \CHp spectrum is fitted with a single Gaussian function shown in shaded yellow (\OHp and CO(9-8) spectral fits are as indicated as described in Fig. \ref{fig:CO98OHvelfit}). The \CHp line is blue-shifted $\sim250$ \kms with respect to the bulk molecular gas, tracing the same kinematic component as the \OHp absorption. }
				\label{fig:CHvelfit}
			\end{center}
		\end{figure}
		
		\cite{Falgarone2017} presented results finding both \CHp absorption and emission in G09v1.40, as part of a sample of six $z\sim2$ DSFGs. They conclude that the broad ($>$1000 \kms) \CHp emission arises in shocked gas associated with galactic winds driven by the central starbursts, whilst the narrower \CHp absorption lines must trace gas outside the galaxy. \CHp absorption traces the cool diffuse gas sitting in turbulent halos around these galaxies, which are mechanically fueled by the outflowing gas from the central starburst. In the sample of six DSFGs, there are in total 4 blueshifted \CHp absorption lines (only 3 are reported by \cite{Falgarone2017} as the spectrum of G09v1.40 was wrongly displayed due to an incorrect redshift).
			
	\subsection{Comparison of \OHp and \CHp}\label{sec:comp}
		Given the similarities between \OHp and \CHp chemistry, we perform a simple analysis of the 'low' and 'high' spatial resolution $\CHp\!$(1-0) observations. 
		
		The low-resolution observations were previously studied by \cite{Falgarone2017} and included self-calibration in their reduction of the data. \cite{Falgarone2017} report overlapping \CHp absorption and emission lines at central velocities of $111\pm7\ \kms$ and $28\pm34\ \kms$ and velocity dispersions of $1124\pm87$ \kms and $361\pm24$ \kms, respectively, using an incorrect redshift of 2.0894 (quoted in \citealt{Bussmann2013}). In our re-analysis of the data, we make use of the automatic ALMA pipeline products and adopt the precise redshift of $z=2.0924\pm0.0001$ derived by \cite{Yang2017} from multi-J CO spectra, which agrees within errors with the redshift derived from our CO(9-8) of z= 2.093. With this correction, the \CHp absorption has a central velocity of $\sim-250$ \kms, blue-shifted with respect to the bulk molecular gas and closely follows the \OHp absorption, as shown in Fig. \ref{fig:CHvelfit}. We find $\sim4\times$ weaker \CHp emission compared to that reported by \cite{Falgarone2017} when we stack the spectra over the source. 
		
		\begin{figure}
			\begin{center}
				\includegraphics[width=\linewidth]{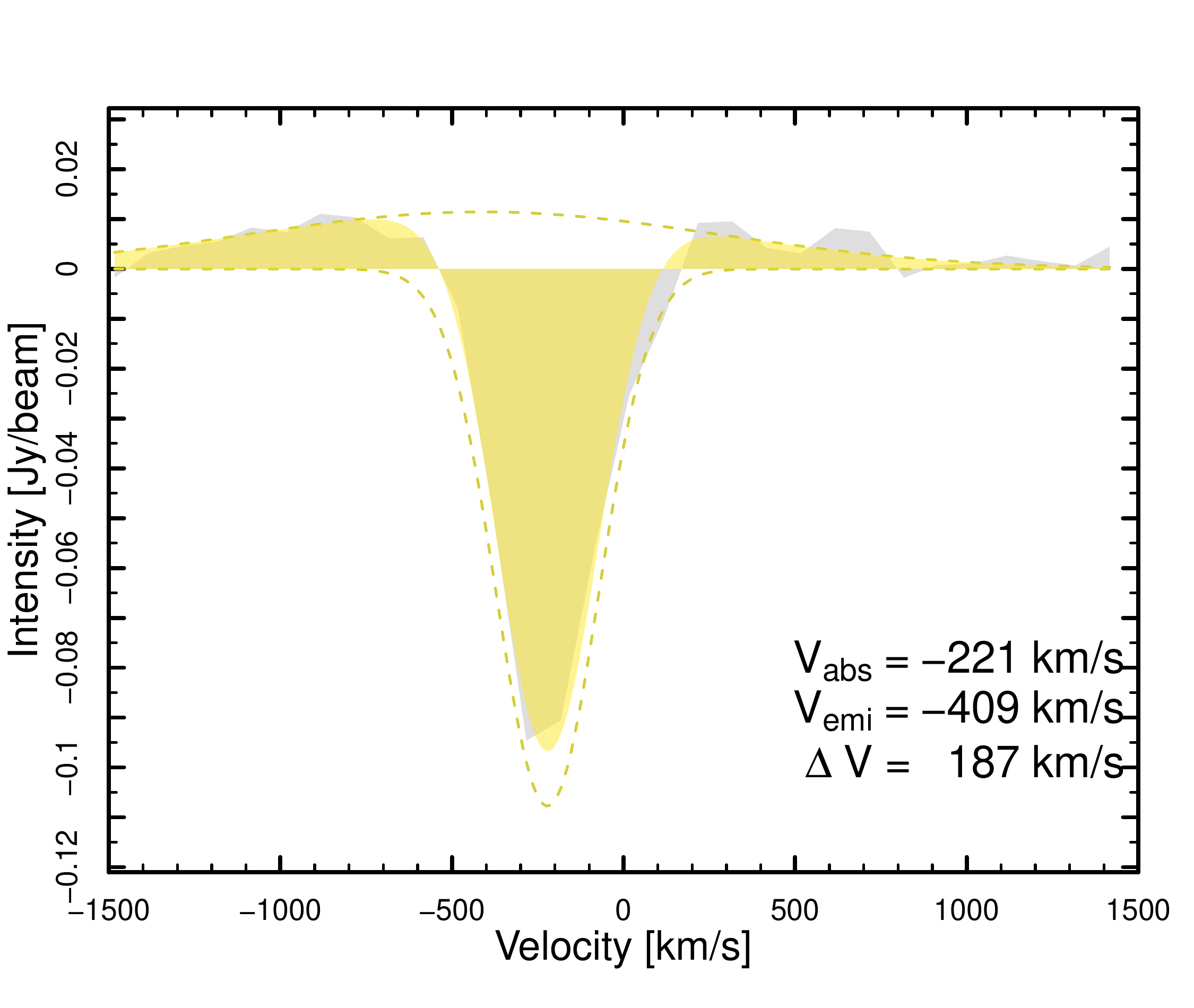}
				\caption{Stacked $\CHp\!$(1-0) spectra of all spaxels with continuum signal to noise $>3$, in the high spatial resolution data set. The spectra is fitted with two Gaussian components, shown separately with the dashed yellow lines and combined with the shaded yellow region. The blue-shifted absorption and emission lines reveal outflowing neutral and shocked gas with line of sight velocities of ${\rm V_{abs}=-221\ \kms}$ and ${\rm V_{emi}=-409\ \kms}$, respectively. The strength of the \CHp emission is much weaker, and its central velocity much more blue-shifted than previously reported by \cite{Falgarone2017} using the low spatial resolution data set.}
				\label{fig:CHstackfreq}
			\end{center}
		\end{figure}
		
		To investigate this further we analyse the high-resolution $\CHp\!$(1-0) observations. We image this data set using a robust weighting of 0 and select spectral channels greater than 1500 \kms away from the line center for the continuum modelling and subtraction so as to avoid any contamination of the emission. We then stack all spaxels with a continuum S/N$>3$ and again find only a weak signature of the wide \CHp emission line, approximately $3\times$ lower than that reported by \cite{Falgarone2017}. 
		
		To analyse the spatial distributions of the $\CHp\!$(1-0) absorption and emission in G09v1.40 we return to the low-resolution data set as these observations provide higher signal to noise in the spectra of the \CHp line in each spaxel, and a more comparable beam size to the \OHp observations.
		
\begin{figure}
			\begin{center}
				\includegraphics[width=\linewidth]{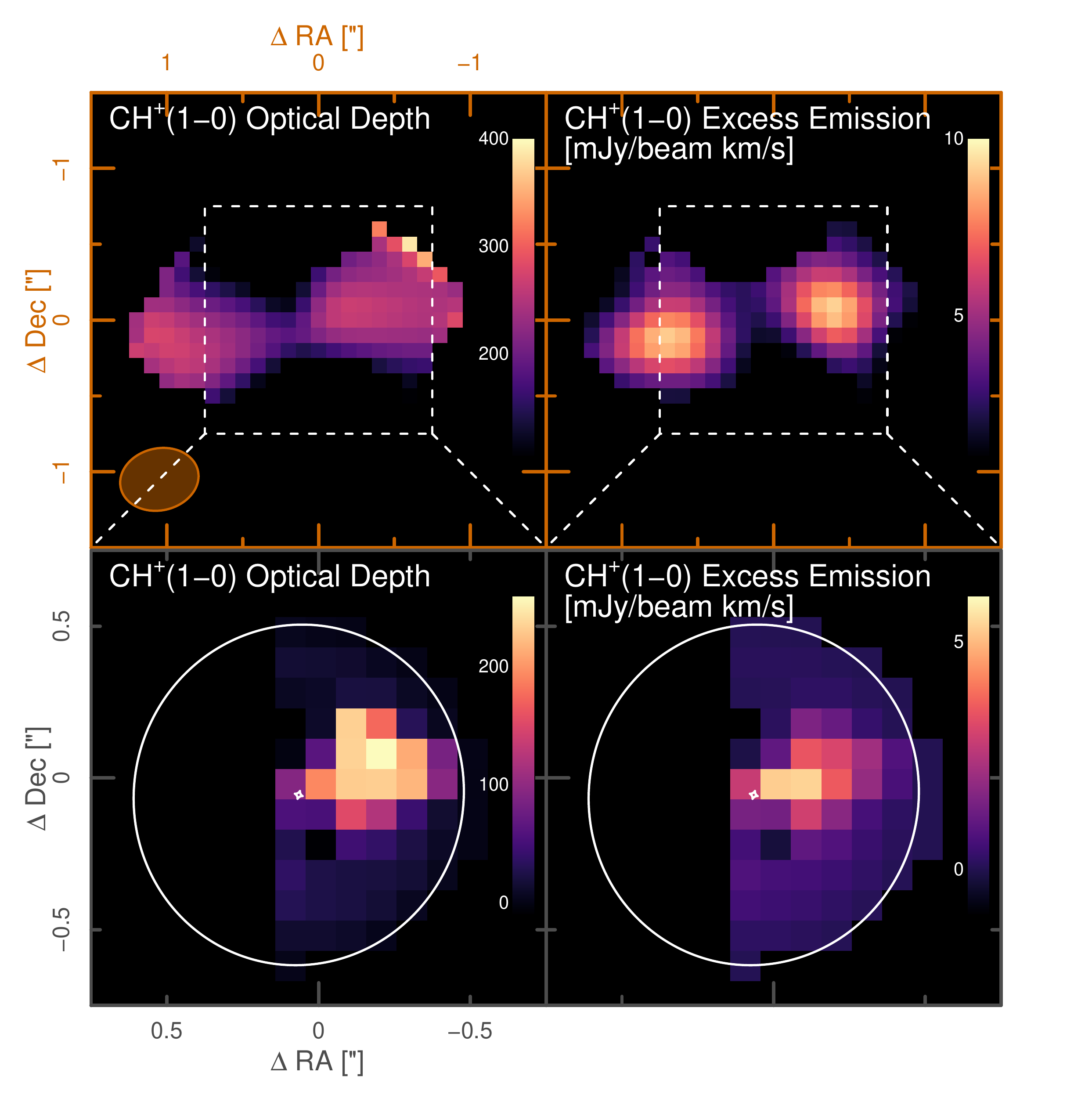}
				\caption{\textbf{Top Row:} Image plane maps of the $\CHp\!$(1-0) absorption line optical depth and excess $\CHp\!$(1-0) emission in the low-resolution \CHp data set. The beam ($0.59^{\prime\prime}\times0.46^{\prime\prime}$) is displayed in the bottom left corner of the left-hand panel. \textbf{Bottom Row:} Source plane reconstructions of the above panels. Whilst the \CHp optical depth map follows the \OHp optical depth map morphology, the \CHp emission follows more closely the dust continuum intensity distribution. Pixel area ($0.1^{\prime\prime}\times0.1^{\prime\prime}$) is the same in both the image and source planes.}
				\label{fig:CHpmaps}
			\end{center}
		\end{figure}

		The same spectral fitting routine introduced in section \ref{sec:Results} is applied to $\CHp\!$(1-0) absorption line in each spaxel of the low-resolution data set. We do not attempt to fit simultaneously for the emission and absorption due to the low S/N of the emission line. From each spaxel, we subtract the best-fit gaussian to the \CHp absorption and sum the residuals. We interpret a positive summed residual as excess $\CHp\!$(1-0) emission and provides an approximate distribution of the $\CHp\!$(1-0) in both absorption and emission. The weak \CHp emission appears to be compact and co-spatial with the dust continuum whilst the absorption covers a more extended area, cospatial with that of the \OHp absorption in our data (Fig. \ref{fig:CHpmaps}). 
		
		The kinematic and spatial coincidence of the \CHp and \OHp absorption lines indicate these molecules are tracing the same diffuse, turbulent, and predominantly atomic gas reservoir. The compact and central spatial distribution of the \CHp excess emission (to be interpreted cautiously), is consistent with the scenario of \CHp emission tracing shocked regions of dense gas in close proximity to a strong ultraviolet radiation source (i.e. the central starburst region). 
		
\section{Derived Outflow Properties} \label{sec:MOFR}
	In this section we derive key outflow properties, (total neutral gas mass, mass outflow rate (${\rm\dot{M}_{OF}}$), kinetic energy flux $\dot{\rm E}$ and momentum flux $\dot{\rm p}$: Table \ref{tab:OFprops}), to further investigate its ejection mechanism and to quantify the impact of the outflow on the evolution of its host galaxy. 

	\subsection{Outflow Mass}
		The $\OHp\!(1_1-1_0)$ column density ${\rm N}_{\rm OH^+}$ and integrated optical depth are related through 
		
		\begin{eqnarray}
		\int\tau_\nu dv=&\frac{\lambda^3}{8\pi}{\rm AN}_{\rm OH^+}
		\end{eqnarray}
		where ${\rm A}=2.11\times10^{-2}{\rm\ s}^{-1}$ is the Einstein coefficient and $\lambda\approx0.029{\rm\ cm}$ is the wavelength of the transition (see Appendix \ref{sec:N} for derivation). This gives an average ${\rm N}_{\rm OH^+}$ value in the source plane of $2.4\times10^{14}\ {\rm cm}^{-2}$. 
		
		To derive a total neutral H column density we must now adopt an \OHp abundance, ${\rm X}_{\rm OH^+\!}={\rm N_{OH^+}/N_{H}}$. The formation of \OHp is sensitive to the cosmic-ray ionisation rate of atomic hydrogen (see Eq. \ref{CROHp}), allowing its abundance relative to that of the ${\rm H_2O^+}$ ion to be used as a constraint on the radiation field (Gerin et al. 2010; Neufeld et al. 2010; Porras et al. 2014; Indriolo et al. 2015; Zhao et al. 2015). Observational studies utilizing this method have uncovered a wide range in ionisation rates from as low as $10^{-17}{\rm s^{-1}}$ in dense gas of the Milky Way, to rates of $10^{-13}{\rm s^{-1}}$ in the nuclear regions of ULIRGs \citep{GonzalezAlfonso2013,GonzalezAlfonso2018}. Naively we may expect observations such as the latter, or those targeting the disks of nearby starburst galaxies \citep{vdTak2016} to match the conditions of high z DSFGs such as G09v1.40, as is indeed reflected in the ionisation rate estimate of the star-forming region in SMM J2135-0102 ($10^{-13}-10^{-11}{\rm s^{-1}}$, \citealt{Danielson2013}). Observations of \OHp and ${\rm H_2O^+}$ absorption in the same galaxy (and in that of SDP 17b), however, show ionisation rates several orders of magnitude lower \citep{Indriolo2018}, reflecting ionisation levels and locations in ${\rm N}(\OHp)/{\rm N(H)}$ vs. ${\rm N}_{\OHp}/{\rm N_{H_2O^+}}$ parameter space seen in diffuse clouds of the Milky Way \citep{Indriolo2015,Neufeld2017}. This supports the scenario where \OHp absorption traces the same extended, turbulent haloes of neutral gas surrounding high-z DSFGs traced by \CHp absorption \citep{Falgarone2017}, where the large distances from the central starburst region supplying the cosmic-ray flux, results in a dramatic decrease in ionization rate.
		
		Following the evidence provided above, we therefore adopt an \OHp abundance of $\log_{10}({\rm X}_{\rm OH^+\!})=\log_{10}({\rm N_{OH^+}/N_{H}})\approx-7.8\pm^{0.05}_{0.075}$ from \cite{Bialy2019}. This is the mean value derived in their best-fitting and most turbulent isothermal magnetohydrodynamic (MHD) simulation of the diffuse neutral medium which they compare with values observed in Milky Way sightlines. They stress that the dispersions derived in their best-fit models do not reflect the scatter measured in the Milky Way observations, and so we do not adopt the formal uncertainties of their models in our derivations. We instead consider the highest and lowest observed abundances, ($\log({\rm X}_{\rm OH^+})\approx-7.4$ and $-8.2$ ), from the comparison sample of Milky Way sightlines as extreme cases. Outflow properties derived using the extreme observed abundances are listed in Table \ref{tab:OFprops} and shown as dashed curves in Figs. \ref{fig:IncFuncs}, \ref{fig:Energyflux} and \ref{fig:Momflux}. We further note that the average H column densities found in G09v1.40 are $\sim2\times$ higher than in the models of \cite{Bialy2019} and caution that all abundances derived above from both models and observations do not contain non-equilibrium chemistry and may not match other physical properties of outflowing gas and gas at high redshift.

		With ${\rm X}_{\rm OH^+\!}$ in hand, the total neutral gas mass of the outflow, by summing over all the pixels (i,j) in the source plane, is given by,
		\begin{eqnarray}
			{\rm M_{neut}} =  1.36\mHI\sum_{\rm i,j}{\rm N}_{\rm OH^+}{\rm A_{pix}}/{\rm X}_{\rm OH^+},
		\end{eqnarray}
		where the 1.36 factor is the correction for the helium abundance, \mHI is the mass of a hydrogen atom and ${\rm A_{pix}}$ is the area of a single pixel in ${\rm cm^2}$.
		
		Following this method we measure a total neutral gas mass of the outflow in G09v1.40 of ${\rm M_{neut}} = 6.7\times10^9\ {\rm M_{\odot}}$. This is more than $25\%$ as massive as the molecular gas mass (corrected for the helium abundance) in the host galaxy, ${\rm2.5\times10^{10}M_{\odot}}$: derived via non-LTE radiative transfer modelling of multi-J CO lines \citep{Yang2017}. Whilst the uncertainty associated with the \OHp abundance likely dominates, we also note that this measurement excludes any part of the outflow that extends past, or lies behind the dust continuum but cannot be observed in absorption.
		
		As mentioned earlier, the diffuse turbulent and predominantly atomic gas component traced by \OHp absorption is believed to be the same component seen in \CHp absorption \citep{Falgarone2017}. Using a turbulent framework to analyse the global \CHp absorption spectra, \cite{Falgarone2017} derived a radius of the full turbulent reservoir around G09v1.40 of ${\rm r_{TR}=12}$ kpc. They then extrapolate the column densities observed over the dust continuum (0.41 kpc), to both sides of the galaxy and out to 12 kpc, finding a total neutral gas mass of the full turbulent reservoir of $1.1\times10^{10}\ {\rm M_{\odot}}$. Thus to compare with the mass derived from \OHp, we scale the mass full reservoir mass by a factor of ${\rm 0.41\ kpc/(2\times12\ kpc)}$, giving a value of $0.19\times10^9\ {\rm M_{\odot}}$, $3\%$ that of our derived mass. We note however that the turbulent framework used by \cite{Falgarone2017} to convert the observed \CHp absorption into a total neutral gas mass is a very different approach to our own.
		
		For a more direct comparison with our work, we additionally convert the \CHp optical depth map (Fig. \ref{fig:CHpmaps}) into a total neutral outflow gas mass in the same way that we have done for the $\OHp\!$. We first convert the source plane \CHp optical depth map into \CHp column density following the equation presented by \cite{Falgarone2017} in their supplementary methods. We then convert this into a neutral H column density using an average observed \CHp abundance of $7.6\times10^{-9}$ \citep{Godard2014}. Summing over the entire outflow we find a total neutral outflow gas mass of $4.2\times10^9\ {\rm M_{\odot}}$, comparable with that derived from \OHp. 
		
	\subsection{Mass Outflow Rate}	
		\begin{figure}
			\centering
			\includegraphics[width=0.80\linewidth]{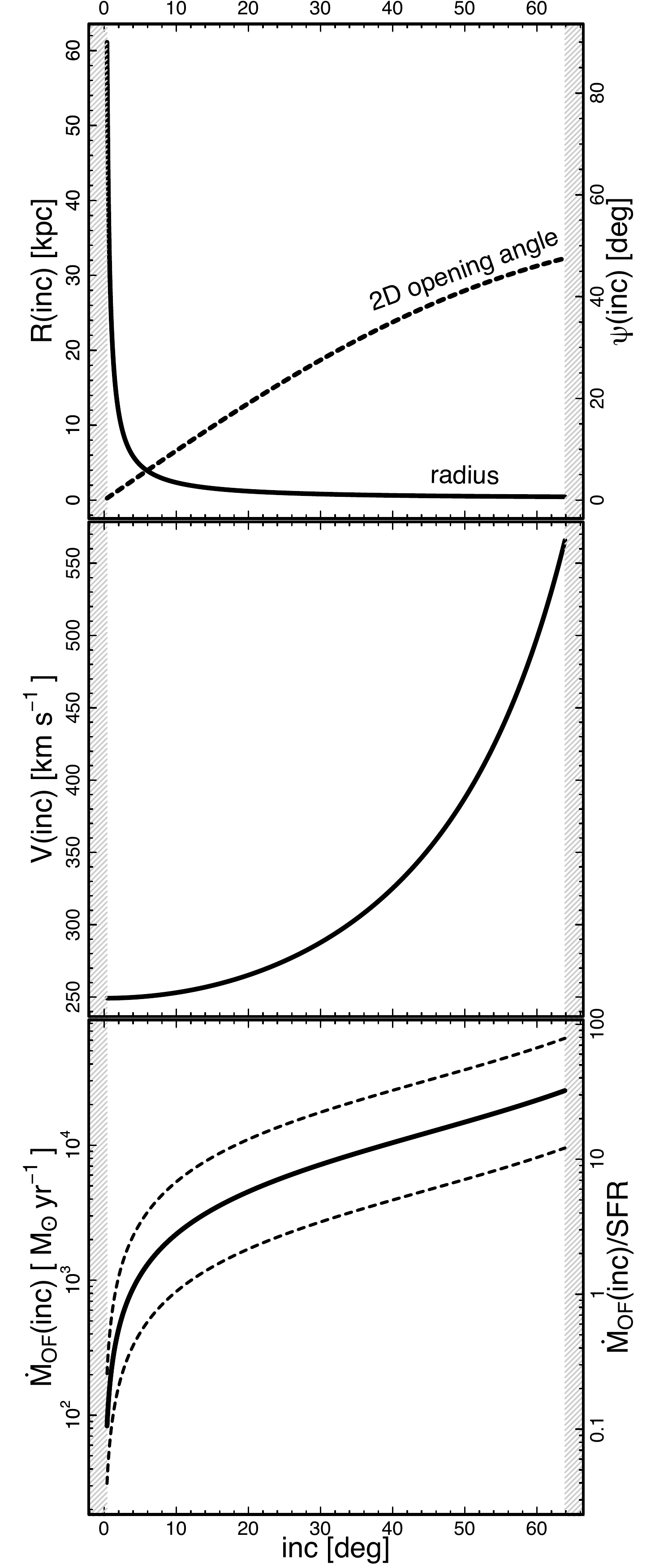}
			\caption{Deprojected outflow parameters as a function of inclination where $inc=90^\circ$ corresponds to an outflow central axis perpendicular to the line of sight. \textbf{Top:} Radius (solid line) and 2D opening angle (dashed line). \textbf{Middle:} Velocity. \textbf{Bottom:} ${\rm\dot{M}_{OF}}$ (solid line) with 1$\sigma$ spread of the modelled \OHp abundance from \cite{Bialy2019} shown by the shaded region. The upper and lower dashed lines indicate the ${\rm\dot{M}_{OF}}$ if the highest and lowest observed \OHp abundances from Milky Way sightlines are assumed, respectively \citep{Bialy2019}. The right-hand axis displays the ${\rm\dot{M}_{OF}}$ normalised by the SFR indicating that if the inclination of the outflow is $>3.6^\circ$ then the neutral gas ${\rm\dot{M}_{OF}}$ exceeds the SFR in G09v1.40. Hatched regions indicate regions outside of the considered inclination range.}
			\label{fig:IncFuncs}
		\end{figure}
		
		\begin{figure*}
			\centering
			\includegraphics[width=1\linewidth]{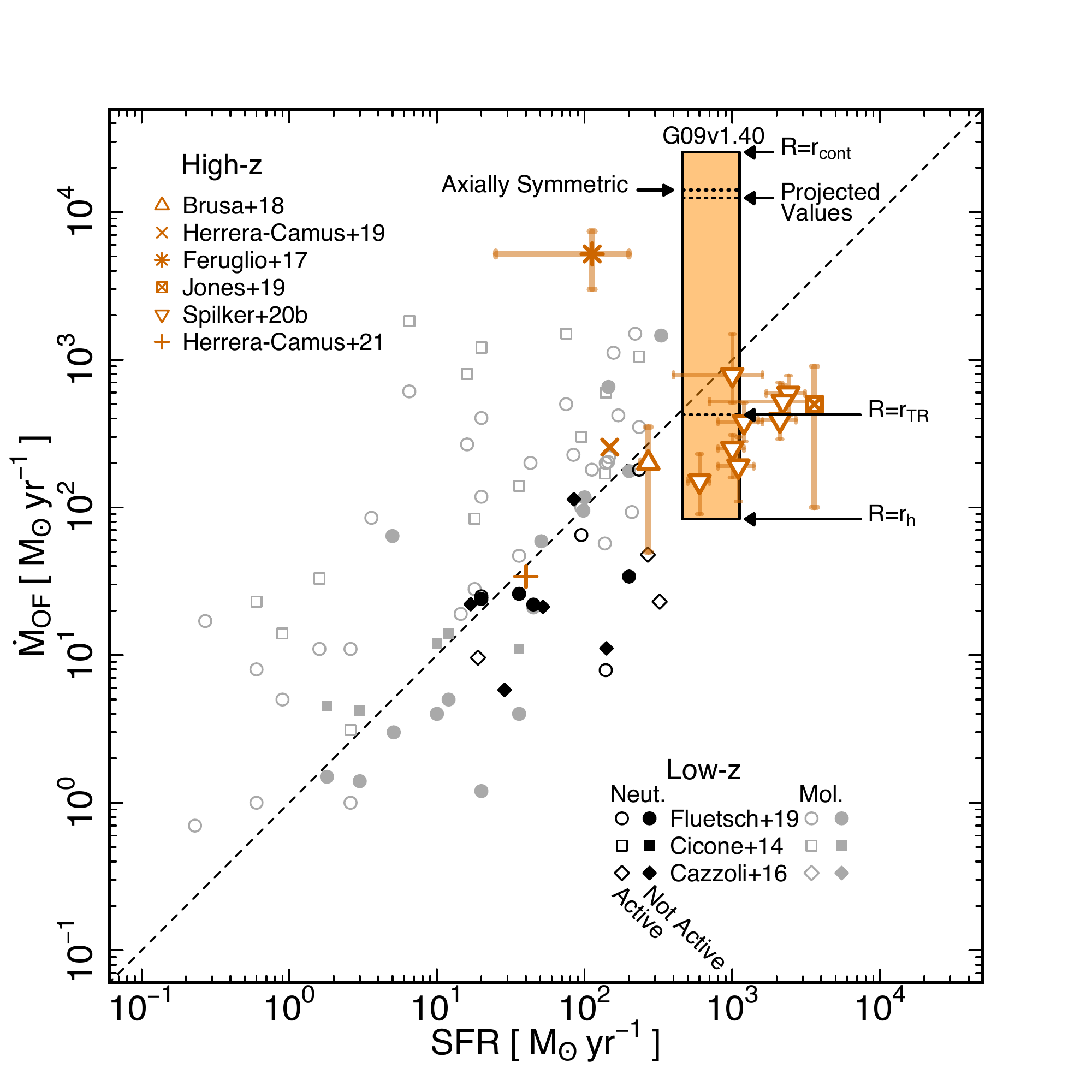}
			\caption{Mass outflow rate as a function of star formation rate. The shaded light orange region indicates the parameter space of the neutral gas outflow in G09v1.40 over the possible inclination range discussed in Sec. \ref{sec:MOFR}. We additionally indicate ${\rm\dot{M}_{OF}}$s estimated for an axially symmetric galaxy with a perpendicular outflow (upper dashed line), using only projected outflow parameters (middle dashed line), and for an outflow with a radius equal to that of the turbulent reservoir surrounding G09v1.40 \citep{Falgarone2017} (lower dashed line). For comparison we plot molecular and neutral gas outflows measured in local galaxy samples with grey and black symbols, respectively \citep{Fluetsch2019,Cicone2014,Cazzoli2016}. Open and filled symbols indicate sources with and without AGN activity, respectively. With orange symbols we show high redshift molecular outflows measured in the following sources, listing their galaxy type, outflow tracer and redshift in brackets: XID2028 (QSO, CO, z = 1.593; \citealt{Brusa2018}), zC400528 (AGN, CO, ${\rm z = 2.387}$; \citealt{Herrera-Camus2019}), APM08279+5255 (QSO, CO, ${\rm z = 3.912}$; \citealt{Feruglio2017}), SPT 0346-52 (DSFG, ${\rm H}_2{\rm O}$, z=5.656; \citealt{Jones2019}), and various SPT sources (DSFG, OH, ${\rm z>4}$; \citealt{Spilker2020b}). Also in orange, we include the neutral gas outflow in HZ4 (main-sequence star-forming galaxy, [CII], $z\sim5$ \citealt{Herrera-Camus2021}).} 
			\label{fig:MOFRvSFR}
		\end{figure*}	
	
		Methods of estimating mass outflow rate, ${\rm\dot{M}_{OF}}$, vary among the literature and depend on the assumed geometry of the outflow \citep{Veilleux2005,Veilleux2020}. As concluded in Sec. \ref{sec:Geo}, the most likely basic geometry of the neutral outflow observed in G09v1.40 is conical. The ${\rm\dot{M}_{OF}}$ is therefore given by 
		\begin{equation}
			{\rm\dot{M}_{OF} =  3\frac{MV}{R}}
		\end{equation}
		where M is the total mass, V the velocity, R the maximum radius of the cone. The factor of 3 accounts for two assumptions: 1) the cone is filled and not just a thin shell; 2) the density of the cone is constant with radius. These assumptions are based on the observed morphology of the \OHp optical depth presented in Fig. \ref{fig:ConeModel}, which is best reproduced by a toy model implementing these assumptions.
		
		As previously discussed, we only observe and measure the 2D projected values of these parameters and require the inclination ($inc$) of the outflow, with respect to the observer's line of sight, to correct for this effect. As our data is not of high enough quality to measure the inclination of the outflow with confidence, we instead opt to place sensible bounds on the inclination and derive a range of possible deprojected geometrical parameters and ${\rm\dot{M}_{OF}}$s. 
		
		We begin by measuring the 2D projected opening angle, ${\rm\psi_{obs}\approx60^\circ}$ rad, of the outflow using the \OHp optical depth map (Fig. \ref{fig:ConeModel}). If the outflow is flowing perpendicular to the LOS, $inc=90^\circ$ , the de-projected 2D opening angle, ${\rm\psi}$, is indeed ${\rm\psi}={\rm\psi_{obs}}$, however if $inc<90^\circ$ then ${\rm\psi}<{\rm\psi_{obs}}$. If we then assume that all the outflowing gas is situated in front of the galaxy (i.e. no part of the cone may have an inclination larger than $90^\circ$), then the maximum possible inclination of the cone's central axis is given by $\psi/2=90^\circ-inc\simeq64^\circ$. 
		
		We cannot determine a lower limit to the inclination based only on the measured projected parameters. Instead, we assume that the outflow does not extend farther than the halo virial radius ${\rm r_h}$, thus putting a limit on the radius of the outflow which increases rapidly at small inclinations (Fig. \ref{fig:IncFuncs}). Taking the stellar radius-halo radius value (${\rm SRHR\simeq0.018}$, defined as the ratio of galaxy radius to halo virial radius) measured by \cite{Somerville2018} and the effective half-light radius of the reconstructed NIR distribution in G09v1.40 (Table \ref{tab:OldPars}) we estimate a halo radius of ${\rm r_h=0.018r_{eff,NIR}}=61$ kpc. The minimum inclination of the outflow possible is then given by $inc_{\rm min}={\rm asin}({\rm R_{obs}/R_{h}})\simeq0.38^\circ$.

		Thus, we take an inclination range of $inc=0.4^\circ-64^\circ$ that we believe the observed conical outflow may have. This corresponds to a range of possible outflow radii of $R={\rm R_{obs}/sin(inc)=0.45-61\ kpc}$, 2D opening angles of $\psi=\arctan(\tan(\psi_{min})\sin(inc))=0.2^\circ-24^\circ$, where ${\rm\psi_{min}}=2(90^\circ-inc_{\rm max})$, and outflow velocities of ${\rm V}={\rm V_{obs}}/\cos(inc)=250-570\ \kms$ where $V_{obs}$ is the maximum LOS velocity in the source plane \OHp velocity map. The neutral gas ${\rm\dot{M}_{OF}}$ in G09v1.40 may then have a value within the range of ${\rm {\rm\dot{M}_{OF}}=83-25400\ M_{\odot}\ yr^{-1}}$, which exceeds the SFR$={\rm788\pm300\ M_{\odot}\ yr^{-1}}$ if the inclination is above $3.6^\circ$ (see Fig. \ref{fig:IncFuncs}). This corresponds to mass-loading factors of the neutral gas outflow between 0.11-32. 
		
		In Fig. \ref{fig:MOFRvSFR} we compare the neutral gas ${\rm\dot{M}_{OF}}$ of G09v1.40 with local neutral and molecular outflow samples from both active and purely star-forming galaxies \citep{Fluetsch2019,Cicone2014,Cazzoli2016}, along with the small sample of measured molecular and neutral outflows at high redshift \citep{Brusa2018,Herrera-Camus2019,Feruglio2017,Jones2019,Spilker2020b,Herrera-Camus2021} as a function of SFR. For similar SFRs, it is evident in the low redshift samples that molecular outflows and outflows driven by AGNs have higher ${\rm\dot{M}_{OF}}$s than neutral outflows and outflows driven by star formation. For G09v1.40, the range in deprojected ${\rm\dot{M}_{OF}}$ spans more than two orders of magnitude, comparable to the scatter seen in the full comparison sample. 
		
		Since it is difficult to compare the range of ${\rm\dot{M}_{OF}}$ derived for G09v1.40 with those in the literature, where deprojection of outflow parameters is either not attempted or derived for a single assumed inclination, we consider three additional derivations of ${\rm\dot{M}_{OF}}$: 1) using only projected values, 2) assuming an axially symmetric disk and perpendicular outflow and 3) assuming an alternative maximum outflow radius equal to the radius of the diffuse turbulent halo surrounding G09v1.40. 
		
		To derive ${\rm\dot{M}_{OF}}$ using projected outflow parameters we take the LOS outflow velocity, ${\rm V_{obs}}$, and the effective radius of the dust continuum, ${\rm r_{eff,\ cont.\ model}}$, provided by the \texttt{visilens} model. This is a comparable method to that used by, e.g. \cite{Spilker2020b} and gives a ${\rm\dot{M}_{OF}}$ of ${\rm12500\ M_{\odot}\ yr^{-1}}$, which we mark in Fig. \ref{fig:MOFRvSFR}, placing it at the most extreme end of observed outflows at all SFRs. 
		
		Assuming an axially symmetric disk model, we estimate an inclination of $49^\circ$ for G09v1.40 using $\sin(inc)=(1-(b/a)^2)/(1-q_0)$, where the axial ratio, $(b/a)=0.679$, is provided by the best-fit \texttt{visilens} model, and a typical intrinsic thickness of $q_0 = 0.2$ is assumed \citep{ForsterSchreiber2020}. This is the method used by \cite{Herrera-Camus2021} for the neutral outflow in HZ4 and corresponds to a deprojected velocity of 380 \kms, an outflow radius of 540 pc and a ${\rm\dot{M}_{OF}}$ of ${\rm14100\ M_{\odot}\ yr^{-1}}$, comparable to that derived for the projected case.
		
		As an alternative maximum outflow radius, we consider the radius of the turbulent halo of diffuse neutral gas seen in \CHp \citep{Falgarone2017}, believed to be the same reservoir containing \OHp \citep{Indriolo2018}. \cite{Falgarone2017} analysed the \CHp halo surrounding G09v1.40 using a turbulent framework, estimating a radius of ${\rm r_{TR}=12}$ kpc. This corresponds to lower limits on the outflow velocity and ${\rm\dot{M}_{OF}}$ of 250 \kms and ${\rm 420\ M_{\odot}\ yr^{-1}}$, respectively, comparable with other high-z molecular outflows.
		
		Despite the large range in ${\rm\dot{M}_{OF}}$ that can be derived for G09v1.40, it appears that the neutral outflow is at least comparable, if not considerably more extreme than that of molecular outflows observed at high redshift. In a study of eight nearby AGN and star formation driven outflows observed across their ionised, atomic and molecular phases, the molecular gas was found, on average, to dominate the total mass and ${\rm\dot{M}_{OF}}$ of their outflows \citep{Fluetsch2020}. The neutral gas dominates in only two purely star-forming galaxies. The authors suggest that the more powerfully driven AGN outflows are compacted more by the ambient CGM, leading to the observed higher gas densities and thus higher molecular gas fractions in these outflows. This, however, is certainly not the rule and comparable or greater neutral gas fractions have also been found in sources hosting an AGN, such as in the low redshift Seyfert systems Mrk 231, Mrk 273 and IRAS F08572+3915 (see compilation by \citealt{Herrera-Camus2020}). It is therefore perhaps not surprising that the ${\rm\dot{M}_{OF}}$ of the neutral outflowing gas component found in G09v1.40 is comparable to or greater than the average molecular outflows observed in similar galaxies. 
		
		In the case of a highly inclined outflow in G09v1.40, the extreme ${\rm\dot{M}_{OF}}$ derived can be reconciled theoretically with a highly obscured QSO scenario \citep{Costa2018}. When multi-scattering radiation pressure from IR radiation on dust grains is taken into account, outflows of predominantly cool gas, and peak ${\rm\dot{M}_{OF}}$ on the order of ${\rm 10^3-10-^4\ M_{\odot}\ yr^{-1}}$, can be produced. This phase is short-lived ($<10$ Myr) and requires the QSO to be heavily enshrouded in dense gas; a possible scenario for a compact DSFG like G09v1.40. We venture further into the required driving mechanisms in the following subsections. 
		
	\subsection{Outflow Energetics}
		\begin{figure}
			\centering
			\includegraphics[width=\linewidth]{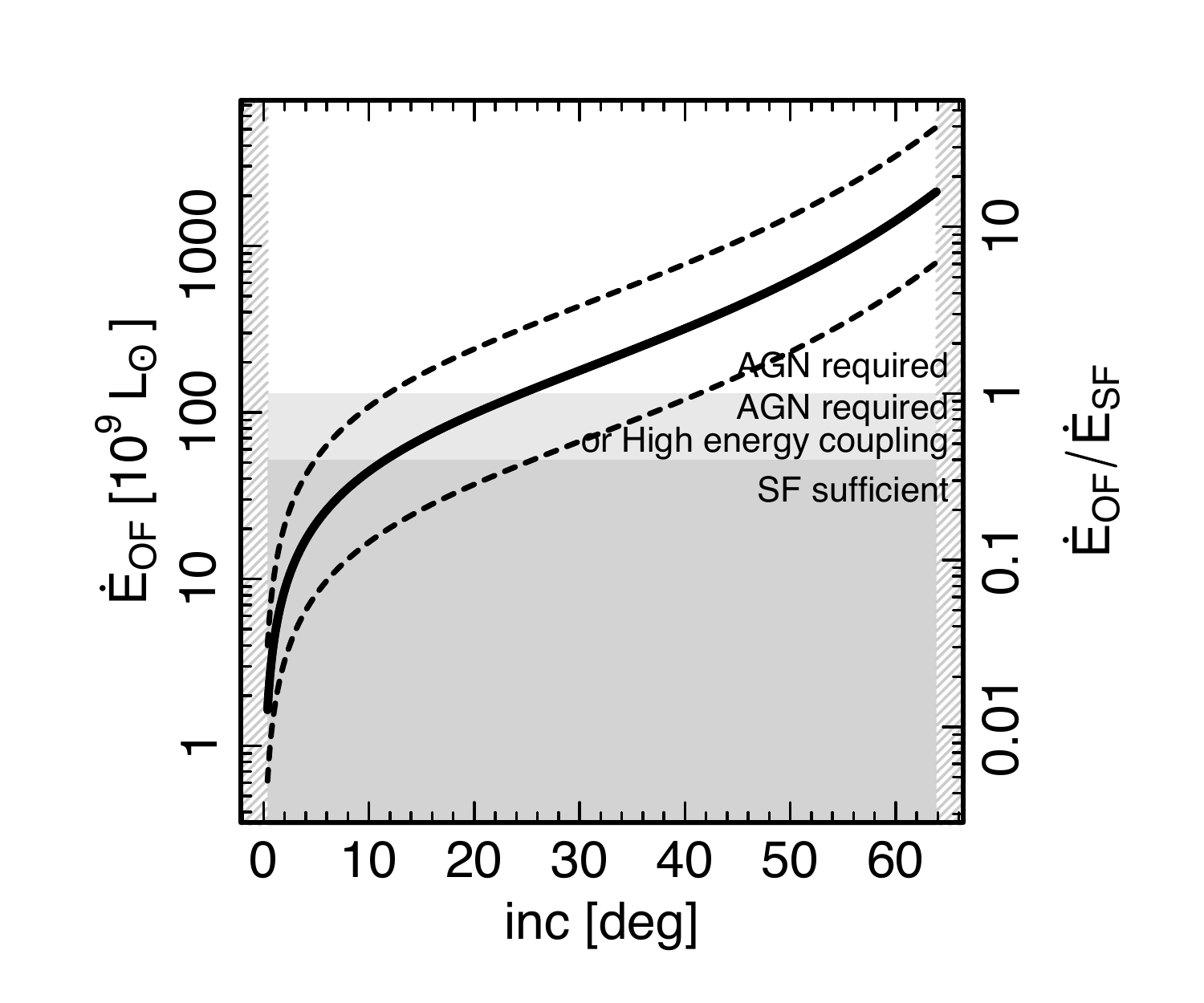}
			\caption{Deprojected outflow kinetic energy flux (kinetic luminosity) of the outflow ${\rm\dot{E}_{OF}}$ as a function of inclination (left axis) and normalised by the kinetic energy flux injected by SNe ${\rm\dot{E}_{SF}}$ (a.k.a. energy loading, right axis). The bottom grey region indicates the energy transferred to the ISM assuming an energy coupling up to $40\%$, which can occur if SNe are clustered. Higher ${\rm\dot{E}_{OF}}$ require either unusually high energy coupling or an AGN contribution.The upper and lower dashed lines indicate the ${\rm\dot{E}_{OF}}$ if the highest and lowest observed \OHp abundances from Milky Way sightlines are assumed, respectively \citep{Bialy2019}. An outflow flowing perpendicular to the line of sight has an $inc=90^\circ$. Hatched regions indicate regions outside of the considered inclination range.}
			\label{fig:Energyflux}
		\end{figure}
	
		In this section, we consider the possible sources and mechanisms required to drive energy and momentum flux of the observed neutral outflow in G09v1.40. We again approach this problem considering the range of possible inclinations (see Sec. \ref{sec:MOFR}), as both the momentum and energy flux of the outflow depend strongly on the inclination assumed and therefore presents multiple possible feedback and outflow scenarios.
		
		We begin by deriving the kinetic energy flux (a.k.a. kinetic luminosity/ kinetic power) of the outflowing neutral gas ${\rm\dot{E}_{OF}}$:
		
		\begin{eqnarray}
		{\rm\dot{E}_{OF}}&=&\frac{1}{2}{\rm\dot{M}_{OF}}({\rm V}^2+3\sigma^2),
		\end{eqnarray}
		
		using the minimum value in the source plane velocity dispersion map, $\sigma=77$ \kms. We find possible values ranging between ${\rm\dot{E}_{OF}}=1.6-2100\times10^{9}\ {\rm L_\odot}$ which we compare to the expected kinetic energy flux injected by SNe, ${\rm\dot{E}_{SF}}=7\times10^{41}({\rm SFR/M_{\odot} yr^{-1}})\ {\rm erg\ s^{-1}}=130\times10^{9}\ {\rm L_\odot}$ \citep{Veilleux2005} (Fig. \ref{fig:Energyflux}) providing ratios between ${\rm\dot{E}_{OF}/\dot{E}_{SF}=0.012-16}$. The fraction of ${\rm\dot{E}_{SF}}$ that is ultimately coupled to the ISM, and therefore used in driving the outflow, depends strongly on the clustering of SNe in the galaxy and gas-phase metallicity and structure of the ISM. If clustering is strong, SN-driven super-bubbles can retain as much as 40\% of the input energy (e.g. \citealt{Sharma2014,Fielding2018}) which would provide enough energy to drive the neutral outflow in G09v1.40 if the inclination is $\leq12^\circ$. If $inc>12^\circ$ then the outflow requires either an extremely high and unusual energy coupling efficiency or an additional source of energy flux, i.e. a past, low-luminosity or obscurred AGN, to be driven. If $inc>24^\circ$, a coupling efficiency of $>100\%$ is required and an AGN contribution must certainly be playing a role to preserve energy conservation. 
		
		Kinetic coupling efficiencies of AGN driven outflows can be up to $\sim0.1$ (see Fig. 2 in compilation by \citealt{Harrison2018}), which would imply an AGN with luminosity of at least $\sim8\times10^{46}\ erg\ s^{-1}$ in G09v1.40 for the most extreme scenario. This is well within the range of observed AGN luminosities and does not exclude any of the high inclination outflow scenarios.
		
		Next, we consider whether the outflow is consistent with a momentum or energy-driven scenario. Momentum-driven outflows occur when the thermal energy of the shocked SN ejecta is efficiently radiated away. The observed momentum flux of the outflow must then be supplied by the momentum flux deposited directly by SNe ejecta or by radiation pressure from young stars on dust grains in the outflow.
		
\begin{figure}
			\centering
			\includegraphics[width=\linewidth]{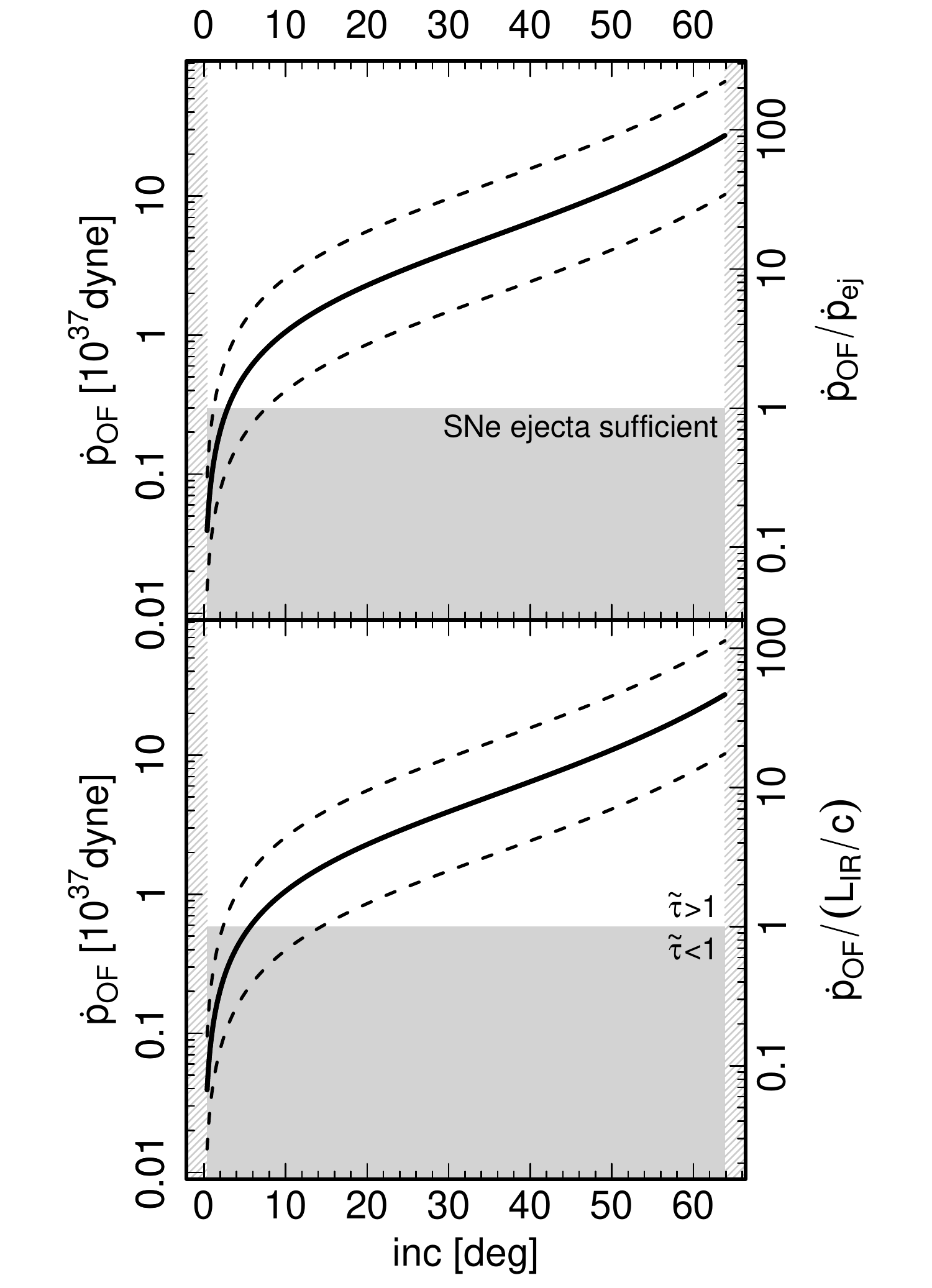}
			\caption{Deprojected momentum flux of the outflow ${\rm\dot{p}_{OF}}$ (left axes) as a function of inclination: \textbf{Top:} Comparing ${\rm\dot{p}_{OF}}$ to the momentum flux injected by SNe ejecta ${\rm\dot{p}_{ej}}$ (right axis) where the grey region indicates the momentum flux available from the estimated SNe rate ${\rm\sim15\ SNe\ yr^{-1}}$. \textbf{Bottom:} Comparing ${\rm\dot{p}_{OF}}$ to the possible momentum flux contributed by radiation pressure from young stars on outflow dust grains. Values on the right axis indicate the required effective IR optical depth for a given ${\rm\dot{p}_{OF}}$ where the optically thin regime is indicated by the grey region. The 1$\sigma$ spread in the \OHp abundance modelled by \cite{Bialy2019} is shown by the dark gray shading and the upper and lower dashed lines indicate the ${\rm\dot{E}_{OF}}$ if the highest and lowest observed \OHp abundances from Milky Way sightlines are assumed, respectively \citep{Bialy2019}. An outflow flowing perpendicular to the line of sight has an $ inc=90^\circ$. Hatched regions indicate regions outside of the considered incliniation range.}
			\label{fig:Momflux}
		\end{figure}

		First, we derive the momentum flux of the outflowing neutral gas ${\rm\dot{p}_{OF}}$:
		
		\begin{eqnarray}
		{\rm\dot{p}_{OF}}&=&{\rm \dot{M}_{OF}V},
		\end{eqnarray}
	
		finding a possible range of ${\rm\dot{p}_{OF}}=0.040-27\times10^{37}$ dyne which we display on left axes in Fig. \ref{fig:Momflux}. We then estimate the momentum flux expected to be deposited by SN ejecta as the product of the supernova rate (${\rm\sim15\ SNe\ yr^{-1}}$) and the momentum associated with the ejecta of a single SN. For an ejecta mass of ${\rm10\ M_{\odot}}$ and launch velocity v = 3000 \kms\ (see, e.g. Section 2.2 in \citealt{Murray2005}), the total injected momentum flux is ${\rm\dot{p}_{ej} = 3.0\times10^{36}}$ dyne, giving ${\rm\dot{p}_{OF}/\dot{p}_{ej}=0.13-91}$. This is sufficient to drive the neutral outflow through a momentum-driven phase if $inc\leq2.9^\circ$.
		
		In addition to the momentum flux deposited by SNe ejecta, momentum flux supplied by UV radiation pressure from young stars onto dust grains in the outflow, may also be contributing. The effectiveness of this mechanism depends on the optical depth of the outflow:
		
		\begin{eqnarray}
		{\rm\dot{p}_{rad}}&=&{\rm \tilde{\tau}L_{bol}/c},
		\end{eqnarray}
		
		where ${\rm \tilde{\tau}=(1-e^{-\tau_{single}})(1+\tau_{eff,IR})}$ which includes both single and multiple scattering events \citep{Hopkins2014, Hopkins2020}. ${\rm\tilde{\tau}}$ therefore ranges from ${\rm \tau_{single}=\tau_{UV/optical}<<1}$ when the outflow is optically thin, to ${\rm\sim(1+\tau_{eff,IR})\ }$ when optically thick, where ${\rm \tau_{UV/optical}}$ and ${\rm\tau_{eff,IR}}$ are the optical depths in the UV/optical and IR regimes, respectively \citep{Murray2005}. If we make the conservative assumption that $L_{bol}\approx L_{IR}$ (i.e. all the UV stellar radiation is absorbed and reradiated in the IR) then ${\rm\dot{p}_{OF}/(L_{IR}/c)}=0.067-47$ and radiation pressure could deposit a momentum flux of the order $3.9\times10^{36}$ dyne for an optically thick outflow (${\rm \tilde{\tau}\approx1}$). In addition to the momentum flux deposited by SN ejecta, this is sufficient to drive the neutral gas as a momentum-driven outflow if $inc\leq8.4^\circ$. In the case of an optically thin outflow, single scattering radiation could provide a maximum momentum of ${\rm (1-e^{-\tau_{single}})L_{IR}/c}$. 
	
\begin{figure}
			\centering
			\includegraphics[width=\linewidth]{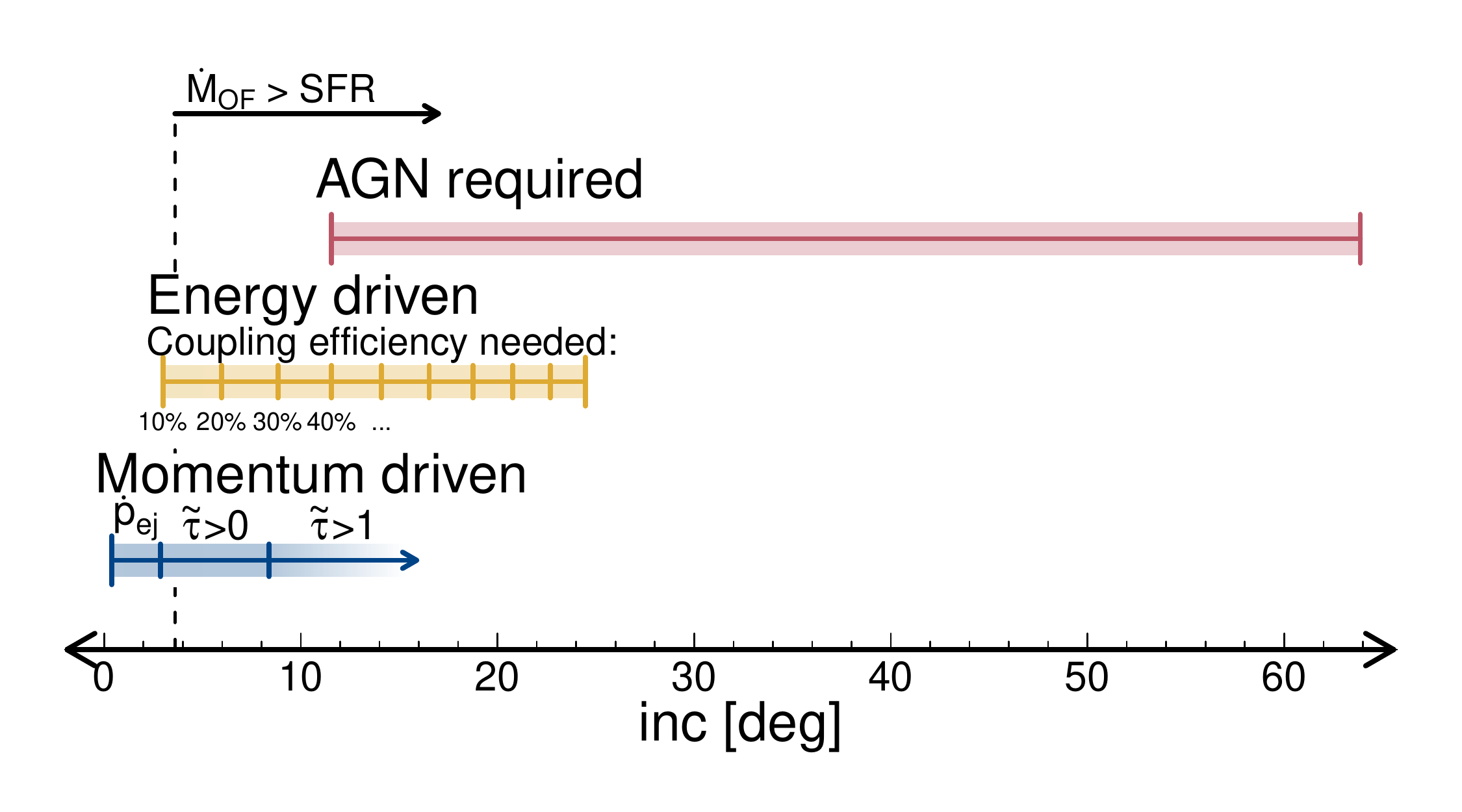}
			\caption{Possible outflow driving mechanisms as a function of outflow inclination. If the inclination is $<2.9^\circ$ the momentum provided by SNe ejecta is sufficient to drive the outflow through a momentum-driven phase (blue bar). If inc$>2.9^\circ$ a momentum-driven phase is still possible, if radiation pressure from young stars on dust grains in the outflow is non-negligible. This requires ${\rm\tau_{eff,IR}>0}$ and ${\rm\tau_{eff,IR}>1}$ if inc$>8.4^\circ$. In the absence of radiation pressure, the outflow requires a momentum boost if inclinations are $>2.9^\circ$, which can be provided by an energy-driven phase of the shocked SNe ejecta if thermal energy is not immediately and completely radiated away. The energy coupling required to drive the outflow via SNe feedback is indicated below the yellow bar. If the inclination is $>12^\circ$ the outflow requires either an unusually high energy coupling of $>40\%$, which is higher than that expected from a clustered SNe scenario, or a contribution from a low-luminosity, obscurred or fossil AGN (red bar). In reality, all driving mechanisms may be contributing simultaneously and if the full multiphase and double-sided outflow were to be taken into account the inclination ranges shown here would be shifted to smaller inclinations.}
			\label{fig:DrivingMechs}
		\end{figure}

		If, however, the thermal energy of the shocked SN ejecta is not efficiently radiated away, it may be used up in doing work against the ambient medium, driving an energy-driven outflow. This provides a boost in the momentum flux of the outflow in addition to the momentum flux deposited by the ejecta and thus drives a stronger outflow. If radiation pressure is negligible and the inclination of the outflow is $>2.9^\circ$, it is possible that we are observing an energy-driven outflow. The effectiveness of this mechanism depends on the coupling efficiency of energy to the ISM (see Fig. \ref{fig:DrivingMechs}) which is unlikely to exceed $40\%$ in the case of highly clustered SNe. Thus if the inclination is $>12^\circ$ an unusually high energy coupling efficiency or past, low-luminosity or obscurred AGN activity is needed to explain both the energy and momentum flux of the outflow. 
		
		We summarise the possible driving scenarios of the neutral gas outflow in G09v1.40 in Fig. \ref{fig:DrivingMechs} but again note that if the full multiphase and double-sided outflow is taken into account, the ranges presented here would be shifted to smaller inclinations. We also note that our results are sensitive to the \OHp abundance assumed and provide alternative curves (dashed lines in Fig. \ref{fig:Energyflux}/\ref{fig:Momflux}/\ref{fig:DepletionTime}) using the extreme observational \OHp abundance taken from \cite{Bialy2019} which would again significantly shift the inclination ranges of the driving mechanism scenarios summarised in Fig. \ref{fig:DrivingMechs}.

	\subsection{Impact on the Host Galaxy and Fate of the Outflowing Neutral Gas}\label{sec:Fate}
\begin{figure}
			\centering
			\includegraphics[width=\linewidth]{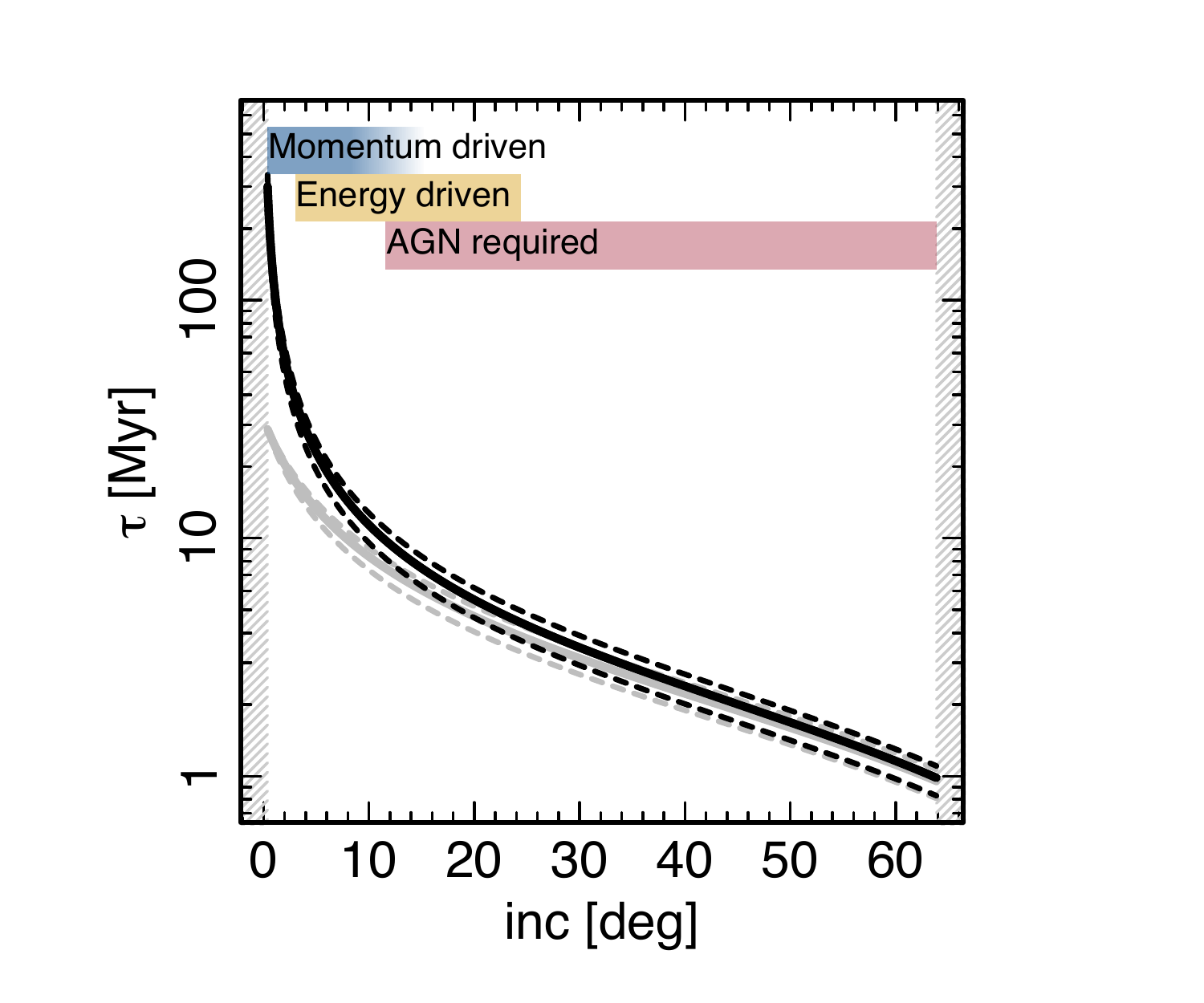}
			\caption{Depletion time of the host galaxy's gas reservoir given the observed neutral gas mass outflow rate (black lines), and in combination with the SFR (grey lines), as a function of possible outflow inclination. The upper and lower dashed lines (black and grey) indicate the depletion time if the highest and lowest observed \OHp abundances from Milky Way sightlines are assumed, respectively \citep{Bialy2019}. The outflow driving mechanisms from Fig. \ref{fig:DrivingMechs} are shown at the top. An outflow flowing perpendicular to the line of sight has an $inc=90^\circ$. Hatched regions indicate regions outside of the considered inclination range.}
			\label{fig:DepletionTime}
		\end{figure}

	Cool gas outflows remove the direct fuel for star formation, and therefore must have an impact on the future growth and activity of the host galaxy. Disregarding the possibility of gas accretion or a change in SFR and/or ${\rm\dot{M}_{OF}}$ in G09v1.40, we estimate a depletion time of the host galaxy's gas reservoir due to the observed neutral gas mass outflow rate using: $\tau_{\rm OF} = {\rm M_{gas}}/{\rm\dot{M}_{OF}}$. We derive this timescale over the range of possible inclinations, finding a $\tau_{\rm OF}$ between 300 Myr for the lowest possible inclination (i.e. the star-formation driven end of the spectrum), and 0.98 Myr at the highest possible inclination (i.e. the AGN driven end of the spectrum). If depletion due to star formation is also taken into account, $\tau_{\rm OF+SFR}$, these timescales reduce to 29 and 0.95 Myr, respectively (Fig. \ref{fig:DepletionTime}), where the depletion time due to star formation alone is $\tau_{\rm SFR}=32$ Myr. Thus, if the inclination is low, star formation likely plays a major role in the depletion of the host galaxy gas reservoir.
		
	Typical depletion times derived from SFRs in compact star-forming galaxies and quasars at $z\sim2-4$ are on the order of $\sim 50$ and $\sim 100$ Myr, respectively \citep{Stacey2021,Spilker2016}, consistent with the timescales derived for the low inclination scenarios in G09v1.40. The $\sim 1$ Myr depletion times derived for high inclinations are instead consistent with timescales predicted for DSFGs to transition into unobscurred gas-poor QSOs, via a far-infrared bright QSO phase \citep{Simpson2012,Costa2018}. Thus, if a high inclination scenario in G09v1.40 is assumed, this would suggest the galaxy is currently in an evolutionary stage just prior to or at the beginning of a highly obscured QSO phase.
		
		Ejected gas may, however, be re-accreted back onto the galaxy at a later time, replenishing the galaxy's gas reservoir and prolonging $\tau_{\rm OF}$. We therefore investigate the likelihood of the neutral gas outflow in G09v1.40 escaping the galaxy's potential well by considering the mass required, ${\rm M_{req}}$, to gravitationally bind an outflow with velocity V and radius R,
		\begin{equation}
			{\rm M_{req} }= \frac{V^2R}{2G},
		\end{equation}
		where G is the gravitational constant. 
		
\begin{figure}
			\centering
			\includegraphics[width=\linewidth]{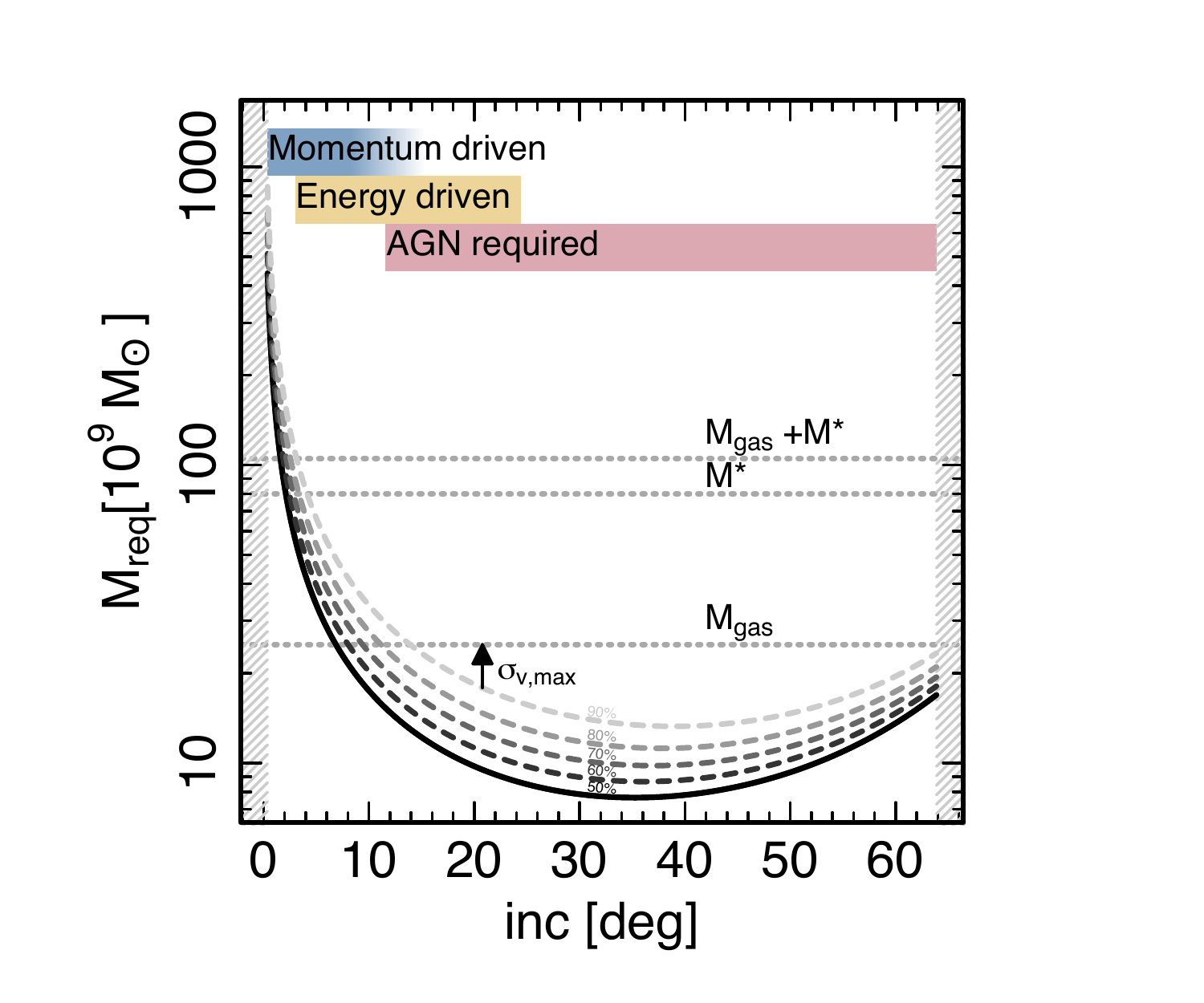}
			\caption{The mass required to gravitationally bind the outflowing neutral gas, as a function of possible inclination. The solid black curve is derived using the central velocity of the blueshifted \OHp line (i.e. indicating the mass required to bind 50\% of the outflowing neutral gas at a given inclination). The dashed curves are derived using an outflow velocity at the 60th, 70th, 80th and 90th percentiles (i.e. indicating the mass required to bind 60\%, 70\%, 80\% and 90\% of the outflowing neutral gas), using a velocity dispersion of 77 \kms. The dotted horizontal lines indicate the gas mass $M_{gas}$ \citep{Yang2017}, stellar mass ${\rm M*}$ \citep{Ma2015} and total mass ${\rm M_{gas}+M*}$ of the host galaxy. The outflow driving mechanisms from Fig. \ref{fig:DrivingMechs} are shown at the top. An outflow flowing perpendicular to the line of sight has an $inc=90^\circ$. Hatched regions indicate regions outside of the considered inclination range.}
			\label{fig:EscapeMass}
		\end{figure}

		For an outflow with a Gaussian velocity distribution, this equation will provide the ${\rm M_{req}}$ capable of containing half the outflowing material if the central velocity is used. We therefore derive required binding masses, over the range of possible inclinations, using deprojected velocities at the 50th, 60th, 70th, 80th and 90th percentiles (again assuming a velocity dispersion of 77 \kms), corresponding to a ${\rm M_{req}}$ capable of containing 50\%, 60\%, 70\%, 80\%, and 90\% of the outflow (Fig. \ref{fig:EscapeMass}). The results do not significantly change if the maximum velocity dispersion $\sigma_{\rm v, max}=130\ \kms$ of the outflowing neutral gas is taken instead (indicated by the arrow in Fig. \ref{fig:EscapeMass}). 
		
		${\rm M_{req}}$ is largest ($8.6\times10^{11}\ M_{\odot}$) if the outflow inclination is small, due to the very large deprojected radii in this regime. If the inclination is $>14^\circ$, the gas mass of the galaxy alone is capable of containing 90\% or more of the neutral outflow. If we include the stellar mass ${\rm M*}=0.8\pm0.1\times10^{11}\ {\rm M}_{\odot}$ \citep{Ma2015}, G09v1.40 is capable of retaining this fraction of the outflow down to an inclination of $3^\circ$, implying that for the majority of possible outflow scenarios, most or all of the outflowing neutral gas will remain bound to the galaxy. This material is then available to be re-accreted by the galaxy at a later time unless additional feedback processes, such as thermal feedback from an unobscured QSO phase \citep{Costa2018}, causes the gas to remain in the circum-galactic medium. For an inclination larger than $11^\circ$, where an AGN is required to drive the outflow and heating of the circum-galactic medium is likely, the galaxy may still be expected to quench on timescales of $\tau_{\rm OF}\sim1$ Myr. 
		
		\begin{table*}
			\centering
			\caption{Derived properties of the Neutral Outflow in G09v1.40 using the mean \OHp abundance, $\log({\rm X}_{\rm OH^+\!})$, modelled by \cite{Bialy2019} and the highest and lowest observed abundances in the Milky Way sightlines used as comparison in their analysis.} \label{tab:OFprops}
			\begin{tabular}{ccccccccc}
					\hline\hline
					${\rm X}_{\rm OH^+}$&${\rm M_{neut}}$ &${\rm\dot{M}_{OF}}$&${\rm\dot{E}_{OF}}$&${\rm\dot{E}_{OF}/\dot{E}_{SF}}$&${\rm\dot{p}_{OF}}$&${\rm\dot{p}_{OF}/\dot{p}_{ej}}$&${\rm\dot{p}_{OF}/(L_{IR}/c)}$&${\rm \tau_{OF}}$\\
					&$[10^9 M_{\odot}]$&$[M_{\odot}\ {\rm yr}^{-1}]$&$[10^{9}\ L_\odot]$&&$[10^{37}\ {\rm dyne}]$&&&[Myr]\\
					\hline
					Model: Mean &6.7&83-25400&1.6-2100&0.013-16&0.040-27&0.13-91&0.067-47&300-0.98\\ 
					Observed: High &16&200-6200&4.0-5200&0.031-40&0.096-66&0.32-220&0.16-110&250-0.83\\
					Observed: Low &2.5&31-9550& 0.62-790&0.0047-6.1&0.015-10&0.049-34&0.025-18&340-1.1\\
					\hline
				\end{tabular}
		\end{table*}
	
\section{Conclusions} \label{sec:Conc}
	We have presented resolved ($0.52''\times0.41''$) ALMA Band 6 observations of a massive $z=2.09$ neutral outflow in the gravitationally lensed DSFG G09v1.40 (HATLASJ085358.9+015537). We detect the outflow in absorption with the 1033 GHz $\OHp\!(1_1-1_0)$ transition, exploiting its close proximity to the CO(9-8) transition to observe both lines and the underlying 1034 GHz dust continuum, simultaneously with a single ALMA tuning. We obtain a spatially and spectrally resolved view of the cool neutral gas in the outflow as traced by the $\OHp\!$, blue-shifted with respect to the warm dense gas at systemic velocities as traced by CO(9-8). We perform spectral fitting on all spaxels to obtain clean intensity and velocity maps of the outflowing \OHp absorption and systemic CO(9-8) emission. In addition, we use ancillary data from ALMA tracing the CH+(1-0) absorption and underlying continuum at both low and high-angular resolution, and from the Keck K-band at 2.2 microns tracing the stellar emission of the background galaxy.
	
	The CO(9-8) displays a strikingly different image plane morphology to that of the dust continuum, following more closely that of the stellar distribution \citep{Calanog2014}. The image plane optical depth distribution of the \OHp absorption follows the continuum by construction but displays a comparatively more elongated morphology, falling off dramatically to the North and South, and does not display an Einstein ring, as seen in the dust, CO(9-8) emission and stars. 
	
	We obtain a lens model, exploiting the high-resolution ($0.17''\times0.13''$) dust continuum observations, with \texttt{visilens} and reconstruct all our 2D data maps into the source plane using the pixelated reconstruction code \texttt{LENSTOOL}. The dust continuum reveals itself as a compact ellipse with the CO(9-8) and stellar components offset to the East and displaying more extended distributions. The blue-shifted \OHp forms an extended triangular morphology flaring out towards the West. There is a dip in the stellar emission at the position of peak dust continuum which we believe is most likely due to extreme extinction in this region.
	
	Three simple outflow geometries, (a sheet, spherical-bubble and cone) are considered and compared with the observed and reconstructed \OHp morphology and kinematics. We find that a conical outflow geometry, where outflowing gas is ejected from the central dusty star-forming region towards the West, is the most suitable choice.
	
	The physical conditions necessary for forming \OHp in the ISM suggests that \OHp absorption traces the diffuse and predominantly atomic gas component of a turbulent outflow. Comparing the absorption of OH+ with that of the CH+, which similarly probes diffuse atomic gas (Falgarone et al. 2017), we find that both absorptions lines are cospatial in G09v1.40, tracing the same blueshifted kinematic component, confirming these lines trace the same gas phase. We therefore adopt an \OHp to HI abundance from \cite{Bialy2019} of $\log_{10}(n_{\rm OH^+}/n_{H})\approx-7.8\pm^{0.05}_{0.075}$, finding a total atomic gas mass of the outflow of ${\rm M_{neut} = 6.7\times10^9 M_{\odot}}$, which is more than 25\% as massive as the molecular gas component in the host galaxy \citep{Yang2017}. 
	
	We consider a range of possible 2D projections of the conical outflow, deriving possible inclinations of the central axis with respect to the observer's line of sight between $inc=0.4^\circ-64^\circ$. Over this inclination range we derive possible deprojected outflow radii between ${\rm R=0.45-61\ kpc}$, 2D opening angle between $inc=0.2^\circ-24^\circ$, and velocity between ${\rm V=250-570\ \kms}$. 
	
	Physical properties of the conical outflow are also derived as functions of possible incination. The total neutral gas ${\rm\dot{M}_{OF}}$ is between ${\rm 83-25400\ M_{\odot}\ yr^{-1}}$, which exceeds the SFR of ${\rm788\pm300\ M_{\odot}\ yr^{-1}}$ if the inclination is greater than $3.6^\circ$. We find ranges in the kinetic and momentum fluxes of ${\rm\dot{E}_{OF}=1.6-2100 \times10^{9}\ L_\odot}$ and ${\rm\dot{p}_{OF}}=0.040-27 \times10^{36}$ dyne, respectively. 
	
	We compare these values to the kinetic energy (${\rm\dot{E}_{SF}}={\rm130\times10^{9}}\ L_\odot$) and momentum flux injected by SNe (${\rm\dot{p}_{ej} = 3.0\times10^{36}}$ dyne) and radiation from young stars to determine the likely driving mechanism of the outflow, finding that this depends strongly on the inclination assumed. If the inclination is $\leq2.9^\circ$, the outflow may be momentum-driven by SNe ejecta. If $inc>2.9^\circ$, the outflow may still be momentum-driven provided radiation pressure from young stars onto dust grains in the outflow is taken into account. For $inc>8.4^\circ$, this requires the outflow to be optically thick. 
	
	In the case where thermal energy deposited by SNe into the ISM is not efficiently radiated away, it may be used to do work on the ambient medium, providing a momentum boost for an energy-driven outflow. If radiation pressure is negligible then the outflow may be energy-driven if the inclination is $inc>2.9^\circ$ up to a maximum inclination of $12^\circ$ where a coupling efficiency of the thermal energy to the ISM of $40\%$ is needed. If the $inc>12^\circ$, either an extremely high coupling efficiency or an additional driving mechanism is needed, e.g. an AGN. 
	
	Depletion times of the host galaxy gas reservoir, due to the SFR and neutral outflow range from 29 Myr in the regime of a stellar driven outflow, down to 0.95 Myr at the extreme end of the AGN-driven regime. This is consistent with timescales derived for other intensely star-forming galaxies at the same redshift \citep{Stacey2021}, and with timescales predicted for DSFGs to transition into unobscured gas-poor QSOs, via a far-infrared bright QSO \citep{Simpson2012,Costa2018}. In the latter case, this would imply that G09v1.40 is in a phase just prior to a highly obscurred QSO phase.
	
	Most or all of the gas in the neutral outflow, however, is likely to remain bound to the galaxy in all but the least inclined scenarios, where the deprojected radii are large. This gas may then be re-accreted by the galaxy at a later time, replenishing the gas reservoir, unless additional feedback, such as thermal feedback from a previously obscured QSO, causes the gas to remain in the CGM.
		
	Whilst the current observations provide sufficient information to determine global properties and offsets between the dust, gas and stellar components, analysis of the detailed morphological and kinematic structures will require new observations at higher spatial resolution, including a determination of the true inclination of the neutral outflow in G09v1.40.
		
	Finally, we note that our analysis of the outflow in G09v1.40 using \OHp/\CHp only probes the diffuse neutral component, on one side of the galaxy. The full multiphase, double-sided outflow will carry even more mass, momentum and energy, likely shifting the conclusions of this paper to more extreme scenarios. Future observations, targeting other phases of the outflow (e.g., molecular and ionised) in both absorption and emission lines, are needed to fully constrain the impact of this outflow on the evolution of G09v1.40. 
	
\acknowledgments
	The authors thank the referee for their many, and very appreciated, suggestions, questions and corrections. The authors also thank Jae Calanog for providing the reduced and lens subtracted Keck NIR images introduced in Sec. \ref{sec:ObsKeck}. D.R. acknowledges support from the National Science Foundation under grant numbers AST-1614213 and AST-1910107. D.R. also acknowledges support from the Alexander von Humboldt Foundation through a Humboldt Research Fellowship for Experienced Researchers. M.J.M.~acknowledges the support of the National Science Centre, Poland through the SONATA BIS grant 2018/30/E/ST9/00208. M.R. acknowledges the support of the Veni research programme with project number 202.225 and the Vidi research programme with project number 639.042.611, which are (partly) financed by the Dutch Research Council (NWO). This paper makes use of the following ALMA data: \newline${\rm ADS/JAO.ALMA\#2015.1.01042.S}$,\newline ${\rm ADS/JAO.ALMA\#2013.1.00164.S}$,\newline and ${\rm ADS/JAO.ALMA\#2016.1.00282.S}$. ALMA is a partnership of ESO (representing its member states), NSF (USA) and NINS (Japan), together with NRC (Canada), MOST and ASIAA (Taiwan), and KASI (Republic of Korea), in cooperation with the Republic of Chile. The Joint ALMA Observatory is operated by ESO, AUI/NRAO and NAOJ. 
%

\vspace{5mm}
\facilities{ALMA, Keck}


\software{\texttt{CASA} (v4.53; \citealt{McMullin2007}), \texttt{visilens} \citep{Hezaveh2013,Spilker2016} \& \texttt{LENSTOOL}
\citep{Kneib1996,Jullo2007,Jullo2009}.}



\appendix
\section{Beam Smearing Effects on Source Reconstruction of Gravitational Lenses} \label{sec:beam}
	\begin{figure}
		\centering
		\includegraphics[width=\linewidth]{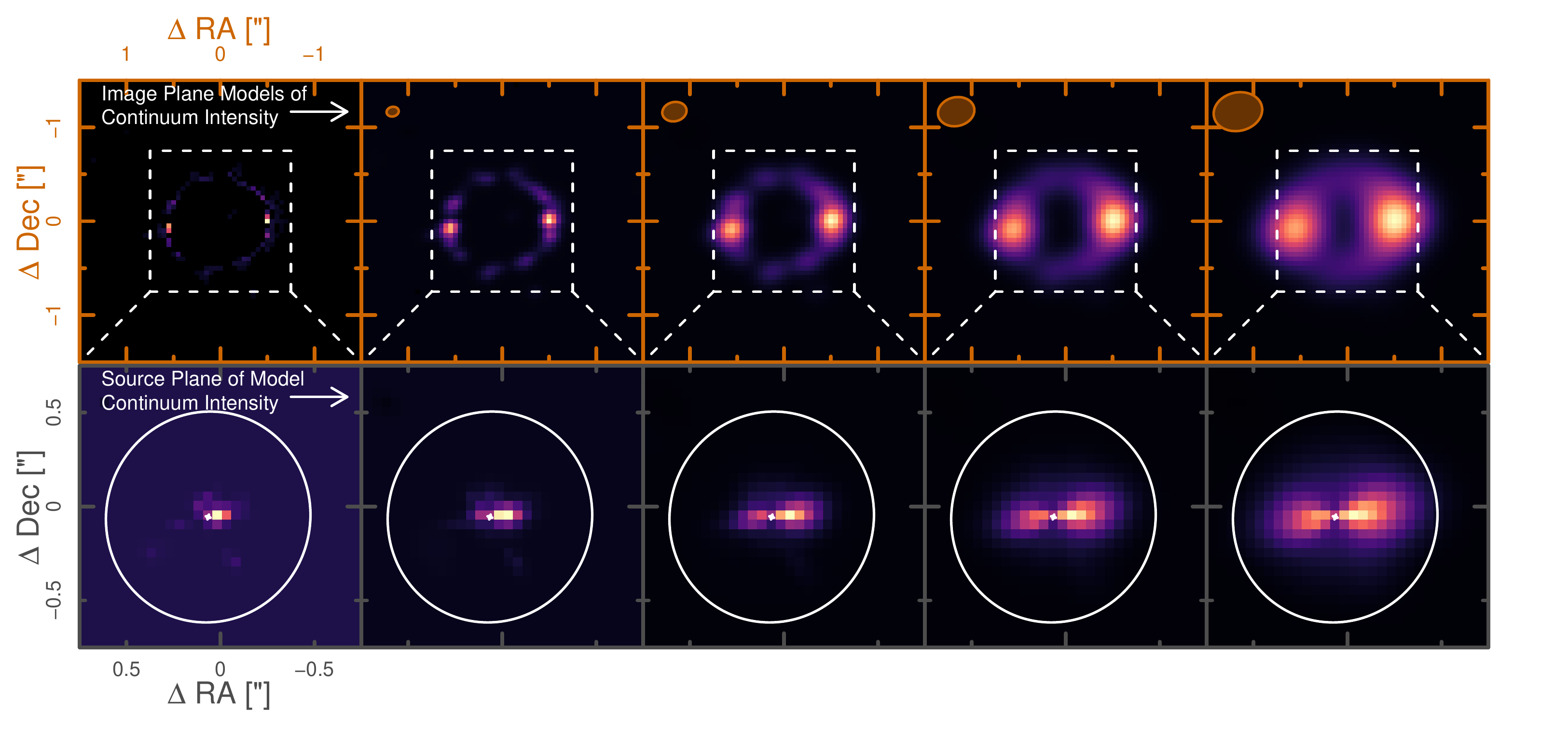}
		\caption{Illustrating the effect of beam smearing on our observations and reconstruction of the dust continuum in G09v1.40. \textbf{Top Row:} Image plane model of the dust continuum, derived from the cleaning method during data reduction, with no beam convolution (far left panel) and convolved with a beam FWHM at 25\%, 50\%, 75\% and 100\% that of the beam in our observations (consecutive panels to the right). Beam sizes are shown by the shaded orange ellipse in the upper left of each panel and the dash white region indicates the region enlarged in the lower panels. \textbf{Bottom Row:} Source plane reconstructions of the panels above with lensing caustics shown in white. As the data is convolved with increasingly larger beam sizes, more of the flux in the image plane is smeared past the Einstein ring and is then reconstructed on the wrong side of the inner caustic in the source plane. This causes the artifact to the East of the inner caustic in the source plane to become more severe.}
		\label{fig:BeamSmearingCont}
	\end{figure}
	
	\begin{figure}
		\centering
		\includegraphics[width=\linewidth]{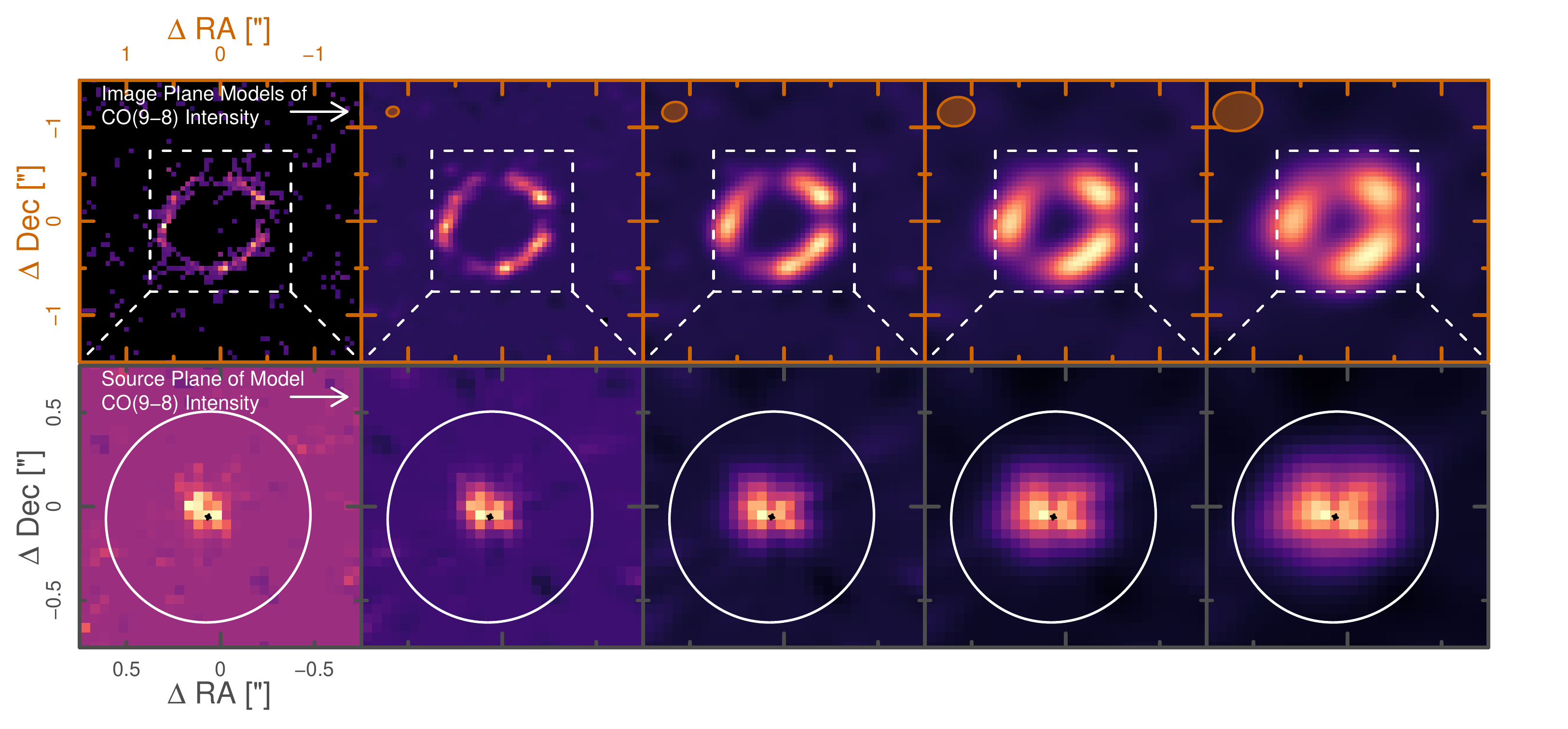}
		\caption{Illustrating the effect of beam smearing on our observations and reconstruction of the CO(9-8) emission in G09v1.40. \textbf{Top Row:} Image plane model of the CO(9-8) emission, derived from the cleaning method during data reduction, with no beam convolution (far left panel) and convolved with a beam FWHM 25\%, 50\%, 75\% and 100\% that of the beam in our observations (consecutive panels to the right). Beam sizes are shown by the shaded orange ellipse in the upper left of each panel and the dash white region indicates the region enlarged in the lower panels. \textbf{Bottom Row:} Source plane reconstructions of the panels above with the inner lensing caustics shown in black and Einstein ring shown in white. As the data is convolved with increasingly larger beam sizes, more of the flux in the image plane is smeared across the Einstein ring and is then reconstructed on the wrong side of the inner caustic in the source plane. This causes flux from locations in the source plane on opposite sides of the caustic (North East and South West) to be blended together after reconstruction and results in the spurious North-Western artifact which becomes more severe with larger beam sizes.}
		\label{fig:BeamSmearingCO98}
	\end{figure}

	When we observe the sky, the spatial distribution of the sky emission is convolved with the shape of the beam. This has the effect of smearing the light emitted from structures smaller than the beam over its point spread function. The consequence of this in our observations is that our sources appear larger and fuzzier than they intrinsically are (see upper rows of Fig. \ref{fig:BeamSmearingCont} and \ref{fig:BeamSmearingCO98}). For a gravitationally lensed source sitting on or very close to the inner caustic of the gravitational lens, the light from one side of the caustic will be smeared over and onto the other side of the caustic. When mapping this light back into the source plane it will not be reconstructed in the correct position.
	
	This is most obviously demonstrated by the dust continuum in G09v1.40, whose simple double image configuration and faint Einstein ring in the image plane (Fig. \ref{fig:MapStack}), implies an intrinsic source plane geometry of a single extended source, partially overlapping the inner caustic but with its peak situated just to the West. However, when the image plane intensity map is reconstructed, two peaks appear in the source plane, one to the West of the inner caustic as expected and another weaker mirror-image of the peak on the opposite side of the caustic. 
	
	To investigate the role of beam smearing in this process we reconstruct the model of the dust continuum intensity produced by the cleaning procedure during data reduction. This provides an indication of the continuum intensity distribution without beam convolution. We also reconstruct maps of the model map convolved with artificial beams with axes 25\%, 50\%, 75\% and 100\% as long as the beam in our observations. The source plane reconstruction of the fully deconvolved model reveals a single source to the West of the inner caustic with no mirroring image on the East. As the beam size is increased, the artifact to the East of the caustic appears and grows (Fig. \ref{fig:BeamSmearingCont}). 
	
	We repeat this experiment on the moment 0 map of the model CO(9-8) emission. Again we note that the model is only an indication of the true deconvolved intensity distribution. Again, the fully deconvolved map produces a single source when reconstructed in the source plane. The CO(9-8) emission is more extended and elongated than the dust continuum and sits directly over the inner caustic and extending to the North East and South West of the caustic. A source overlapping the inner caustic should form a quadruply imaged lens configuration in the image plane, which is not obvious in our observations, but is evident in the middle-upper panel of Fig. \ref{fig:BeamSmearingCO98}. The South West image is in fact two merged images of the source which are not separately resolved by our beam. As the model is convolved with larger beam sizes, flux emitted from positions within the source on either side of the caustic are blended and a spurious artifact appears and grows to the North-West of the caustic (see Fig. \ref{fig:BeamSmearingCO98}).

\section{\OHp $1_0-1_1$ Optical Depth to Column Density} \label{sec:N}
	We trace the neutral gas outflow in G09v1.40 with the 1033 GHz \OHp $1_0-1_1$ line in absorption. Both levels are split into 2 by hyperfine structure, resulting in four energy levels which we label 0, 1, 2 and 3, in order of increasing energy. The line that we detect is thus the sum of 4 different absorption lines: $0\rightarrow2$,$0\rightarrow3$, $1\rightarrow2$ and $1\rightarrow2$. The Einstein A coefficients, level energies and quantum numbers of each transition can be found at Splatalogue (\href{https://splatalogue.online/}{https://splatalogue.online/}). 
	
	For a single absorption line (ignoring stimulated emission) the optical depth is given by
		\begin{eqnarray}
			{\rm \tau_\nu^{\rm single}=N_l\frac{\lambda^3}{8\pi}\frac{g_u}{g_l}\frac{A_{ul}}{\sqrt{2\pi\sigma}}e^{-\nu^2/2\sigma^2}}
		\end{eqnarray}
	where ${\rm g_u}$ and ${\rm g_l}$ are the statistical weights of the upper and lower levels respectively, $A_{ul}$ the Einstein coefficient of the transition from the upper to lower energy state, ${\rm N_l}$ the column density of the lower level and $\lambda_{ul}$ is the wavelength of the line which is assumed to be Gaussian centered on $\nu$ with velocity dispersion $\sigma$. Assuming all the \OHp molecules are in the ground state (${\rm N_l>>N_u}$) the quantity of interest is $N_l$, the column density in the lower, absorbing level which in this case is the combination of the two hyperfine ground state levels ${\rm N_1 + N_0}$.
	
	Integrating the optical depth over velocity we obtain for a single line
		\begin{eqnarray}
			{\rm\int\tau_\nu^{\rm single} dv}&=&{\rm N_l\frac{\lambda^3}{8\pi}\frac{g_u}{g_l}\frac{A_{ul}}{\sqrt{2\pi\sigma}}\int e^{-v^2/2\sigma^2}dv}	\\
			&=&{\rm N_l\frac{\lambda^3}{8\pi}\frac{g_u}{g_l}A_{ul}}
		\end{eqnarray}
	
	Summing all four lines we therefore obtain
		\begin{eqnarray}
			{\rm\int\tau_\nu dv=}&{\rm\frac{1}{8\pi}\big( \frac{g_3}{g_0}A_{30}\lambda_{30}^3N_0+\frac{g_3}{g_1}A_{31}\lambda_{31}^3N_1+} \nonumber \\
			&{\rm\frac{g_2}{g_0}A_{20}\lambda_{20}^3N_0+\frac{g_2}{g_1}A_{21}\lambda_{21}^3N_1 \big)}
		\end{eqnarray}
	
	From Splatalogue: ${\rm A_{20} = 7.03\times10^{-3}s^{-1}}$, ${\rm A_{30} = 1.76\times10^{-2}s^{-1}}$, ${\rm A_{21} = 1.41\times10^{-2}s^{-1}}$ and
	${\rm A_{31} = 3.53\times10^{-3}s^{-1}}$, ${\rm g_1 = g_2 = 2}$ and ${\rm g_0 = g_3 = 4}$ and $\lambda_{30}\approx\lambda_{31}\approx\lambda_{20}\approx\lambda_{21}\approx0.29{\rm\ cm}$. This leads to
		\begin{eqnarray}
		{\rm\int\tau_\nu dv}&=&{\rm\frac{\lambda^3}{8\pi}\big( \frac{1}{2}A_{20}N_0+A_{30}N_0+A_{21}N_1+2A_{31}N_1\big)}\nonumber \\
			&=&{\rm\frac{\lambda^3}{8\pi}\big(2.11\times10^{-2}N_0+2.11\times10^{-2}N_1\big)	}
		\end{eqnarray}
	
	which can simply written as 
		\begin{eqnarray}
			{\rm \int\tau_\nu dv=}&{\rm\frac{\lambda^3}{8\pi}{\rm AN}_{\rm OH^+}}
		\end{eqnarray}
	where the Einstein coefficient is given by ${\rm A}=2.11\times10^{-2}{\rm s}^{-1}$.


\begin{thebibliography}{}
\bibitem[Aalto et al.(2015)]{Aalto2015} Aalto, S., Garcia-Burillo, S., Muller, S., et al.\ 2015, \aap, 574, A85. doi:10.1051/0004-6361/201423987
\bibitem[Alatalo et al.(2011)]{Alatalo2011} Alatalo, K., Blitz, L., Young, L.~M., et al.\ 2011, \apj, 735, 88
\bibitem[Benson et al.(2003)]{Benson2003} Benson, A.~J., Bower, R.~G., Frenk, C.~S., et al.\ 2003, \apj, 599, 38
\bibitem[Berta et al.(2021)]{Berta2021} Berta, S., Young, A.~J., Cox, P., et al.\ 2021, A\&A 646, A122
\bibitem[Bialy et al.(2017)]{Bialy2017} Bialy, S., Burkhart, B., \& Sternberg, A.\ 2017, \apj, 843, 92
\bibitem[Bialy et al.(2019)]{Bialy2019} Bialy, S., Neufeld, D., Wolfire, M., et al.\ 2019, \apj, 885, 109
\bibitem[Bower et al.(2012)]{Bower2012} Bower, R.~G., Benson, A.~J., \& Crain, R.~A.\ 2012, \mnras, 422, 2816
\bibitem[Bregman (1980)]{Bregman1980} Bregman, J.~N.\ 1980, \apj, 236, 577. doi:10.1086/157776
\bibitem[Bregman et al.(2013)]{Bregman2013} Bregman, J.~N., Miller, E.~D., Seitzer, P., et al.\ 2013, \apj, 766, 57. doi:10.1088/0004-637X/766/1/57
\bibitem[Brusa et al.(2018)]{Brusa2018} Brusa, M., Cresci, G., Daddi, E., et al.\ 2018, \aap, 612, A29
\bibitem[Bussmann et al.(2013)]{Bussmann2013} Bussmann, R.~S., P{\'e}rez-Fournon, I., Amber, S., et al.\ 2013, \apj, 779, 25
\bibitem[Calanog et al.(2014)]{Calanog2014} Calanog, J.~A., Fu, H., Cooray, A., et al.\ 2014, \apj, 797, 138
\bibitem[Calder{\'o}n et al.(2016)]{Calderon2016} Calder{\'o}n, D., Bauer, F.~E., Veilleux, S., et al.\ 2016, \mnras, 460, 3052. doi:10.1093/mnras/stw1210
\bibitem[Calistro Rivera et al.(2018)]{CalistroRivera2018} Calistro Rivera, G., Hodge, J.~A., Smail, I., et al.\ 2018, \apj, 863, 56. doi:10.3847/1538-4357/aacffa
\bibitem[Carilli \& Walter(2013)]{Carilli2013} Carilli, C.~L. \& Walter, F.\ 2013, \araa, 51, 105
\bibitem[Carniani et al.(2017)]{Carniani2017} Carniani, S., Marconi, A., Maiolino, R., et al.\ 2017, \aap, 605, A105
\bibitem[Casey et al.(2014)]{Casey2014} Casey, C.~M., Narayanan, D., \& Cooray, A.\ 2014, \physrep, 541, 45. doi:10.1016/j.physrep.2014.02.009
\bibitem[Cazzoli et al.(2016)]{Cazzoli2016} Cazzoli, S., Arribas, S., Maiolino, R., et al.\ 2016, \aap, 590, A125. doi:10.1051/0004-6361/201526788
\bibitem[Chen et al.(2017)]{Chen2017} Chen, C.-C., Hodge, J.~A., Smail, I., et al.\ 2017, \apj, 846, 108. doi:10.3847/1538-4357/aa863a
\bibitem[Chisholm et al.(2016)]{Chisholm2016} Chisholm, J., Tremonti, C.~A., Leitherer, C., et al.\ 2016, \mnras, 457, 3133
\bibitem[Cicone et al.(2014)]{Cicone2014} Cicone et al., 2014, A\&A, 562, 21
\bibitem[Cicone et al.(2015)]{Cicone2015} Cicone, C., Maiolino, R., Gallerani, S., et al.\ 2015, \aap, 574, A14. doi:10.1051/0004-6361/201424980
\bibitem[Cochrane et al.(2021)]{Cochrane2021} Cochrane, R.~K., Best, P.~N., Smail, I., et al.\ 2021, \mnras, 503, 2622. doi:10.1093/mnras/stab467
\bibitem[Cole et al.(2000)]{Cole2000} Cole, S.~M., Baugh, C., Frenk, C., et al.\ 2000, Astronomy, Physics and Chemistry of H$^{+}$$_{3}$, 2093
\bibitem[Costa et al.(2018)]{Costa2018} Costa, T., Rosdahl, J., Sijacki, D., et al.\ 2018, \mnras, 479, 2079. doi:10.1093/mnras/sty1514
\bibitem[Danielson et al.(2011)]{Danielson2011} Danielson, A.~L.~R., Swinbank, A.~M., Smail, I., et al.\ 2011, \mnras, 410, 1687
\bibitem[Danielson et al.(2013)]{Danielson2013} Danielson, A.~L.~R., Swinbank, A.~M., Smail, I., et al.\ 2013, \mnras, 436, 2793. doi:10.1093/mnras/stt1775
\bibitem[Decarli et al.(2018)]{Decarli2018} Decarli, R., Walter, F., Venemans, B.~P., et al.\ 2018, \apj, 854, 97. doi:10.3847/1538-4357/aaa5aa
\bibitem[El-Badry et al.(2019)]{ElBadry2019} El-Badry, K., Ostriker, E.~C., Kim, C.-G., et al.\ 2019, \mnras, 490, 1961
\bibitem[Enia et al.(2018)]{Enia2018} Enia, A., Negrello, M., Gurwell, M., et al.\ 2018, \mnras, 475, 3467. doi:10.1093/mnras/sty021
\bibitem[Fan et al.(2018)]{Fan2018} Fan, L., Knudsen, K.~K., Fogasy, J., et al.\ 2018, \apjl, 856, L5
\bibitem[Falgarone et al.(2010)]{Falgarone2010} Falgarone, E., Ossenkopf, V., Gerin, M., et al.\ 2010, \aap, 518, L118
\bibitem[Falgarone et al.(2017)]{Falgarone2017} Falgarone, E., Zwaan, M.~A., Godard, B., et al.\ 2017, \nat, 548, 430
\bibitem[Feruglio et al.(2010)]{Feruglio2010} Feruglio, C., Maiolino, R., Piconcelli, E., et al.\ 2010, \aap, 518, L155
\bibitem[Feruglio et al.(2017)]{Feruglio2017} Feruglio, C., Ferrara, A., Bischetti, M., et al.\ 2017, \aap, 608, A30
\bibitem[Fielding et al.(2018)]{Fielding2018} Fielding, D., Quataert, E., \& Martizzi, D.\ 2018, \mnras, 481, 3325. doi:10.1093/mnras/sty2466
\bibitem[Fischer et al.(2010)]{Fischer2010} Fischer, J., Sturm, E., Gonz{\'a}lez-Alfonso, E., et al.\ 2010, \aap, 518, L41
\bibitem[Fluetsch et al.(2019)]{Fluetsch2019} Fluetsch, A., Maiolino, R., Carniani, S., et al.\ 2019, \mnras, 483, 4586
\bibitem[Fluetsch et al.(2020)]{Fluetsch2020} Fluetsch, A., Maiolino, R., Carniani, S., et al.\ 2020, arXiv e-prints, arXiv:2006.13232
\bibitem[Foreman-Mackey et al.(2013)]{Foreman-Mackey2013} Foreman-Mackey, D., Hogg, D.~W., Lang, D., et al.\ 2013, \pasp, 125, 306
\bibitem[F{\"o}rster Schreiber \& Wuyts(2020)]{ForsterSchreiber2020} F{\"o}rster Schreiber, N.~M. \& Wuyts, S.\ 2020, \araa, 58, 661. doi:10.1146/annurev-astro-032620-021910
\bibitem[Fujimoto et al.(2017)]{Fujimoto2017} Fujimoto, S., Ouchi, M., Shibuya, T., et al.\ 2017, \apj, 850, 83. doi:10.3847/1538-4357/aa93e6
\bibitem[Gallerani et al.(2018)]{Gallerani2018} Gallerani, S., Pallottini, A., Feruglio, C., et al.\ 2018, \mnras, 473, 1909. doi:10.1093/mnras/stx2458
\bibitem[Garc\'{\i}a-Burillo et al.(2014)]{GarciaBurillo2014} Garc\'{\i}a-Burillo et al., 2014, A\&A, 567, A1 25
\bibitem[Garc\'{\i}a-Burillo et al.(2015)]{GarciaBurillo2015} Garc\'{\i}a-Burillo et al., 2015, A\&A, 580, A3 5
\bibitem[Gentry et al.(2017)]{Gentry2017} Gentry, E.~S., Krumholz, M.~R., Dekel, A., et al.\ 2017, \mnras, 465, 2471
\bibitem[Gerin et al.(2016)]{Gerin2016} Gerin, M., Neufeld, D.~A., \& Goicoechea, J.~R.\ 2016, \araa, 54, 181
\bibitem[Ginolfi et al.(2020)]{Ginolfi2020} Ginolfi, M., Jones, G.~C., B{\'e}thermin, M., et al.\ 2020, \aap, 633, A90. doi:10.1051/0004-6361/201936872
\bibitem[Godard et al.(2009)]{Godard2009} Godard, B., Falgarone, E., \& Pineau Des For{\^e}ts, G.\ 2009, \aap, 495, 847
\bibitem[Godard \& Cernicharo(2013)]{Godard2013} Godard, B., \& Cernicharo, J.\ 2013, \aap, 550, A8
\bibitem[Godard et al.(2014)]{Godard2014} Godard, B., Falgarone, E., \& Pineau des For{\^e}ts, G.\ 2014, \aap, 570, A27
\bibitem[Gonz{\'a}lez-Alfonso et al.(2013)]{GonzalezAlfonso2013} Gonz{\'a}lez-Alfonso, E., Fischer, J., Bruderer, S., et al.\ 2013, \aap, 550, A25. doi:10.1051/0004-6361/201220466
\bibitem[Gonz{\'a}lez-Alfonso et al.(2017)]{GonzalezAlfonso2017} Gonz{\'a}lez-Alfonso, E., Fischer, J., Spoon, H.~W.~W., et al.\ 2017, \apj, 836, 11. doi:10.3847/1538-4357/836/1/11
\bibitem[Gonz{\'a}lez-Alfonso et al.(2018)]{GonzalezAlfonso2018} Gonz{\'a}lez-Alfonso, E., Fischer, J., Bruderer, S., et al.\ 2018, \apj, 857, 66
\bibitem[Governato et al.(2010)]{Governato2010} Governato, F., Brook, C., Mayer, L., et al.\ 2010, \nat, 463, 203
\bibitem[Harrison et al.(2018)]{Harrison2018} Harrison, C.~M., Costa, T., Tadhunter, C.~N., et al.\ 2018, Nature Astronomy, 2, 198. doi:10.1038/s41550-018-0403-6
\bibitem[Heckman et al.(1990)]{Heckman1990} Heckman, T.~M., Armus, L., \& Miley, G.~K.\ 1990, \apjs, 74, 833
\bibitem[Heckman, \& Borthakur(2016)]{Heckman2016} Heckman, T.~M., \& Borthakur, S.\ 2016, \apj, 822, 9
\bibitem[Henriques et al.(2020)]{Henriques2020} Henriques, B.~M.~B., Yates, R.~M., Fu, J., et al.\ 2020, \mnras, 491, 5795
\bibitem[Herrera-Camus et al.(2019)]{Herrera-Camus2019} Herrera-Camus, R., Tacconi, L., Genzel, R., et al.\ 2019, \apj, 871, 37
\bibitem[Herrera-Camus et al.(2020)]{Herrera-Camus2020} Herrera-Camus, R., Janssen, A., Sturm, E., et al.\ 2020, \aap, 635, A47. doi:10.1051/0004-6361/201936434
\bibitem[Herrera-Camus et al.(2020)]{Herrera-Camus2020} Herrera-Camus, R., Janssen, A., Sturm, E., et al.\ 2020, \aap, 635, A47. doi:10.1051/0004-6361/201936434
\bibitem[Herrera-Camus et al.(2021)]{Herrera-Camus2021} Herrera-Camus, R., F{\"o}rster Schreiber, N., Genzel, R., et al.\ 2021, arXiv:2101.05279
\bibitem[Hezaveh et al.(2013)]{Hezaveh2013} Hezaveh, Y.~D., Marrone, D.~P., Fassnacht, C.~D., et al.\ 2013, \apj, 767, 132
\bibitem[Hodge et al.(2015)]{Hodge2015} Hodge, J.~A., Riechers, D., Decarli, R., et al.\ 2015, \apjl, 798, L18. doi:10.1088/2041-8205/798/1/L18
\bibitem[Hollenbach et al.(2012)]{Hollenbach2012} Hollenbach, D., Kaufman, M.~J., Neufeld, D., et al.\ 2012, \apj, 754, 105
\bibitem[Hopkins et al.(2014)]{Hopkins2014} Hopkins, P.~F., Kere{\v{s}}, D., O{\~n}orbe, J., et al.\ 2014, \mnras, 445, 581
\bibitem[Hopkins et al.(2020)]{Hopkins2020} Hopkins, P.~F., Grudi{\'c}, M.~Y., Wetzel, A., et al.\ 2020, \mnras, 491, 3702
\bibitem[Imara et al.(2018)]{Imara2018} Imara, N., Loeb, A., Johnson, B.~D., et al.\ 2018, \apj, 854, 36
\bibitem[Indriolo et al.(2015)]{Indriolo2015} Indriolo, N., Neufeld, D.~A., Gerin, M., et al.\ 2015, \apj, 800, 40
\bibitem[Indriolo et al.(2018)]{Indriolo2018} Indriolo, N., Bergin, E.~A., Falgarone, E., et al.\ 2018, \apj, 865, 127
\bibitem[Ivison et al.(2010)]{Ivison2010} Ivison, R.~J., Smail, I., Papadopoulos, P.~P., et al.\ 2010, \mnras, 404, 198
\bibitem[Jones et al.(2019)]{Jones2019} Jones, G.~C., Maiolino, R., Caselli, P., et al.\ 2019, \aap, 632, L7
\bibitem[Jullo et al.(2007)]{Jullo2007} Jullo, E., Kneib, J.-P., Limousin, M., et al.\ 2007, New Journal of Physics, 9, 447
\bibitem[Jullo \& Kneib(2009)]{Jullo2009} Jullo, E., \& Kneib, J.-P.\ 2009, \mnras, 395, 1319
\bibitem[Kennicutt(1998)]{Kennicutt1998} Kennicutt, R.~C.\ 1998, \araa, 36, 189. doi:10.1146/annurev.astro.36.1.189
\bibitem[Kere{\v{s}} et al.(2009)]{Keres2009} Kere{\v{s}}, D., Katz, N., Dav{\'e}, R., et al.\ 2009, \mnras, 396, 2332
\bibitem[Kneib et al.(1996)]{Kneib1996} Kneib, J.-P., Ellis, R.~S., Smail, I., et al.\ 1996, \apj, 471, 643
\bibitem[Lehnert, \& Heckman(1996)]{Lehnert1996} Lehnert, M.~D., \& Heckman, T.~M.\ 1996, \apj, 462, 651
\bibitem[Lutz et al.(2020)]{Lutz2020} Lutz, D., Sturm, E., Janssen, A., et al.\ 2020, \aap, 633, A134. doi:10.1051/0004-6361/201936803
\bibitem[Ma et al.(2015)]{Ma2015} Ma, B., Cooray, A., Calanog, J.~A., et al.\ 2015, \apj, 814, 17. doi:10.1088/0004-637X/814/1/17
\bibitem[Madau \& Dickinson(2014)]{Madau2014} Madau, P., \& Dickinson, M.\ 2014, \araa, 52, 415
\bibitem[Maiolino et al.(2012)]{Maiolino2012} Maiolino, R., Gallerani, S., Neri, R., et al.\ 2012, \mnras, 425, L66
\bibitem[Maiolino \& Mannucci (2019)]{Maiolino2019} Maiolino, R. and Mannucci, F.,\ 2019, \aapr , 27, 3
\bibitem[Martin(2005)]{Martin2005} Martin, C.~L.\ 2005, \apj, 621, 227
\bibitem[McMullin et al.(2007)]{McMullin2007} McMullin, J.~P., Waters, B., Schiebel, D., et al.\ 2007, Astronomical Data Analysis Software and Systems XVI, 376, 127
\bibitem[Meyer, \& York(1987)]{Meyer1987} Meyer, D.~M., \& York, D.~G.\ 1987, \apjl, 315, L5
\bibitem[Mitchell et al.(2020)]{Mitchell2020} Mitchell, P.~D., Schaye, J., \& Bower, R.~G.\ 2020, \mnras, 497, 4495. doi:10.1093/mnras/staa2252
\bibitem[Murray et al.(2005)]{Murray2005} Murray, N., Quataert, E., \& Thompson, T.~A.\ 2005, \apj, 618, 569
\bibitem[Negrello et al.(2010)]{Negrello2010} Negrello, M., Hopwood, R., De Zotti, G., et al.\ 2010, Science, 330, 800
\bibitem[Negrello et al.(2017)]{Negrello2017} Negrello, M., Amber, S., Amvrosiadis, A., et al.\ 2017, \mnras, 465, 3558. doi:10.1093/mnras/stw2911
\bibitem[Nelson et al.(2019)]{Nelson2019} Nelson, D., Pillepich, A., Springel, V., et al.\ 2019, \mnras, 490, 3234
\bibitem[Neufeld \& Wolfire(2017)]{Neufeld2017} Neufeld, D.~A. \& Wolfire, M.~G.\ 2017, \apj, 845, 163. doi:10.3847/1538-4357/aa6d68
\bibitem[Pereira-Santaella et al.(2018)]{PereiraSantaella2018} Pereira-Santaella, M., Colina, L., Garc{\'\i}a-Burillo, S., et al.\ 2018, \aap, 616, A171. doi:10.1051/0004-6361/201833089
\bibitem[Pereira-Santaella et al.(2020)]{PereiraSantaella2020} Pereira-Santaella, M., Colina, L., Garc{\'\i}a-Burillo, S., et al.\ 2020, \aap, 643, A89. doi:10.1051/0004-6361/202038838
\bibitem[Pillepich et al.(2019)]{Pillepich2019} Pillepich, A., Nelson, D., Springel, V., et al.\ 2019, \mnras, 490, 3196
\bibitem[Planck Collaboration et al.(2016)]{Planck2016} Planck Collaboration, Ade, P.~A.~R., Aghanim, N., et al.\ 2016, \aap, 594, A13
\bibitem[Rangwala et al.(2011)]{Rangwala2011} Rangwala, N., Maloney, P.~R., Glenn, J., et al.\ 2011, \apj, 743, 94
\bibitem[Rees \& Ostriker(1977)]{Rees1977} Rees, M.~J., \& Ostriker, J.~P.\ 1977, \mnras, 179, 541
\bibitem[Riechers et al.(2010)]{Riechers2010} Riechers, D.~A., Capak, P.~L., Carilli, C.~L., et al.\ 2010, \apjl, 720, L131. doi:10.1088/2041-8205/720/2/L131
\bibitem[Riechers et al.(2011)]{Riechers2011} Riechers, D.~A., Hodge, J., Walter, F., et al.\ 2011, \apjl, 739, L31
\bibitem[Riechers et al.(2013)]{Riechers2013} Riechers, D.~A., Bradford, C.~M., Clements, D.~L., et al.\ 2013, \nat, 496, 329. doi:10.1038/nature12050
\bibitem[Riechers et al.(2021a)]{Riechers2021} Riechers, D.~A., Cooray, A., Perez-Fournon, I., et al.\ 2021, arXiv:2101.11006
\bibitem[Riechers et al.(2021b)]{Riechers2021b} Riechers, D.~A., Nayyeri, H., Burgarella, D., et al.\ 2021, \apj, 907, 62. doi:10.3847/1538-4357/abcf2e
\bibitem[Ritchey et al.(2015)]{Ritchey2015} Ritchey, A.~M., Welty, D.~E., Dahlstrom, J.~A., et al.\ 2015, \apj, 799, 197
\bibitem[Rubin et al.(2014)]{Rubin2014} Rubin, K.~H.~R., Prochaska, J.~X., Koo, D.~C., et al.\ 2014, \apj, 794, 156
\bibitem[Rupke et al.(2002)]{Rupke2002} Rupke, D.~S., Veilleux, S., \& Sanders, D.~B.\ 2002, \apj, 570, 588
\bibitem[Rupke et al.(2005)]{Rupke2005} Rupke, D.~S., Veilleux, S., \& Sanders, D.~B.\ 2005, \apjs, 160, 115
\bibitem[Rupke, \& Veilleux(2013)]{Rupke2013} Rupke, D.~S.~N., \& Veilleux, S.\ 2013, \apj, 768, 75
\bibitem[Rupke et al.(2017)]{Rupke2017} Rupke, D.~S.~N., G{\"u}ltekin, K., \& Veilleux, S.\ 2017, \apj, 850, 40
\bibitem[Rybak et al.(2015)]{Rybak2015} Rybak, M., Vegetti, S., McKean, J.~P., et al.\ 2015, \mnras, 453, L26. doi:10.1093/mnrasl/slv092
\bibitem[Schaye et al.(2015)]{Schaye2015} Schaye, J., Crain, R.~A., Bower, R.~G., et al.\ 2015, \mnras, 446, 521
\bibitem[Sharma et al.(2014)]{Sharma2014} Sharma, P., Roy, A., Nath, B.~B., et al.\ 2014, \mnras, 443, 3463
\bibitem[Shopbell \& Bland-Hawthorn(1998)]{Shopbell1998} Shopbell, P.~L. \& Bland-Hawthorn, J.\ 1998, \apj, 493, 129. doi:10.1086/305108
\bibitem[Simcoe et al.(2004)]{Simcoe2004} Simcoe, R.~A., Sargent, W.~L.~W., \& Rauch, M.\ 2004, \apj, 606, 92
\bibitem[Simpson et al.(2012)]{Simpson2012} Simpson, J.~M., Smail, I., Swinbank, A.~M., et al.\ 2012, \mnras, 426, 3201. doi:10.1111/j.1365-2966.2012.21941.x
\bibitem[Simpson et al.(2017)]{Simpson2017} Simpson, J.~M., Smail, I., Wang, W.-H., et al.\ 2017, \apjl, 844, L10. doi:10.3847/2041-8213/aa7cf2
\bibitem[Somerville, \& Primack(1999)]{Somerville1999} Somerville, R.~S., \& Primack, J.~R.\ 1999, \mnras, 310, 1087
\bibitem[Somerville et al.(2018)]{Somerville2018} Somerville, R.~S., Behroozi, P., Pandya, V., et al.\ 2018, \mnras, 473, 2714
\bibitem[Spilker et al.(2016)]{Spilker2016} Spilker, J.~S., Marrone, D.~P., Aravena, M., et al.\ 2016, \apj, 826, 112
\bibitem[Spilker et al.(2018)]{Spilker2018} Spilker, J.~S., Aravena, M., B{\'e}thermin, M., et al.\ 2018, Science, 361, 1016
\bibitem[Spilker et al.(2020a)]{Spilker2020} Spilker, J.~S., Phadke, K.~A., Aravena, M., et al.\ 2020, \apj, 905, 85. doi:10.3847/1538-4357/abc47f
\bibitem[Spilker et al.(2020b)]{Spilker2020b} Spilker, J.~S., Aravena, M., Phadke, K.~A., et al.\ 2020b, \apj, 905, 86. doi:10.3847/1538-4357/abc4e6
\bibitem[Spoon et al.(2013)]{Spoon2013} Spoon et al., 2013, ApJ, 775, 127
\bibitem[Stacey et al.(2021)]{Stacey2021} Stacey, H.~R., McKean, J.~P., Powell, D.~M., et al.\ 2021, \mnras, 500, 3667. doi:10.1093/mnras/staa3433
\bibitem[Stone et al.(2016)]{Stone2016} Stone, M., Veilleux, S., Mel{\'e}ndez, M., et al.\ 2016, \apj, 826, 111. doi:10.3847/0004-637X/826/2/111
\bibitem[Strickland et al.(2004)]{Strickland2004} Strickland, D.~K., Heckman, T.~M., Colbert, E.~J.~M., et al.\ 2004, \apjs, 151, 193
\bibitem[Sturm et al.(2011)]{Sturm2011} Sturm, E., Gonz{\'a}lez-Alfonso, E., Veilleux, S., et al.\ 2011, \apjl, 733, L16
\bibitem[Swinbank et al.(2011)]{Swinbank2011} Swinbank, A.~M., Papadopoulos, P.~P., Cox, P., et al.\ 2011, \apj, 742, 11
\bibitem[van der Tak et al.(2016)]{vdTak2016} van der Tak, F.~F.~S., Wei{\ss}, A., Liu, L., et al.\ 2016, \aap, 593, A43. doi:10.1051/0004-6361/201628120
\bibitem[Thompson et al.(2005)]{Thompson2005} Thompson, T.~A., Quataert, E., \& Murray, N.\ 2005, \apj, 630, 167
\bibitem[Thompson et al.(2015)]{Thompson2015} Thompson, T.~A., Fabian, A.~C., Quataert, E., et al.\ 2015, \mnras, 449, 147
\bibitem[Vayner et al.(2017)]{Vayner2017} Vayner, A., Wright, S.~A., Murray, N., et al.\ 2017, \apj, 851, 126
\bibitem[Veilleux et al.(2005)]{Veilleux2005} Veilleux, S., Cecil, G., \& Bland-Hawthorn, J.\ 2005, \araa, 43, 769
\bibitem[Veilleux et al.(2013)]{Veilleux2013} Veilleux, S., Mel{\'e}ndez, M., Sturm, E., et al.\ 2013, \apj, 776, 27
\bibitem[Veilleux et al.(2020)]{Veilleux2020} Veilleux, S., Maiolino, R., Bolatto, A.~D., et al.\ 2020, \aapr, 28, 2
\bibitem[Walter et al.(2002)]{Walter2002} Walter, F., Weiss, A., \& Scoville, N.\ 2002, \apjl, 580, L21
\bibitem[van der Werf et al.(2010)]{vdWerf2010} van der Werf, P.~P., Isaak, K.~G., Meijerink, R., et al.\ 2010, \aap, 518, L42
\bibitem[Wei{\ss} et al.(2012)]{Weiss2012} Wei{\ss}, A., Walter, F., Downes, D., et al.\ 2012, \apj, 753, 102
\bibitem[Westmoquette et al.(2012)]{Westmoquette2012} Westmoquette, M.~S., Clements, D.~L., Bendo, G.~J., et al.\ 2012, \mnras, 424, 416
\bibitem[White \& Rees(1978)]{White1978} White, S.~D.~M., \& Rees, M.~J.\ 1978, \mnras, 183, 341
\bibitem[Yang et al.(2017)]{Yang2017} Yang, C., Omont, A., Beelen, A., et al.\ 2017, \aap, 608, A144
\bibitem[Zhang et al.(2018)]{Zhang2018} Zhang, Z.-Y., Ivison, R.~J., George, R.~D., et al.\ 2018, \mnras, 481, 59. doi:10.1093/mnras/sty2082


\end{thebibliography}
\end{document}